\documentclass[traditabstract,longauth]{aa}  

\usepackage{graphicx}
\usepackage{txfonts}
\usepackage{float}
\usepackage{times,epsfig,amsmath}
\usepackage{epstopdf}
\usepackage{pdflscape}
\usepackage{adjustbox}
\usepackage{amssymb, amsmath} 
\usepackage[usenames,dvipsnames]{color}
\usepackage{txfonts}
\usepackage{microtype}
\usepackage{url}
\usepackage[bookmarks, colorlinks, breaklinks]{hyperref}  
\hypersetup{linkcolor=blue,citecolor=MidnightBlue,filecolor=black,urlcolor=MidnightBlue} 
\usepackage{booktabs}
\usepackage{subfig}
\usepackage{commath}
\usepackage{natbib}
\bibpunct{(}{)}{;}{a}{}{,} 
\usepackage[bookmarks, colorlinks, breaklinks, ]{hyperref}  
\usepackage{tablefootnote}
\usepackage{pdfpages}

\newcommand{\chandra}{{\it Chandra}}
\newcommand{\asca}{{\it ASCA}}
\newcommand{\sax}{{\it BeppoSAX}}

\newcommand{\planck}{{\it Planck}}
\newcommand{\xmm}{{XMM-{\it Newton}}}
\newcommand{\xmms}{{\it XMM}}

\newcommand{\suzaku}{{\it Suzaku}}

\begin{document} 
\title{CHEX-MATE: Robust reconstruction of temperature profiles in galaxy clusters with XMM-Newton}
 \author{M. Rossetti
          \inst{1}
          \and D. Eckert \inst{2}
          \and F. Gastaldello \inst{1} 
         \and E. Rasia \inst{3,4}
         \and  G.W. Pratt \inst{5}
        \and S. Ettori \inst{6,7}
          \and S. Molendi \inst{1}
          \and M. Arnaud \inst{5}
          \and M. Balboni \inst{1,8}
          \and I. Bartalucci \inst{1}
          \and R.M. Batalha \inst{5}
          \and S. Borgani \inst{3,4,9}
          \and H. Bourdin \inst{10,11}
          \and S. De Grandi \inst{12}
          \and F. De Luca \inst{10,11}
          \and M. De Petris \inst{13}
          \and W. Forman \inst{14}
          \and M. Gaspari \inst{15}
          \and S. Ghizzardi \inst{1}
          \and A. Iqbal \inst{5}
          \and S. Kay\inst{16}
          \and L. Lovisari \inst{1,14}
          \and B.J. Maughan \inst{17}
          \and P. Mazzotta\inst{10,11}
          \and E. Pointecouteau \inst{18}
          \and G. Riva \inst{1,19}
          \and J. Sayers\inst{20}
          \and M. Sereno\inst{6,7}
          }
\authorrunning{Rossetti et al.}

   \institute{INAF, IASF-Milano, via A.Corti 12, 20133 Milano, Italy\\
              \email{mariachiara.rossetti@inaf.it}
         \and
         Department of Astronomy, University of Geneva, Ch. d’Ecogia 16, CH-1290 Versoix, Switzerland\label{inst2}
    \and INAF, Osservatorio di Trieste, via Tiepolo 11, 34131, Trieste, Italy\label{inst3} 
   \and IFPU, Institute for Fundamental Physics of the Universe, Via Beirut 2, 34014 Trieste, Italy \label{inst4} 
   \and Universit\'e Paris-Saclay, Universit\'e Paris Cit\'e, CEA, CNRS, AIM, 91191, Gif-sur-Yvette, France \label{inst5}
    \and INAF, Osservatorio di Astrofisica e Scienza dello Spazio, via Piero Gobetti 93/3, 40129 Bologna, Italy \label{inst6}  
    \and INFN, Sezione di Bologna, viale Berti Pichat 6/2, 40127 Bologna, Italy \label{inst7} 
    \and DiSAT, Universit\`a degli Studi dell’Insubria, via Valleggio 11, I-22100 Como, Italy \label{inst8}   
    \and Astronomy Unit, Department of Physics, University of Trieste, Via Tiepolo 11, 34131 Trieste, Italy \label{inst9}
    \and Universit\`a degli studi di Roma ‘Tor Vergata’, Via della ricerca scientifica, 1, 00133 Roma, Italy \label{inst10}
    \and INFN, Sezione di Roma 2, Universit\`a degli studi di Roma Tor Vergata, Via della Ricerca Scientifica, 1, Roma, Italy \label{inst11}
    \and INAF, Osservatorio Astronomico di Brera, Via E. Bianchi 46, 23807 Merate, Italy \label{inst12}
    \and Dipartimento di Fisica, Sapienza Universit\`a di Roma, Piazzale Aldo Moro 5, 00185 Roma, Italy \label{inst13}
    \and Center for Astrophysics | Harvard \& Smithsonian, 60 Garden Street, Cambridge, MA 02138, USA \label{inst14}
    \and Department of Astrophysical Sciences, Princeton University, Princeton, NJ 08544, USA \label{inst15}
    \and Jodrell Bank Centre for Astrophysics, Department of Physics and Astronomy, The University of Manchester, Alan Turing Building, Manchester M13 9PL, UK \label{inst16}
    \and HH Wills Physics Laboratory, University of Bristol, Tyndall Ave, Bristol, BS8 1TL, UK \label{inst17}
    \and IRAP, Universit\'e de Toulouse, CNRS, CNES, UPS, 9 av. du Colonel Roche, BP44346, 31028 Toulouse Cedex 4, France \label{inst18}
    \and Dipartimento di Fisica, Universit\`a degli Studi di Milano, Via G. Celoria 16, 20133 Milano, Italy \label{inst19}
    \and California Institute of Technology, 1200 East California Boulevard, Pasadena, California 91125, USA \label{inst20}
             }
             
\abstract{
The ``Cluster HEritage project with \xmm: Mass Assembly and Thermodynamics at the Endpoint of structure formation'' (CHEX-MATE) is a multi-year Heritage program, to obtain homogeneous \xmm\ observations of a representative sample of 118 galaxy clusters. The observations are tuned to reconstruct the distribution of the main thermodynamic quantities of the ICM up to $R_{500}$ and to obtain individual mass measurements, via the hydrostatic-equilibrium equation, with a precision of 15-20\%. Temperature profiles are a necessary ingredient for the scientific goals of the project and it is thus crucial to derive the best possible temperature measurements from our data. This is why we have built a new pipeline for spectral extraction and analysis of \xmm\ data, based on a new physically motivated background model and on a Bayesian approach with Markov Chain Monte Carlo (MCMC) methods, that we present in this paper for the first time. We applied this new method to a subset of 30 galaxy clusters representative of the CHEX-MATE sample and show that we can obtain reliable temperature measurements up to regions where the source intensity is as low as 20\% of the background, keeping systematic errors below 10\%. We compare the median profile of our sample and the best fit slope at large radii with literature results and we find a good agreement with other measurements based on \xmm\ data. Conversely, when we exclude from our analysis the most contaminated regions, where the source intensity is below 20\% of the background, we find significantly flatter profiles, in agreement with predictions from numerical simulations and  independent measurements with a combination of Sunyaev-Zeldovich and X-ray imaging data.}
   \maketitle
\section{Introduction}
Galaxy clusters represent the endpoint of structure formation in the Universe: they are the most massive virialized structures to have formed thus far and are located at the nodes of the cosmic web, continuously accreting matter from the filaments. The intra-cluster medium (ICM) constitutes their main baryonic component and is heated to X-ray emitting temperatures ($10^7-10^8$ K) during the formation and accretion processes, driven by dark matter. As such, the ICM carries important information on the physical processes of structure formation and on the global cluster properties: ICM thermodynamic properties correlate well with the total gravitational mass and properly scaled radial profiles of thermodynamic quantities are nearly universal, reflecting the properties of the underlying dark matter structure \citep{kaiser86}. Deviations from these relations do exist and are a clear signspot of important effects such as cooling, non-gravitational feedback from Active Galactive Nuclei (AGN) or supernovae, bulk motions and turbulence induced by accretion processes  (see \citealt{lovisarimaughan} for a recent review, 
\citealt{gaspari14, bulbul19,sereno20,poon23}). \\
The ICM temperature is one of the key observable quantities that can be derived from X-ray observations of galaxy clusters, along with density and metal abundance. It is derived from the analysis of X-ray spectra, mainly through the position of the {\it Bremsstrahlung} exponential cut-off. In the last 30 years, the previous and present generation of X-ray satellites have allowed spatially resolved spectral measurements, mapping the distribution of ICM temperatures from the cores to the external regions.  Temperature profiles in radial annuli are a necessary ingredient to derive the radial distributions of thermodynamic quantities, such as pressure and entropy, and to reconstruct the total mass, through the hydrostatic equilibrium equation. Early temperature profiles with \asca\ and \sax\ \citep[e.g.][]{fabian94, degrandimolendi01} unambiguously demonstrated that some clusters do show a temperature decline in their inner regions, consistent with a short cooling time of the high-density central regions and the prediction of the cooling flow model \citep{fabian84,fabian94_rev}. With the fall of this model, after high resolution spectra with \xmm\ RGS showed that the gas does not cool below a critical value \citep[e.g.][]{peterson01,peterson03}, these systems were dubbed Cool Cores (CC, \citealt{molendipizzo01}) and showed to be typically associated with clusters in a relaxed dynamical state. Conversely, in the external regions, \asca\ and \sax\ observations, lead to controversial results on the presence of a temperature gradient \citep[e.g.][]{markevitch98,irwin99,white00,degrandimolendi01}. This is due to technical challenges in measuring temperature profiles in the external regions of galaxy clusters, where the intensity of the source is low with respect to contamination from other brighter cluster regions because of the large Point Spread Function (PSF) and of several background components (see \citealt{ettori11} for a review). The former issue has been resolved with the advent of the current generation of X-ray telescopes \xmm\ and \chandra, which unambiguously showed a declining trend of the temperature profiles with radius in the external regions \citep[e.g.][]{vikhlinin06,pratt07,snowden08,leccardi08}. However, the low intensity of the cluster emission with respect to the instrumental and celestial background and foreground is still a major challenge, hampering reliable temperature measurements up to the virial radius of galaxy clusters and thus leaving a large fraction of the cluster volume poorly explored. In the last decade, the \suzaku\ satellite, benefiting from a low-earth orbit and thus a lower particle background with respect to \xmm\ and \chandra,  allowed us to extend temperature measurements up to the virial radius, for a limited sample of a dozen clusters. The steep decrease in the temperature profile and flattening in the entropy at the virial radius observed in some clusters are still under debate (see the review by \citealt{walker19} and references therein). \\
An independent method, that avoids the issues of X-ray spectroscopy, for deriving temperature profiles is the combination of the spatially resolved Sunyaev-Zeldovich (SZ, \citealt{sz1972}) signal, which is proportional to the integral of the ICM pressure along the line of sight, with the gas density, which can be derived from X-ray images and is less dependent on the background treatment than the temperature. This approach has been attempted for a few clusters \citep[e.g.][]{basu10,eckert13,tchernin16,pratt16,adam17, sereno18,ghirardini21}. In the X-COP project, \citet{ghirardini19} compared the temperature profiles obtained with the combination of \planck\ and \xmm\ data with those obtained with the purely spectroscopic \xmms\ analysis and found them to be consistent in the radial range where the two methods overlap, i.e. within $R_{500}$. While this approach proved very successful for measuring temperatures in the outskirts of galaxy clusters, it is so far limited to clusters with high enough SZ signal and extension in the sky matching the capabilities of the SZ instruments (nearby extended objects for low resolution instruments like \planck\ or compact systems for high resolution instruments like NIKA2 and ALMA). Moreover, the spatial resolution of spectroscopic X-ray temperature profiles is still superior to that obtained with the joint SZ-X-ray analysis. \\ 
The use of radial profiles for studying the temperature structure of the ICM is of course an approximation: several authors have shown two-dimensional maps of the ICM temperatures, with significant non-radial variations \citep[e.g.][]{bourdin08,franck13,lagana19}. Nonetheless, the spherical symmetry approximation is necessary for reconstructing the total mass with hydrostatic equilibrium equation and in conditions of limited statistics, such as in faint and distant clusters or in the external regions. Recently, \citet{lovisari23} quantified the impact of temperature inhomogeneities on the reconstructed radial profiles to be below 5\% on average, but reaching larger values (10-20\%) in external and peculiar regions. \\
The ``Cluster HEritage project with \xmm: Mass Assembly and Thermodynamics at the Endpoint of structure formation'' (CHEX-MATE, \citealt{paper1}) is a multi-year Heritage program with \xmm\ awarded in 2017. It consists of a homogeneous set of observations of a representative sample of 118 galaxy clusters, extracted from the \planck\ SZ catalogue (PSZ2, \citealt{PSZ2}). It aims at deriving the total masses, through the hydrostatic equilibrium equation, and characterizing the thermodynamic properties of the full sample \citep[see][for more details]{paper1}. The observations are tailored to reach a precision of 15-20\% on the hydrostatic masses within $R_{500}$. In this context, the temperature profile is a fundamental ingredient of the CHEX-MATE analysis and the desired precision on the mass is translated into a 15\% precision on the temperature measurement at $R_{500}$, i.e. the radius within which the mean density is 500 times the critical density of the Universe at the cluster's redshift . Any issue affecting the temperature profiles will be inevitably propagated into the final CHEX-MATE results, and it is thus crucial for our project to obtain the most accurate temperature profiles from our data and assess the impact of our choices and of possible systematics on the temperature measurements. This is why we developed a new pipeline for spectral extraction and analysis (Sec.\ref{sec:pipeline}), with a special focus on the background model (Sec. \ref{sec:physmodel}), that we present in this paper for the first time. We applied this method to a sample of 30 objects (Sec \ref{sec:sample}) that we used to test it and evaluate the impact of our analysis on the temperature measurements (Sec. \ref{sec:sys}). We present the first results on the temperature profiles for this sample in Sec. \ref{sec:results} and we discuss our findings in Sec. \ref{sec:discussion}.\\     
In our paper, we assume a flat $\Lambda$CDM cosmology with $H_0=70\ \rm{km}\ \rm{s}^{-1}\ \rm{Mpc}^{-1}$, $\Omega_M=0.3$ and $\Omega_{\Lambda}=0.7$. Errors are given at 68\% confidence level, unless otherwise stated.  Whenever we use the notation $M_{500}$ in this paper, we refer to the masses derived from the \planck\ SZ signal with the method described in \citet{PSZ2}, using the MMF3 algorithm \citep{melin06}, as discussed in \citet{paper1}.

\section{The sample}
\label{sec:sample}

\begin{figure}[!ht]
    \centering
    \includegraphics[width=0.45\textwidth]{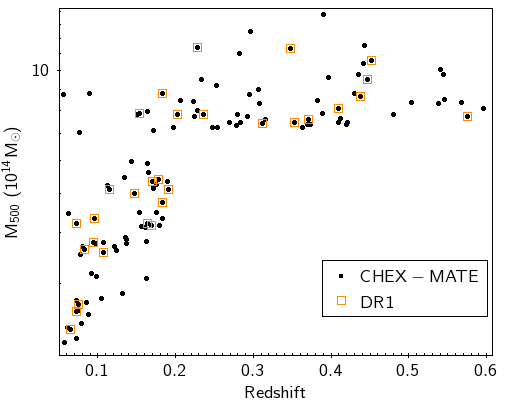}
    \caption{Distribution of the CHEX-MATE (black circles) and DR1 (orange open squares) objects in the mass-redshift plane, showing that our subsample is representative of the original selection.} 
    \label{fig:MZplane}
\end{figure}

We built our sample to fulfill two different requirements: i) being a ``technical'' sample, allowing us to test our pipeline and new methods with a limited number of objects under different analysis conditions (source angular extension, background levels, etc.); ii) being representative of the full CHEX-MATE sample, in terms of its selection quantities (mass, redshift, and {\it Planck} SNR). We dubbed it ``Data Release 1'' (DR1 hereafter) because it is the first sample for which results have been internally released to the CHEX-MATE collaboration.
To build the sample, we selected clusters in the $M-z$ plane (Fig. \ref{fig:MZplane}), looking also at the distribution of other physical parameters, such as the extension in the sky as measured by $R_{500}$
(ranging from $2.9$ to $13.5$ arcminutes) and Galactic absorption ($N_H$ spanning one order of magnitude from $1.03\times 10^{20}\, \rm{cm}^{-2}$ to $1.02\times 10^{21}\, \rm{cm}^{-2}$) with different levels of molecular contribution\citep{bourdin23}.  We also looked at the distribution of the quality assessement indicators that we computed for each observation (see Sec. \ref{sec:data_reduction}), making sure that the selected observations encompass different levels of cosmic-ray induced particle background (Sec. \ref{sec:particlebkg}), following the solar cycle \citep{marelli21,gastaldello22}, and of the residual focused component (Sec. \ref{sec:SPbkg}, \citealt{salvetti17}). We also checked that the clean exposure time of the selected observations allows us to meet the feasibility requirements of the CHEX-MATE project.\\
We confirm our sample definition by checking that the DR1 sample is representative of the original CHEX-MATE selection \citep{paper1}, by performing Kolmogorov-Smirnov (KS) tests on the distribution of mass, redshift, or \planck\ S/N. 
Since we excluded from our selection double clusters that would require an ad-hoc strategy in the definition of regions for the radial profile (Sec. \ref{sec:spectral_extract}), we further checked that we did not preferentially select relaxed clusters, by performing a KS test on the distribution of the $M$ morphological parameter, which combines the information from the light concentration and centroid shift parameters and from the second and third order power ratios (see \citealt{campitiello22} for details). The test returns a very high probability ($>99\%$) that DR1 is representative of the morphological distribution of the original CHEX-MATE sample. \\
The final DR1 sample is composed of 30 clusters, listed in Table \ref{tab:allclusters}, along with their main physical properties and the \xmm{} observations that we used in our analysis. 
We show their distribution in the mass-redshift plane along with the parent CHEX-MATE sample in Fig. \ref{fig:MZplane}. We recall that masses here are estimated from the \planck{} SZ data as described in \citet{PSZ2} using the $Y_{SZ}-M_{500}$ scaling relation calibrated with \xmm{} data and are not corrected for any hydrostatic bias.\\
We show the images of all clusters in our sample in Appendix \ref{app:gallery}.
\begin{table*}
\centering
\caption{List of the clusters of the DR1 sample. We quote: the PSZ2 name, the redshift, the nominal integrated mass from \planck{} MMF3 data, the cluster size $R_{500}$ in arcminutes, as derived from the mass with Eq. 9 in \citet{PSZ1cosmo}, the Tier to which each cluster belongs (see \citealt{paper1} for details) and the \xmm{} observations used in the analysis}

\begin{tabular}{l c c c c l}
\hline
\hline
  \multicolumn{1}{c}{PSZ2 Name} &
  \multicolumn{1}{c}{Redshift} &
  \multicolumn{1}{c}{$M_{500}$} &
  \multicolumn{1}{c}{$R_{500}$} &
  \multicolumn{1}{c}{Tier} &
  \multicolumn{1}{c}{XMM obsid} \\
  \multicolumn{1}{c}{ } &
  \multicolumn{1}{c}{ } &
  \multicolumn{1}{c}{$10^{14} M_\odot$} &
  \multicolumn{1}{c}{arcmin} &
  \multicolumn{1}{c}{ } &
  \multicolumn{1}{c}{center/offset} \\
\hline
PSZ2 G008.31-64.74 & 0.312 & 7.42 & 4.51 & 2 & 0827010901\\
PSZ2 G041.45+29.10 & 0.178 & 5.41 & 6.48 & 1 & 0601080101\\
PSZ2 G042.81+56.61 & 0.072 & 4.22 & 13.50 & 1 & 0202080201/0827361101\\
PSZ2 G046.88+56.48 & 0.115 & 5.10 & 9.40 & 1 & 0827010601\\
PSZ2 G050.40+31.17 & 0.164 & 4.22 & 6.40 & 1 & 0827040101\\
PSZ2 G056.77+36.32 & 0.095 & 4.34 & 10.53 & 1 & 0740900101\\
PSZ2 G056.93-55.08 & 0.447 & 9.49 & 3.70 & 2 & 0503490201\\
PSZ2 G057.78+52.32 & 0.065 & 2.32 & 12.15 & 1 & 0827040301/0827041801\\
PSZ2 G057.92+27.64 & 0.076 & 2.66 & 11.08 & 1 & 0827030301\\
PSZ2 G066.41+27.03 & 0.575 & 7.69 & 2.88 & 2 & 0827320601\\
PSZ2 G072.62+41.46 & 0.228 & 11.43 & 6.73 & 2 & 0605000501\\
PSZ2 G077.90-26.63 & 0.147 & 4.99 & 7.46 & 1 & 0827020101\\
PSZ2 G083.86+85.09 & 0.183 & 4.74 & 6.04 & 1 & 0827030701\\
PSZ2 G113.29-29.69 & 0.107 & 3.57 & 8.86 & 1 & 0827021201\\
PSZ2 G113.91-37.01 & 0.371 & 7.58 & 3.96 & 2 & 0827021001\\
PSZ2 G114.79-33.71 & 0.094 & 3.79 & 10.20 & 1 & 0827320401\\
PSZ2 G149.39-36.84 & 0.170 & 5.35 & 6.71 & 1 & 0827030601\\
PSZ2 G195.75-24.32 & 0.203 & 7.80 & 6.53 & 2 & 0201510101\\
PSZ2 G207.88+81.31 & 0.353 & 7.44 & 4.09 & 2 & 0827020301\\
PSZ2 G224.00+69.33 & 0.190 & 5.11 & 5.99 & 1 & 0827020901\\
PSZ2 G238.69+63.26 & 0.169 & 4.17 & 6.21 & 1 & 0500760101\\
PSZ2 G243.15-73.84 & 0.410 & 8.09 & 3.75 & 2 & 0827011301\\
PSZ2 G243.64+67.74 & 0.083 & 3.62 & 11.23 & 1 & 0827010801\\
PSZ2 G277.76-51.74 & 0.438 & 8.65 & 3.65 & 2 & 0674380301\\
PSZ2 G287.46+81.12 & 0.073 & 2.56 & 11.33 & 1 & 0149900301\\
PSZ2 G313.33+61.13 & 0.183 & 8.77 & 7.42 & 2 & 0093030101\\
PSZ2 G313.88-17.11 & 0.153 & 7.86 & 8.37 & 2 & 0692932001\\
PSZ2 G324.04+48.79 & 0.452 & 10.58 & 3.81 & 2 & 0112960101\\
PSZ2 G340.94+35.07 & 0.236 & 7.80 & 5.76 & 2 & 0827311201\\
PSZ2 G349.46-59.95 & 0.347 & 11.36 & 4.77 & 2 & 0504630101\\
\hline
\end{tabular}
\label{tab:allclusters}
\end{table*}

\section{The CHEX-MATE pipeline for spectral results}
\label{sec:pipeline}
In this Section and in Sec. \ref{sec:physmodel}, we present in details the pipeline that we developed to extract and fit spectra in radial profiles for the CHEX-MATE project. We tested it with other pipelines available in the collaboration and used for other CHEX-MATE projects (\citealt{bourdin23,lovisari23}, De Luca et al. in prep.). In the internal regions of galaxy clusters ($R<0.5R_{500}$) the temperatures estimated with different analysis are consistent within 1-2\%. Differences raise to about 10\% at $R_{500}$, where the details of the background modeling have a large impact on the temperature measurements.

\subsection{Data Reduction}
\label{sec:data_reduction}
We performed the Data Reduction and analysis making use of the \xmm{} Science Analysis System (SAS), version $16.1$ and the Extended Source Analysis Software (ESAS, \citealt{snowden08,kuntz08}) embedded within SAS. We keep the calibration database (CalDB) updated to ensure that we apply the latest calibration files to our data, when we reprocess them starting from the raw ODF files retrieved from the archive.
We tested more recent SAS versions ($19.1$ and $20.0$), but the tasks are either much slower than with SAS $16.1$, thus requiring a too long computing time for a large data-set as CHEX-MATE, either show unresolved bugs. We are supported in our choice by the EPIC calibration status document\footnote{https://xmmweb.esac.esa.int/docs/documents/CAL-TN-0018.pdf}, showing that no major change has been applied in the meantime, with the only exception of the optional cross-calibration with NUSTAR\footnote{https://xmmweb.esac.esa.int/docs/documents/CAL-SRN-0388-1-4.pdf} that we are not using in our pipeline. \\
We refer to \citet{bartalucci23} for details on the first steps of the data reduction, including calibration, standard pattern cleaning, removal of noisy CCDs in the MOS cameras and light-curve filtering. We emphasize that for our analysis we used the same selection of contaminating sources as in \citet{bartalucci23}, and thus we mask both point sources and extended substructures. The latter include known merging substructures, such as in PSZ2G046.88+56.48, a.k.a A2069, where the northern subcluster A2069B \citep[e.g.][]{owers09} has been masked in our analysis, and  PSZ2G195.75-24.32, a.k.a. A520, where we masked the merging subcluster, dubbed as ``the foot'' in \citet{wang16} and leading to the formation of a shock front \citep{markevitch05}, and other extended features whose origin is still unclear (in PSZ2G056.77+36.32, PSZ2G057.78+52.31, PSZ2G057.92+27.64).  Masked sources and regions for all clusters are shown in Appendix \ref{app:gallery}.\\
While reducing each observation, we also compute two indicators that allow us to estimate the level of instrumental background affecting them. The first one is the count-rate in the region of the MOS2 detector not exposed to the sky (i.e. outside the Field Of View, $outFOV$), 
which has been shown to be a good proxy of the background due to the interaction of high-energy ($>10$ MeV) cosmic ray particles, modulated by the solar cycle (Sec. \ref{sec:particlebkg}, see \citealt{gastaldello22}). 
We calculate this parameter following the prescriptions (detector, energy band and region definition) provided in \citet{marelli17}. The second indicator ($inFOV-outFOV$) is the difference between the count-rate in the region exposed to X-ray photons of the MOS2 detector ($inFOV$) and the $outFOV$ value, both computed in a hard band\footnote{As described in \citet{marelli17}, we use events with $E>7$ keV, excluding the energy ranges $[9.4-10]$ keV and $[11-12]$ keV, affected by instrumental lines.} to minimize the celestial contribution. This allows us to estimate the excess background in the FOV with respect to the background level predicted by the $outFOV$ value \citep[e.g.][]{salvetti17}. We will call this component ``Residual Focused Component'', as it is at least partly due to contamination from quiescent soft protons not filtered in the light curve cleaning 
but may also have a different origin \citep[see][]{salvetti17}.
To compute the $inFOV-outFOV$ indicator we could not use the count rate in the full MOS2 FOV as in \citet{marelli17} and \citet{salvetti17} because of the presence of the cluster emission in our observation. We thus used an external annulus between 12 and 14 arcminutes from the aimpoint which, in combination with the hard energy band, minimizes the contribution of the cluster emission. 
These indicators have been very useful to assess the quality of our observations. We will discuss in Sec. \ref{sec:physmodel} how we used them to build our background model.

\subsection{Spectral extraction in annuli}
\label{sec:spectral_extract}
We define regions for the spectral extraction as concentric annuli centered on the peak of the X-ray image. We automatically define the position of the peak as the brightest pixel in the soft band image ($0.7-1.2$ keV), after correcting for the exposure map, masking the point sources, and smoothing with a Gaussian function with $\sigma=5$ pixels (see \citealt{bartalucci23} for more details). We compute the width of each annulus with an adaptive binning method to reach a constant signal to noise ratio ($S/N=50$ in the $0.3-2$ keV energy range) in each spectral bin \citep[e.g.][] {pratt10, chen23}. \\
For each region, we extract spectra, redistribution matrix files (RMF), and ancillary response files (ARF) with the ESAS tools {\tt mos-spectra} and {\tt pn-spectra}. Following the ESAS procedure \citep{kuntz08,snowden08}, these tools also compute the count rate in the unexposed corners ($outFOV$) and select the most similar (in terms of magnitude and hardness ratio, see \citealt{kuntz08} for details) filter-wheel-closed observations in the \xmm{} calibration database. A spectrum for the cosmic-ray induced particle background (CRPB) is then produced for each region with the {\tt mos-back} and {\tt pn-back} tools. We use these ESAS products to build our background model as shown in Sec. \ref{sec:particlebkg}. \\
The regions for spectral extraction can be very small (width as small as $3.3\arcsec$) in the central regions of galaxy clusters with a peaked surface brightness profile. In these conditions, we noticed an error in the {\tt arfgen} and {\tt rmfgen} tools called by {\tt mos/pn-spectra} producing null ARFs and RMFs. To fix this issue, we extract RMFs, ARFs in a circular region with radius 30\arcsec around the peak and we assign them to all regions in the inner 30\arcsec. Since the statistical quality of the background files in small regions is very poor, we also use the appropriately rescaled CRPB spectra in the circular region with $R=30\arcsec$ for all central regions.
Both the effective area and the particle background intensity do not show significant variations on scales smaller than 1 \arcmin.\\
Finally we extracted a spectrum from an external ``background'' region, ideally free of cluster emission, which is needed to calibrate our model of the sky background components (Sec. \ref{sec:skybkg}).
The definition of this region is not trivial in a sample where the extension of the cluster emission in the FOV is very different ($R_{500}$ ranging from $2.9\arcmin$ to $13.5\arcmin$). For all the clusters with $R_{500}\le 9\arcmin$, we define an annulus between $R_{200}= R_{500}/0.7$ and the edge of the FOV. This is also possible 
for the two objects (PSZ2\,G$057.78+53.31$ and PSZ2\,G$042.81+56.61$) with  $R_{500} >12\arcmin$ for which we requested an offset observation for this purpose \citep{paper1}. Conversely, for the six clusters with $9\arcmin<R_{500}<12\arcmin$ we define a ``close background (closebkg)'' region as an annulus between $1.1R_{500}$ and the FOV edge. We separate the residual cluster emission from the celestial background components, making use of ancillary data from the ROSAT All Sky Survey (RASS) diffuse background map \citep{snowden95,snowden97}, as described in detail in Sec. \ref{sec:skybkg}. We will discuss the impact of this strategy in Sec. \ref{sec:sys_RASS}.

\subsection{Spectral fitting}
\label{sec:spectral_fit}
\begin{figure*}
    \centering
    \hspace{-1cm}
    \includegraphics[angle=-90,width=0.41\hsize]{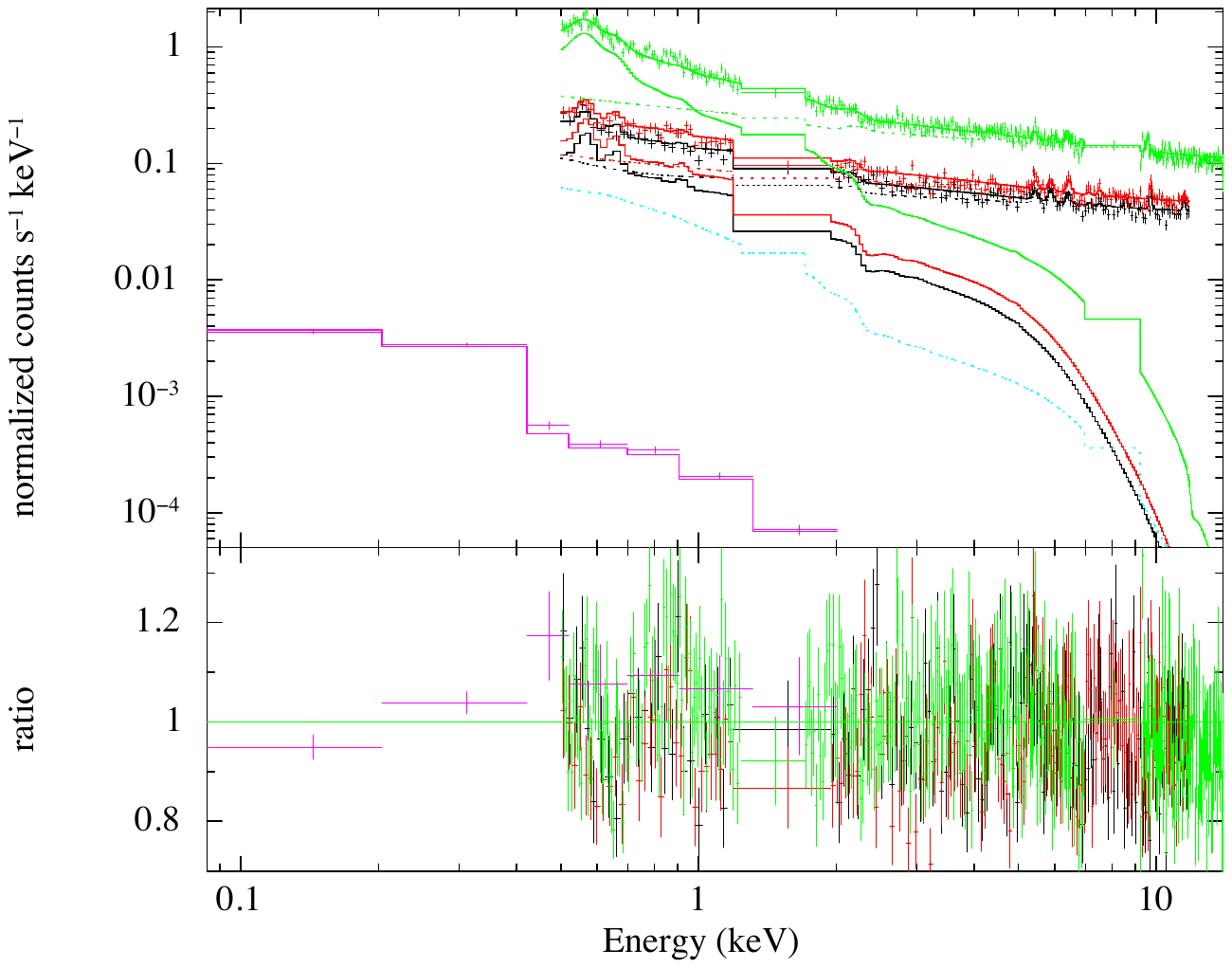}
    \hspace{-2cm}
    \includegraphics[angle=-90,width=0.41\hsize]{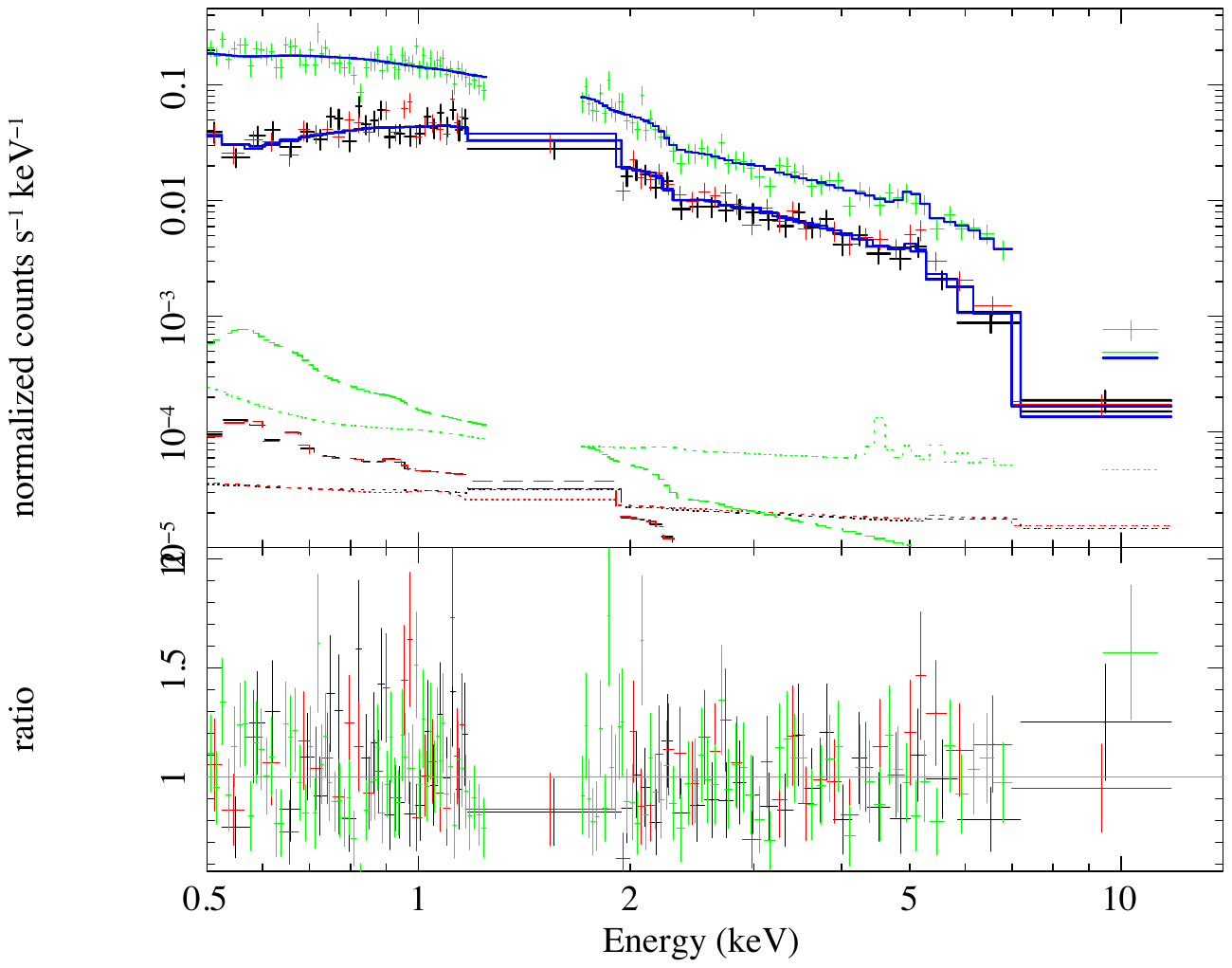}
     \hspace{-2cm}
    \includegraphics[angle=-90,width=0.41\hsize]{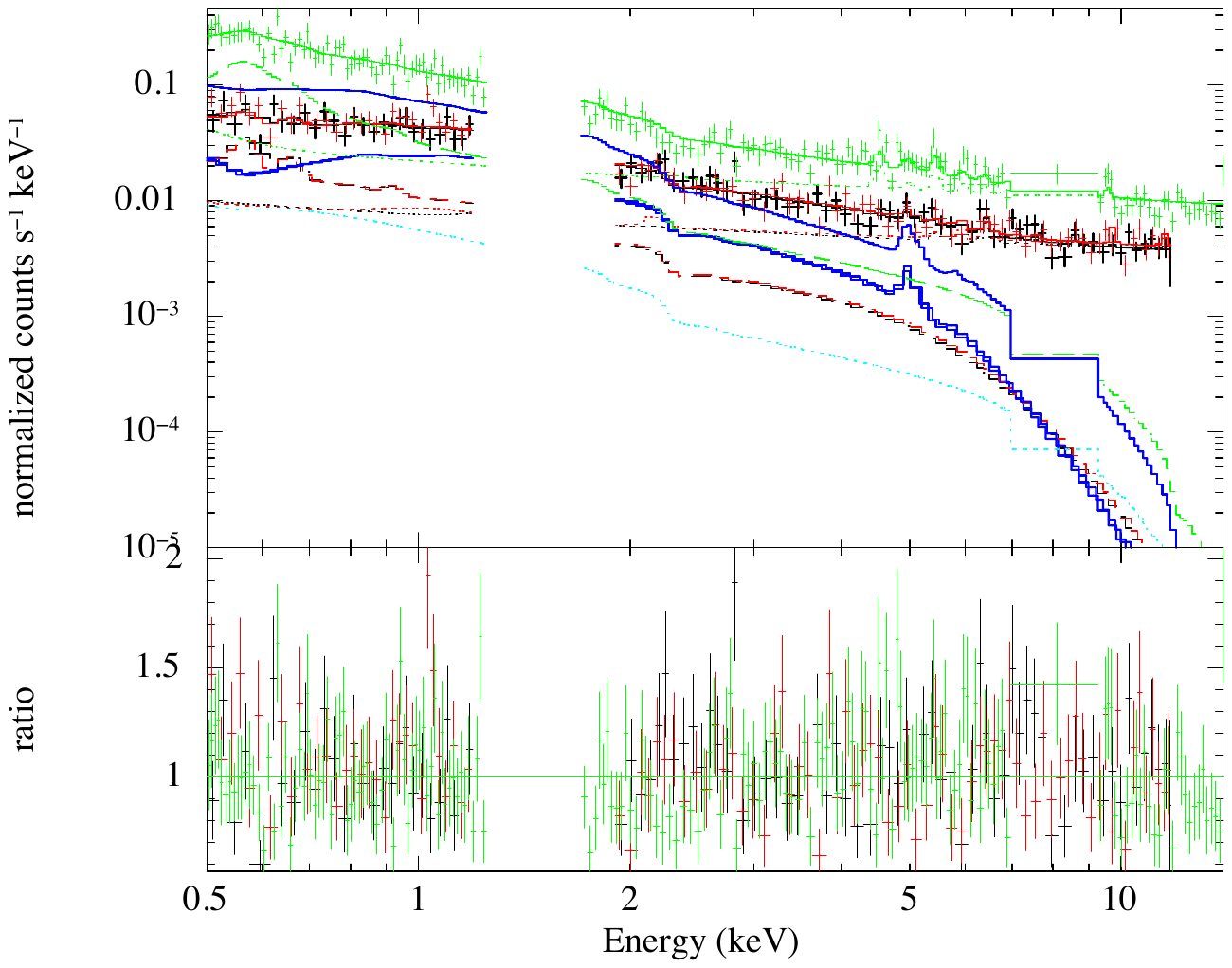} 
    \caption{Examples of our spectral fits for the cluster PSZ2 G349.46-59.95 in different regions: the external region where we fit the sky background ({\it left}), a central region dominated by the cluster emission ({\it middle}), and an external background-dominated region with $SOU/BKG=0.4$ ({\it right}). In all panels, the black, red, and green crosses mark the MOS1, MOS2, and pn spectra, while the lines represent the corresponding models. With the same color scheme, the continuous lines mark the total resulting model for each detector, the dotted lines represent the model for particle background, while the dashed lines the sky background. The dotted cyan line (when visible) mark the contribution of the pn OoT events. In the middle and right panel, we highlight in blue the cluster component. In the left panel, magenta crosses represent the data from the RASS spectrum, which is jointly fitted to the EPIC data in our baseline pipeline. In all plots, the bottom panels mark the ratio of the data with respect to the total model. }
    \label{fig:example_spectra}
\end{figure*}
For each region, we joint-fit the spectra of the three detectors and eventually of multiple observations covering the same region within XSPEC (version 12.11). We show some examples of our fits in Fig. \ref{fig:example_spectra}. We link together the parameters of the models which are applied to the three instruments and to all observations, taking into account the different areas by using a constant factor. Indeed, all the normalization terms in our fits are per unit area (in arcmin$^2$).
Unless otherwise stated, we perform our fits in the band $0.5-12$ keV for the MOS detectors and $0.5-14$ keV for the pn. From these bands, we further exclude the energy ranges corresponding to the more prominent fluorescent lines, namely Al and Si in the range $1.2-1.9$ keV for the MOS detectors and Al in $1.2-1.7$ keV and Cu in $7-9.2$ keV for the pn. 
As in \citet{leccardi08} and \citet{ghirardini19}, we do not subtract any background component in our fit but we model them with a physical model described in Sec. \ref{sec:physmodel} that we apply to the spectra in each region. We estimate the parameters of the celestial components by fitting the spectra in the external background region (see Sec. \ref{sec:spectral_extract}), with the model described in Sec. \ref{sec:skybkg}.
Also the pn Out-of-Time (OoT) events are treated as a further background component: we extract a spectrum from the OoT event files in each region as described in Sec. \ref{sec:spectral_extract} and we fit it with a {\tt phabs(bapec+powerlaw)} model, to mimic the cluster contribution (but with a degraded energy resolution) and the CXB, and another power-law for the particle background. We then properly rescale the best fit model of the OoT contribution, taking into account the operating mode (Full Frame or Extended Full Frame). \\
The baseline method for our fits makes use of Bayesian statistics and of the Markov Chain Monte Carlo (MCMC) algorithm \citep{goodmanweare} within XSPEC. To assess and test it, we also perform more standard fits with a standard Maximum Likelihood approach (ML).
In both cases, we use the Cash statistics \citep{cash79} within XSPEC, as it is more appropriate than $\chi ^2$ for Poissonian data, especially in conditions of high background and poor statistics \citep{leccardi07,humphrey09,kaastra17}. In the ML approach, we build our ``best''  model for the background components (Sec. \ref{sec:physmodel}) and we freeze it, leaving as the only free parameters in our fits those related to the cluster emission (temperature, metal abundance and normalization). In this framework, the best fit parameters are the values minimizing the Cash statistics, and the uncertainties are computed imposing $\Delta C_{stat}=1$.
Conversely, in the Bayesian MCMC we estimate priors for the parameters associated to the background components (Sec. \ref{sec:physmodel}) that are then joint-fitted with the parameters associated to the cluster emission. This method allows us to propagate the uncertainties on the background parameters to the scientific results. In this approach, we compute the best fit values as the mean of the marginal posterior distribution of each parameter and the $1\sigma$ errors by sorting the chain values and then taking the upper and lower limit of the central 68\% of the values.
We run the MCMC fit within XSPEC, with the {\tt chain} command, using the Goodman-Weare algorithm \citep{goodmanweare} with 26 walkers (twice the number of free parameters). We tried different lengths both for the burn-in and for the fit phase on a few typical cases and we noticed to have a good convergence and stable results discarding the first 50000 steps (burn-in phase) and running the chain for other 50000 iterations. \\
Concerning the source component, we model the cluster emission with {\tt apec} (Astrophysical Plasma Emission Code \citealt{smith01}) and the Galactic absorption with {\tt phabs}, assuming {\tt aspl} abundance table \citep{asplund09} and {\tt vern} photo-electric cross sections \citep{verner96}. We discuss in  App.~\ref{app:ab_phabs} the impact of the abundance table on the parameters of cluster emission. 
We fix the redshift at the values provided in \citet{paper1} and the column density to the value $N_{Htot}$ which includes the contribution of both the atomic Hydrogen ($N_{HI}$) and the molecular one (see \citealt{bourdin23} for more details).
We will test the impact of using the total $N_H$ with respect to $N_{HI}$ in Sec. \ref{sec:sys_NH}.


\section{A physical model for the background components}
\label{sec:physmodel}


The typical accuracy of temperature profiles in cluster outskirts is largely determined by a telescope's background level and our ability at predicting the normalization and spectral shape of background components. Here we introduce a physically motivated model for the \xmm{} background which allows us to accurately predict its properties. 
The \xmm{} background is made of several components of different physical origin, in the form of the cosmic-ray induced particle background (CRPB), the residual focused component (RFC), and photons (cosmic X-ray background and foreground emission). We modeled individually these components and calibrated them on a large set of over 500 blank-sky observations from the XMM XXL survey \citep{pierre16}, totaling more than 6 Ms of data. The adopted procedure, its calibration and an estimation of the corresponding background subtraction accuracy are described in the following.

\subsection{Cosmic-ray Particle Induced Background}
\label{sec:particlebkg}
The primary background component in \xmm{} data originates from the interaction between Galactic cosmic rays and the structure of the spacecraft, generating secondary electrons. The secondary electrons are recorded by the CCDs and cannot be individually distinguished from photon-induced events, which produces a background signal that will be referred to as the cosmic-ray particle induced background. The normalization of this component is known to be modulated by the Solar cycle \citep{gastaldello22} and its spectral shape resembles a hard power law with a number of fluorescence lines \citep{kuntz08}. In the MOS detectors, the CRPB intensity can be monitored on-the-fly using the unexposed corners of the detectors, which are shielded from the telescope's FOV and thus record only CRPB-induced events. On the other hand, a large collection of data with the filter wheel set in closed position (hereafter filter-wheel-closed data, FWC) were accumulated by \xmm{} throughout the mission. Again, FWC observations only record CRPB-induced events, allowing us to determine the spectral shape and  the variation of the CRPB across the detector with high accuracy. For each observation, we determine the current CRPB level using the MOS corners and renormalize the FWC data such that they match the level observed in the corner data using the ESAS tasks \texttt{mos-spectra} and \texttt{mos-back}. We then use the rescaled FWC data to extract the CRPB spectrum from the region of interest, and we fit the resulting spectrum with a phenomenological model made of two power laws and a combination of Gaussian fluorescence lines to create an analytical model for the CRPB. 

In the case of the pn detector, the situation is rendered more complicated by the fact that the device does not include an area that is fully shielded from the telescope's FOV \citep{marelli21}. Therefore, it is not possible to monitor the CRPB rate in the same way as for the MOS, especially in time frames that are significantly affected by soft proton (SP) flares. Following \citet{marelli21}, we define an outer region ($R>905\arcsec$ from the aimpoint) which is reasonably shielded from the telescope. We then take advantage of the simultaneous MOS data to predict the level of CRPB in the pn. To this end, we make use of the $inFOV$ and $outFOV$ count rates, as introduced in Sect. \ref{sec:data_reduction}. We recall that $inFOV-outFOV$ can be used as an indicator of the remaining SP contamination \citep{salvetti17}. We then select a subset of our blank-sky observations with low contamination ($inFOV-outFOV<0.04$ cts/s) , for which we expect the pn corner data to be unaffected by soft protons and representative of the CRPB only. We then correlated the [10-14] keV pn count rate in the corners with the MOS2 corner count rate and found an excellent correlation between the two quantities, as already reported by \citet{marelli21}. 

\begin{figure}
    \centering
    \resizebox{\hsize}{!}{\includegraphics{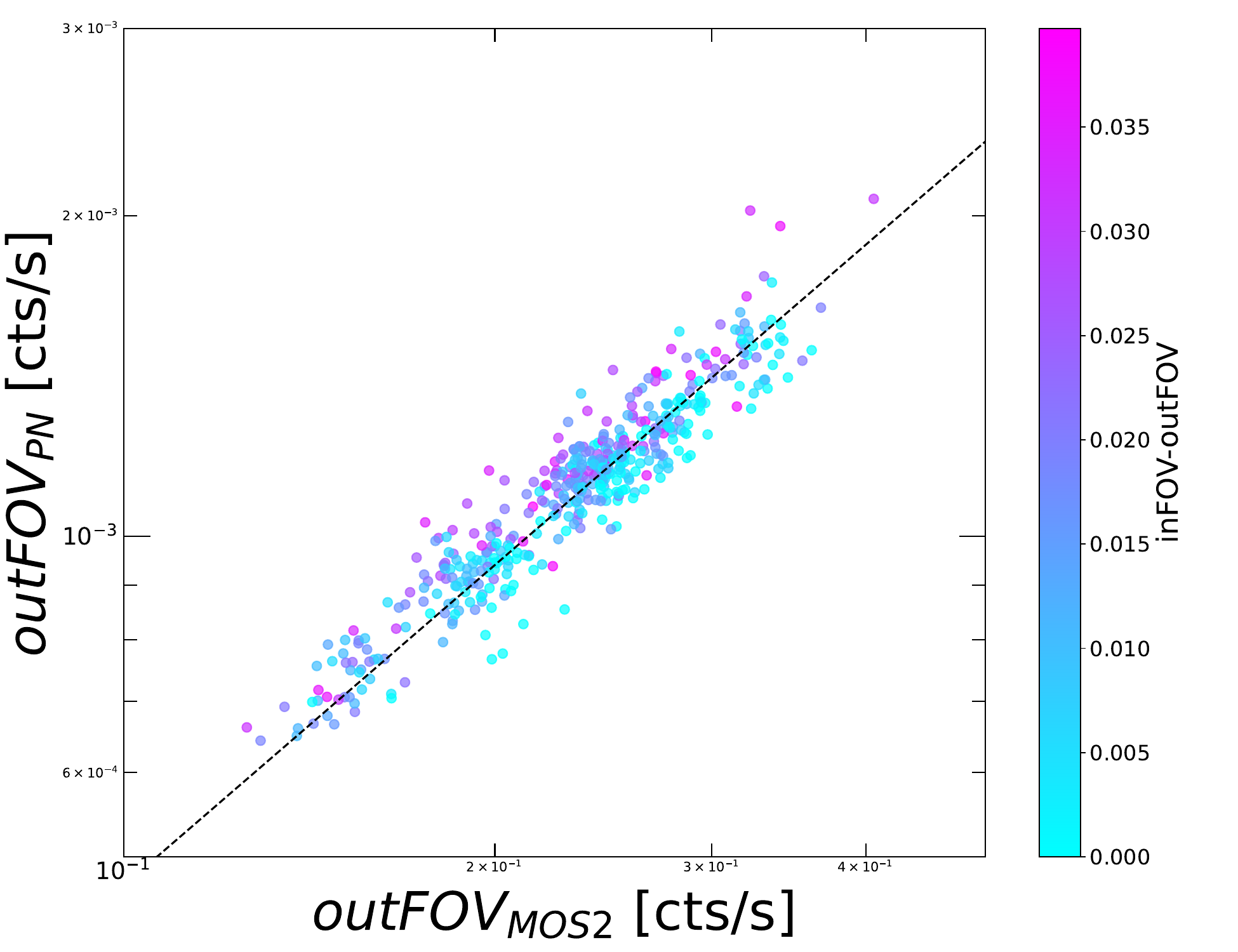}}
    \caption{Relation between the pn and MOS2 high-energy ($10-14$ keV) corner count rates in a large number of blank-sky observations with a low level of residual focused contamination. The color code indicates the level of contamination as traced by the $inFOV-outFOV$ indicator and the dotted line is the best-fit relation (Eq \ref{eq:1}).} 
    \label{fig:pn_outfov}
\end{figure}

In Fig. \ref{fig:pn_outfov} we show the relation between the pn and MOS2 corner count rates for observations with $inFOV-outFOV<0.04$ cts/s. The tight correlation demonstrates that for uncontaminated observations the pn corner count rates provide an accurate estimate of the CRPB level. However, the color code implemented in Fig. \ref{fig:pn_outfov} shows that the scatter of the points around the mean relation correlates with the level of contamination, with the observations falling above the relation exhibiting systematically higher values of $inFOV-outFOV$ than the relations falling below it. To alleviate this issue, we fitted a simple model to the data in which the pn high-energy count rate is described as
\begin{equation}
    outFOV_{PN} = {\rm A_{CRPB}} ~ outFOV_{MOS2} + {\rm A_{SP}}(inFOV-outFOV).
    \label{eq:1}
\end{equation}
In other words the pn $outFOV$ rate is the sum of a CRPB component that is related to the MOS2 corner count rate with a proportionality constant ${\rm A_{CRPB}}$ and of a SP contamination term that is proportional to $inFOV-outFOV$ with a proportionality constant ${\rm A_{SP}}$. We then fitted the relation shown in Fig. \ref{fig:pn_outfov} with this model and determined the best-fitting values of ${\rm A_{CRPB}}$ and ${\rm A_{SP}}$. We are then able to predict the level of the CRPB-induced pn rate per unit area as 
\begin{equation}
    CRPB_{PN} = {\rm A_{CRPB}}~ outFOV_{MOS2}.
    \label{eq:pn_qpb}
\end{equation}
For each observation and each spectrum, we extract the local CRPB spectrum from the FWC data using \texttt{pn-spectra} and renormalize it such that the high-energy normalization matches the prediction of Eq. \ref{eq:pn_qpb}.

While this method has proven to be very effective in reproducing the shape and normalization of the continuum emission, it does not reproduce well the properties of the emission lines, whose central energy and width are sensitive to gain variations and charge-transfer-inefficiency (see \citealt{kuntz08} and \citealt{sanders20} for more details). This is why we decided to exclude the energy ranges of the most prominent spectral lines in our fit, as described in Sec. \ref{sec:spectral_fit}.

\subsection{Residual focused component}
\label{sec:SPbkg}

The \xmm{} orbit is known to host a large number of clouds of low-energy protons ($\sim100$ MeV) that are trapped within the Earth's magnetosphere. When the trajectory of the spacecraft crosses such a cloud, soft protons (SP) can be funneled through the telescope and focused in a similar way as actual photons, albeit with a slightly different vignetting curve \citep{kuntz08}. Soft protons can generate very bright flares of the background intensity, with the background count rates increasing by more than two orders of magnitude. While the light curve filtering process introduced in Sect. \ref{sec:data_reduction} allows us to exclude time periods that are strongly affected by soft proton flares, in some cases there remains a low level of SP contamination even after the most affected time intervals have been clipped, which constitutes a significant part of the residual focused component. We attempted to create a predictive model of the residual SP contamination from the set of $\sim500$ blank-sky pointings discussed above, as described in Appendix \ref{sec:app_bkg}. The SP component is modeled as a power-law $I(E)=N_{QSP}E^{-\Gamma}$ with a fixed slope ($\Gamma=0.6$, Sec \ref{sec:app_bkg1}), while its normalization $N_{QSP}$ is predicted from the $inFOV-outFOV$ indicator, using the calibration expressions described in Sec. \ref{sec:app_bkg2}, one for each instrument. We take into account the variation of the SP component across the field-of-view with its vignetting function, as provided by the ESAS task \texttt{protonscale}. In this way, we can predict the value of $N_{QSP}$ for each region, detector, and observation.

\subsection{Sky background/foreground components}
\label{sec:skybkg}
The sky contribution to our spectra is mainly due to three components \citep[e.g.][]{kuntz00}: the residual Cosmic X-ray Background (CXB), and the foreground emission of the Galactic Halo (GH), and the emission of the Local Hot Bubble (LHB). For the LHB, we use an unabsorbed {\tt apec} thermal model with temperature fixed to $0.11$ keV, 
solar metal abundance, and $z=0$, while for the GH we used an absorbed {\tt apec} with temperature allowed to vary in the range $0.1-0.6$ keV (flat prior in the MCMC fit) as in \citet{mccammon02} and solar metal abundance\footnote{Recently, \citet{ponti23} reported a much lower value for the abundance of the hot circum-galactic medium with eROSITA data. The value of the metal abundance is degenerate with the other free parameters, especially the normalization of the GH and LHB.  We performed simulations of a typical sky spectrum with $Z=0.06Z_\odot$ and found that we can always retrieve a good fit with our procedure, even assuming $Z=Z_\odot$. This suggests that we can still use our baseline model as a good phenomenological representation of the sky background components. }. In both cases, the normalizations are free parameters of the fit.
We model the CXB as an absorbed power-law, with a fixed slope of $1.46$ \citep{moretti09} and free normalization. For the absorption, we always use the {\tt phabs} model and the same $N_H$ value used to fit the cluster emission (Sec. \ref{sec:spectral_fit}). We derive the free parameters of this model by fitting the spectra extracted in an external annulus, ideally free of cluster emission (Sec. \ref{sec:spectral_extract}). While running fits with the Bayesian MCMC approach, we use this approach also to estimate the sky background: we derive posteriors for the four free parameters, that we model with Gaussian distributions to be given as priors in the subsequent fits. \\ 
As discussed in Sec. \ref{sec:spectral_extract}, for six clusters in our sample we could not define a region not contaminated by the cluster emission in the FOV and offset observations are not available. In these conditions, we need to take into account a residual cluster component in the ``closebkg'' region, which is very degenerate with the sky background parameters that we need to estimate. To break this degeneracy and provide better constraints to the free parameters in our model, we follow the method suggested by \citet{snowden08} and we  make use of an ancillary data set: a spectrum extracted from the ROSAT All Sky Survey (RASS) diffuse background map \citep{snowden95,snowden97} in an annulus between one and two degrees from the cluster center, using the HEASARC X-ray background tool\footnote{\url{https://heasarc.gsfc.nasa.gov/cgi-bin/Tools/xraybg/xraybg.pl}}. 
We jointly fit the RASS spectrum with the \xmms\ one, taking into account a cross-correlation factor between the two instruments of 15\% \citep{eckert11}
and binding together the parameters of the sky background components. The normalization of the cluster component is a free parameter in the \xmms\ spectrum while it is fixed to zero in the external RASS data, which essentially do not contain cluster emission. During the fit of the sky background spectra, we fix the temperature of the cluster component beyond $R_{500}$, since leaving it free often lead us to unphysical values for this parameter. The fit is not very sensitive to the exact value of this parameter and we chose to fix it to one third of the mean expected temperature of the cluster, as estimated by the $M-T$ relation in \citet{arnaud05}, taking into account the typical decline of temperature profiles in the literature.\\
We perform the fits in the background region both with and without the RASS ancillary dataset, to study its impact both on the sky parameters estimates and on the temperature profiles (Sec. \ref{sec:sys}).


\subsubsection{Impact of RASS joint fit on sky background parameters}
\label{sec:skybkg_params}

\begin{figure*}
    \centering
    \subfloat{\includegraphics[width=0.45\textwidth]{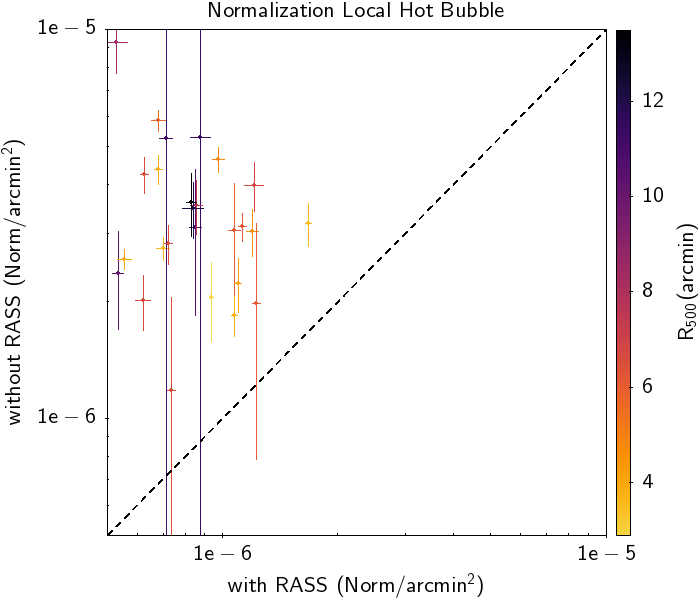}}
    \hspace{0.2cm}
    \subfloat{\includegraphics[width=0.45\textwidth]{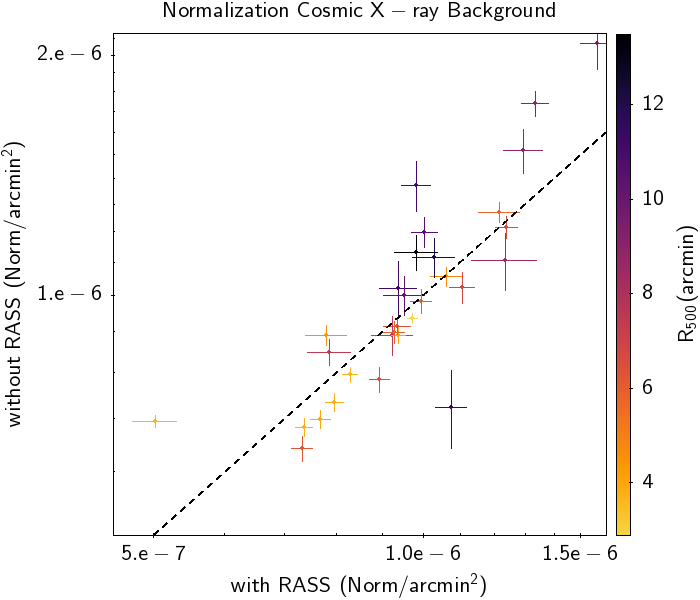}}\\
    \vspace{-0.2cm}
    \subfloat{\includegraphics[width=0.45\textwidth]{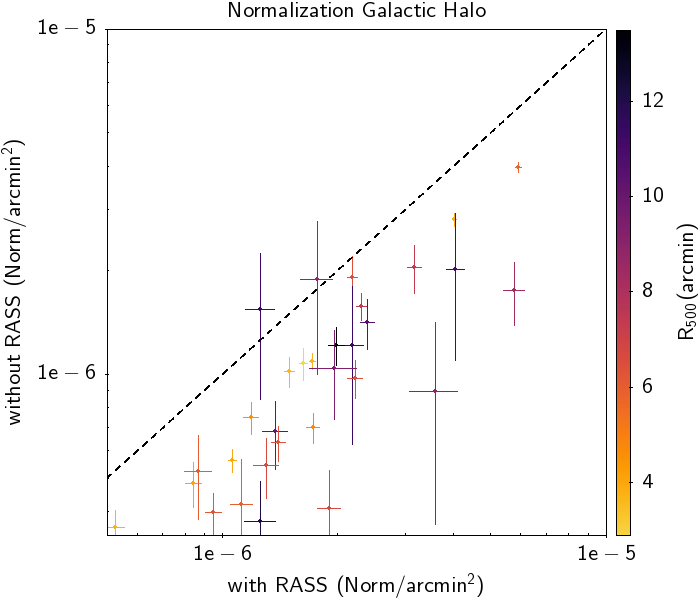}}
    \hspace{0.2 cm}
    \subfloat{\includegraphics[width=0.45\textwidth]{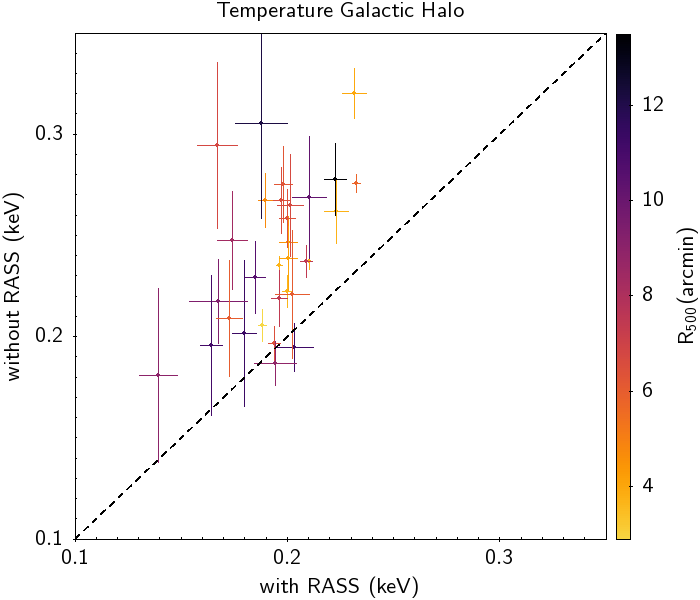}}
    \caption{Comparison of best fit values for sky background parameters obtained with the RASS joint fit (x-axis) and using \xmm\ only (y-axis). We show the normalization of the LHB (upper left), and of the CXB (upper right), while in the bottom panels we report the GH normalization (left) and temperature (right).
    Points are colour-coded by the extension of the cluster, estimated by $R_{500}$, in arcminutes. The dashed line marks the equality line. The LHB and GH components show large systematic variations, reflecting the degeneracies between the components. The CXB normalisation shows only a moderate variation on the full sample, with larger difference for extended clusters.}
    \label{fig:skybkg_compare}
\end{figure*}

We note that the joint fit with the RASS affects significantly the best fit values of some sky components: the normalization of the LHB, which is only poorly constrained in the XMM-only fit, the normalization and temperature of the Galactic Halo (see Fig. \ref{fig:skybkg_compare}). This variation is systematic: with the RASS+XMM joint fit we find lower LHB normalization ($16 \%$, computed as the weighted mean of the ratio XMM+RASS/XMMonly), higher normalization (55 \%) and lower temperature (16 \%) of the GH, reflecting the strong degeneracies between these parameters.\\
For the CXB normalization, we should consider two different effects. For the more extended clusters, where we do not have a clean region in the field of view to estimate the sky background, the fit without RASS should overestimate the CXB normalization, since the residual cluster emission will be attributed to the background components. Indeed, if we consider only the 6 extended clusters ($9\arcmin <R_{500}<12\arcmin $) where we used a ``closebkg'' region (purple points in Fig. \ref{fig:skybkg_compare}), the normalization of the fit without RASS is larger than the one with the joint fit, with a mean difference of 17\%. Besides that, there is another fundamental effect: since the source detection limit in the \xmm{} observations is lower than in the RASS data, the fraction of the resolved CXB is different in the two spectra, leading to different normalizations for this component, while in our joint fit we impose them to be the same. We can quantify this effect, by comparing the best fit normalization with the joint fit and the XMM-only, in the objects where we have a large enough cluster-free region in the \xmm{} observations. Indeed, for the clusters with $R_{500}<6\arcmin$, the normalization in the joint fit is larger by $1.4\%$ (mean, while $3.7\%$ is the median difference). We can thus conclude that with the joint fit we overestimate the normalization by $1-4\%$, which is a much smaller error than the 17\% due to the use of XMM-only data for extended clusters. Moreover this error is smaller than the typical statistical errors on this parameter ($\simeq4\%$), over which we marginalize during the MCMC fit.\\
We will address the impact of the variations of the sky background parameters on the temperature estimates in Sec. \ref{sec:sys_RASS}.

\subsection{Application to the spectral fit}
\label{sec:app_bkgmodel}
In the previous subsections, we have described in detail the properties of our physical background model, how we built and calibrated it. Here, we focus on how we apply it to the spectra during the fitting process, both in the Maximum-Likelihood and in the MCMC approach. As presented in Sec. \ref{sec:spectral_fit}, we joint fit the spectra of the three instruments and to all of them we fit the cluster component and the sky background model. The latter is expressed as {\tt apec+phabs*(apec+powerlaw)}, where the temperature of the Galactic Halo and the normalizations of all three components are estimated from the fit in the external region (Sec. \ref{sec:skybkg}) and appropriately rescaled for the area. All parameters are kept fixed in the standard ML fit, while in the MCMC the normalizations and the GH temperature are treated as Gaussian priors, using the statistical errors on the best-fit parameters for the width (see Table \ref{tab:bkg_params}, where we summarize the properties of the priors for the background components in the MCMC analysis). \\
For the CRPB and the RFC we build a separate model for each detector. The former is expressed as a combination of two power-laws for the continuum emission and several Gaussians to represent the fluorescent lines (Sec. \ref{sec:particlebkg}), overall multiplied by a constant which sets the intensity of this component. This model is first fitted to the spectra of FWC data produced by the ESAS tools {\tt mos/pn-back} and then applied, with its best fit parameters, to the spectra extracted from the observation. When performing the ML fit all parameters are kept fixed, while in the MCMC fit the multiplicative constant is allowed to vary within its uncertainty and treated as a Gaussian prior (Table \ref{tab:bkg_params}). For the pn detector, the procedure is the same but we renormalize the overall constant according to Eq. \ref{eq:pn_qpb}. Concerning the Residual Focused component, we model it with a power-law with fixed slope $-0.6$ and normalization predicted by the $inFOV-outFOV$ value, measured for each detector, with Eq. \ref{eq:nqsp_vs_io} (see Appendix \ref{sec:app_bkg}). For the pn, we estimate it by combining the $inFOV_{PN}$ and $outFOV_{MOS2}$ as described in Eq. \ref{eq:in_out_pn}. While doing the MCMC fit, we treat the normalization as a flat prior, within the intrinsic scatter of the relation between the Normalization of the SP component and the  $inFOV-outFOV$ indicator (Fig. \ref{fig:nqsp_io_mos} and \ref{fig:nqsp_vs_pn_io}).\\

\begin{table*}[]
    \centering
    \caption{Properties of the priors set in the MCMC analysis for the background parameters. For each of them, we report the shape of the prior, how we compute its central value and its width.}
    \begin{tabular}{c c c c }
    \hline
    \hline
    Parameter & Shape & Central value & Width \\
    \hline
     CRPB Norm. (MOS)    &  Gaussian & Best Fit of {\tt mos-back} spectra  & 2\% intrinsic scatter \\
     CRPB Norm. (pn) & Gaussian & Best fit of {\tt pn-back} spectra renormalized by Eq. \ref{eq:pn_qpb} & 6\% intrinsic scatter\\
     RFC Norm. (MOS) & Uniform  & $inFOV-outFOV$, with Eq. \ref{eq:nqsp_vs_io} & intrinsic scatter \\
     RFC Norm. (pn) & Uniform  & $inFOV_{PN}$ and $outFOV_{MOS2}$ with Eq. \ref{eq:nqsp_vs_io} and \ref{eq:in_out_pn} & intrinsic scatter \\
     LHB Norm. & Gaussian &  Best fit in background region & $1\sigma$ errors\\
     GH Temp. & Gaussian &  Best fit in background region & $1\sigma$ errors\\
     GH Norm. &  Gaussian &  Best fit in background region & $1\sigma$ errors\\
     CXB Norm. &  Gaussian &  Best fit in background region & $1\sigma$ errors\\
     \hline
    \end{tabular}
    \label{tab:bkg_params}
\end{table*}

\section{Impact of spectral analysis on temperature measurements}
\label{sec:sys}
\begin{figure}
    \centering
    \includegraphics[width=0.45\textwidth]{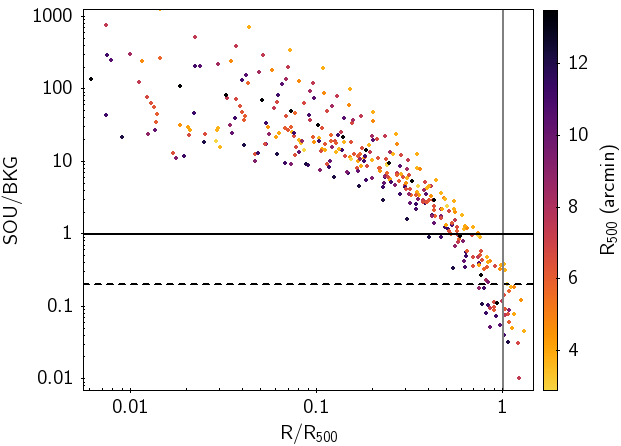}
    \caption{Distribution of the $SOU/BKG$ indicator for MOS2 as a function of scaled radius, for all the 354 regions from which we extracted spectra. The color coding refers to the extension in the sky ($R_{500}$ in arcmin) of the corresponding cluster. The horizontal continuous line shows the value where the source intensity gets lower than the background line, while the dashed line marks the threshold $SOU/BKG=0.2$ used in the analysis of mean and median profiles (Sec. \ref{sec:meanprof}). The vertical line marks $R_{500}$. Most of the regions below the threshold $SOU/BKG=0.2$ at $R<R_{500}$ belong to very extended clusters and are thus located in poorly exposed regions of the FOV.}
    \label{fig:distr_sou2bkg}
\end{figure}

In this section, we evaluate the robustness of our pipeline and the impact of our decisions about the spectral fitting methods on the temperature measurements. While some of these decisions (such as the spectral model, the assumed $N_H$, and the techniques for the spectral fitting) affect all our measurements, other assumptions on the background model will impact mainly the outer cluster regions. Indeed, the effects of the latter strongly depend on the relative intensity of the source with respect to the background. Briefly, even a large error on the background model will not affect significantly the best fit parameters in a region where the source outshines the background, such as in the central regions of galaxy clusters. Conversely, in the external regions where the source intensity is comparable to, or often even smaller than, the background level, even a few percent error in the estimates of the background may affect significantly our temperature measurements. To quantify this, following \citet{leccardi08}, we define the indicator $SOU/BKG$ for all DR1 clusters and for all regions where we extracted spectra. 
We compute it as $(OBS-BKG)/BKG$, where $OBS$ is the observed count rate and $BKG$ is the predicted count-rate by the best fit model of all the background components. We measure all count rates in the energy band $0.7-10$ keV (comparable to the band we used in our spectral fit, see Sec. \ref{sec:spectral_fit}), to be consistent with \citet{leccardi08}. We are aware that the cluster count rate in the  $0.7-10$ keV band depends on temperature, inducing a slight dependence also on the $SOU/BKG$ indicator. We estimated its impact by computing our indicator also in a soft band ($0.7-1.2$ keV), where the count rate is mostly independent of temperature for $kT>2$ keV. We checked that all the results presented later in this paper (e.g. Fig. \ref{fig:tprof_sbthresh} and Fig. \ref{fig:fit_PL}) are robust with respect to the choice of the band in which we compute our indicator and we decided to keep using  the full spectral band, to allow quantitative comparisons with \citet{leccardi08}.
We compute this indicator for each camera but we refer to the values obtained with MOS2.\\
In Fig. \ref{fig:distr_sou2bkg} we show the distribution of the $SOU/BKG$ indicator as a function of radius for all the regions, spanning four orders of magnitudes. While most regions are clearly source dominated, we have a significant fraction ($\sim20\%$) of regions where the background intensity is larger than the source, justifying our need for the most accurate background modeling (Sec. \ref{sec:physmodel}). As we will discuss in Sec. \ref{sec:meanprof}, we trust our temperature measurements for regions with $SOU/BKG>0.2$ and we will thus exclude about 30 regions with $SOU/BKG$ below this threshold in some of the analysis presented later. In Fig. \ref{fig:distr_sou2bkg}, we note that about half of the regions below the threshold are located at $R<R_{500}$, hampering the goal of reliably measuring the temperature profile up to $R_{500}$ for some clusters. As highlighted by the color-coding, most of them are located in extended clusters ($R_{500}>9^\prime$), which by construction of our sample correspond to low-redshift and low-mass objects. These regions are located at relatively large off-axis angles from the aimpoint, where the exposure is lower than in the centre, suppressing the source count-rate more than the background one, whose dominant CRPB component is not affected by vignetting. 


\subsection{MCMC vs Maximum likelihood fitting}
\label{sec:sys_MCMC_ML}
\begin{figure*}
    \centering
    \subfloat{\includegraphics[width=0.49\textwidth]{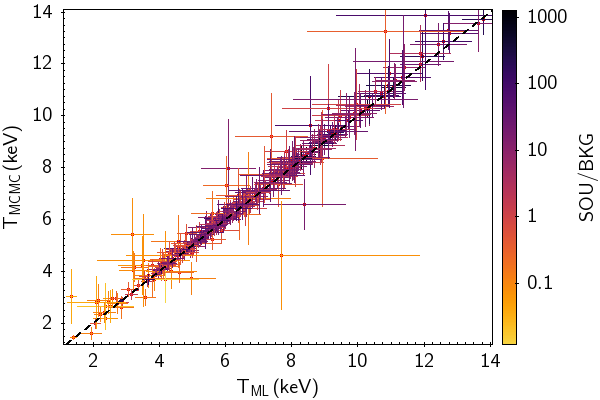}}
    \hspace{0.2 cm}
    \subfloat{\includegraphics[width=0.49\textwidth]{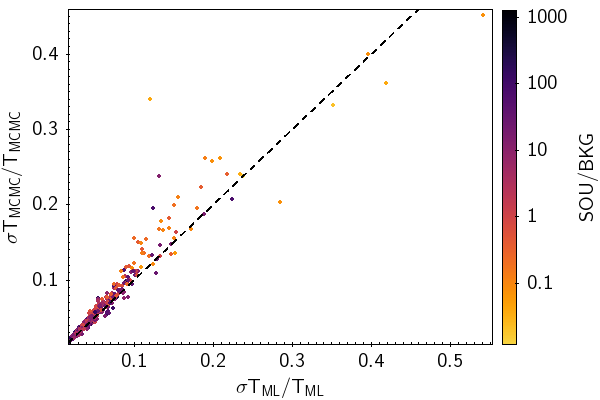}}
    \caption{Comparison of best-fit temperature values ({\it left panel}) and relative symmetric errors ({\it right panel}) for all clusters and all regions obtained with the standard ML technique (x-axis) and with the MCMC approach (y-axis). The colour coding refers to the $SOU/BKG$ ratio in each region, the dashed line represents the equality line. }
     \label{fig:ML_vs_MC}
\end{figure*}
As discussed in Sec. \ref{sec:spectral_fit}, to assess the MCMC framework we have utilized for our analysis, which marginalizes over uncertainties in the background components, we also perform a second set of fits using the standard ML approach, where the background model is fixed. 
It is possible in principle that the best fit parameters obtained with the two methods could differ from one another, since their definition is different. While the best fit temperature in the standard ML fit is the value which minimizes the Cash statistic, in the MCMC it is the mean of the posterior distribution. These values could be different, especially in the case of a skewed distribution. Moreover, during the marginalization process of the MCMC the background parameters can move with respect to the fixed values in the ML fit, and this can lead to a change in the best fit temperature, especially in background-dominated regions. We expect also an effect on the error bars of the best fit parameters, which should be larger in the MCMC fit, encompassing also the uncertainties on the background modeling. \\ 
In the left panel of Fig. \ref{fig:ML_vs_MC}, we compare the best fit temperature values obtained with these two methods: the agreement is very good, with a weighted mean of the difference  $(T_{ML}-T_{MCMC})/T_{MCMC}=(-0.7 \pm 0.3)\%$.
 We expect the MCMC approach to have a small impact in source dominated regions and indeed the difference is consistent with zero, $(T_{ML}-T_{MCMC})/T_{MCMC}=(-0.4\pm0.4)\%$
when $SOU/BKG>10$. Conversely, in background dominated regions, the MCMC approach has a larger impact on the best-fit parameters, since it allows the background parameters to move from their value. Hence, temperatures in this regime are very sensitive to the background models. Indeed, the differences increase to $(-4.6\pm2.5)\%$ in regions with $0.2<SOU/BKG<0.5$ and to $(-7.5\pm3.5)\%$ for $SOU/BKG<0.2$. If we split the regions in three radial bins, we find that the difference is consistent with zero at less than 1-2$\sigma$ for $R<0.4R_{500}$ and $0.4-0.8R_{500}$ and measure a difference of $-5.6\pm2.7\%$ in the last radial bin ($R>0.8R_{500}$). 
While the results are only moderately significant, we can conclude that the standard ML approach leads to an underestimate of the temperature values of about 3-7\% in background-dominated regions. We note that this comparison depends on the estimator used in the MCMC fit to derive the best fit parameters. As described in Sec. \ref{sec:spectral_fit}, we use the mean of the posterior distribution, but we could use also the median or the mode. These values do not differ significantly from the mean on the full sample ($\leq 1\%$) but can vary by a few per cent in background-dominated regions, where the posterior distributions become more skewed, moving closer to the ML estimate. We should thus consider the value provided above as an upper limit for this effect.\\
In the right panel of Fig. \ref{fig:ML_vs_MC}, we compare the $68\%$ relative symmetric error on the temperature measurement with the ML and MCMC approach. Again, in source-dominated regions the error on the measurement is dominated by the statistical quality of the data and is very similar in both approaches. Conversely, in background-dominated regions the error on the ML reflects only the statistical quality on the data, while the one on the MCMC should be larger because it propagates also the uncertainties on the background models. This is indeed observed in Fig.  \ref{fig:ML_vs_MC}, where most of the points with low $SOU/BKG$ lie above the equality line. On average, the MCMC error is $25\%$($31\%$) larger in regions with $SOU/BKG<1$ ( $SOU/BKG<0.5$) with respect to the standard ML approach. 
Conversely, the few points with very low $SOU/BKG$ that lie significantly below the line may indicate that the MCMC did not converge there. We individually investigated the three most deviant outliers: they correspond to radial bins whose center is beyond $R_{500}$ and with a very low $SOU/BKG<0.1$, corresponding to regions that we later excluded from our analysis (see Sec. \ref{sec:meanprof}). The visual inspection of the chains confirms that the MCMC had not converged in these cases and we thus reran the MCMC fitting, increasing the number of steps to $10^5$. After the rerun, the error bars increased significantly. 
We decided not to rerun all our fits with longer MCMC chains and that $5 \times 10^4$ steps is a good compromise between convergence in most cases and computation speed.
\subsection{Impact of total $N_H$}
\label{sec:sys_NH}
As presented in Sec. \ref{sec:spectral_fit}, we use the total $N_Htot$ value, comprising the contribution of both the atomic and molecular hydrogen, computed in \citet{bourdin23}.
We tested the impact of this choice on the temperature measurements by refitting all spectra with the standard atomic absorption ($N_{HI}$) and computing the relative difference of the best fit values obtained with and without the molecular component: $\Delta_{TNH}=(T_{NHI}-T_{NHtot})/T_{NHtot}$. We then divided our sample into three subsamples: a low absorption regime ($N_{HI}<2\times 10^{20}\, \rm{cm}^{-2}$), an intermediate regime (2<$N_{HI}/10^{20}\, \rm{cm}^{-2}<5$), and a (relatively) high absorption sample ($N_{HI}>5\times 10^{20}\, \rm{cm}^{-2}$). We then computed the median value of $\Delta_{TNH}$ on the three subsamples. 
In the low absorption regime differences in the temperature values are negligible and the median difference is only $0.2\%$. In the intermediate regime, we start to see some differences, especially at high temperature, but the median difference is still $1\%$. Finally, in the high absorption sample we systematically find higher temperatures when fitting with the atomic component only, with a median difference of $6.8\%$. These results are consistent with those for the full CHEX-MATE sample described in \citet{bourdin23}.\\
We use the mean difference in the derived temperatures as an estimate of the systematic associated to the assumption of a fixed $N_H$ in our fit. On the full DR1 sample, we estimated $\Delta_{TNH}=(1.5\pm0.3)\%$ and found a consistent value at low $SOU/BKG$ and in different radial ranges.
This is indeed expected since it should not depend on the background level. We emphasize that $N_H$ is a fixed parameter in all our fits and thus the uncertainty on its measurement is not propagated in the temperature estimate even in the MCMC fi
\subsection{Impact of abundance table}
\label{sec:abtable}
Abundance tables are used in XSPEC to compute plasma emission and photoelectric emission models and thus can impact all the free parameters in our fits, not only the metal abundance. The relative changes can be estimated making use of XSPEC simulations, as described in App. \ref{app:ab_phabs}. Here we briefly recall the results concerning the temperature, comparing the ones measured with a given table ($T_{ab}$) with respect to the values obtained with the reference table in our analysis \citep[$T_{aspl}$,][see Sec. \ref{sec:spectral_fit}]{asplund09}. The abundance table is used both in the {\tt apec} model and in the absorption {\tt phabs} model,  and it is the latter which causes the most significant differences in the temperature measurements. Indeed, all the simulations performed with $N_H=0$ show negligible variations ($<1\%$) in the temperature estimate. Conversely, when we use an intermediate absorption, $N_H=5\times 10^{20}\, \rm{cm}^{-2}$, we find significant variations $(T_{ab}-T_{aspl})/T_{ab}$ of $-3\pm1\%$ using  \citet{lodders09} and $-6\pm3\%$ using  \citet{angr}. We interpret this result as being due to the different abundance of Helium in the solar tables. 

\subsection{Impact of background assumption in low $SOU/BKG$ regions}
In this section, we assess the impact of different background treatments on the temperature measurements, especially in outer regions associated with a low $SOU/BKG$ ratio. Since we want to address the impact of background parameters that are free to vary within their priors in the MCMC fit, in this section we show results obtained with the ML fit, where these parameters are kept fixed. 

\subsubsection{Impact of joint fit with RASS}
\label{sec:sys_RASS}
\begin{figure}
    \centering \includegraphics[width=0.45\textwidth]{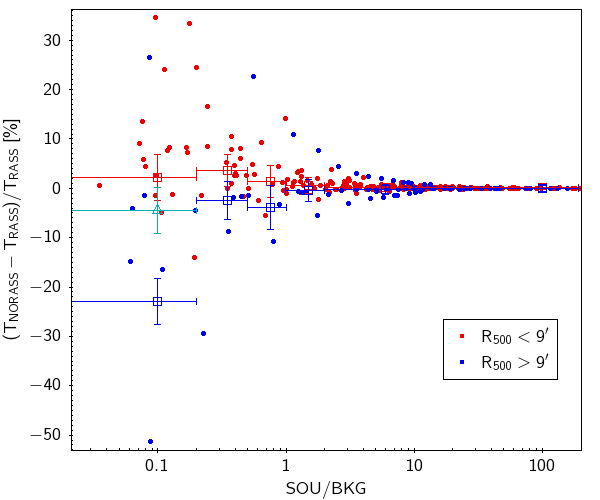}
    \caption{Ratio of the difference between the temperature values obtained using \xmms{} only in the background region ($T_{NORASS}$) and that obtained with our standard pipeline ($T_{RASS}$, joint fit with RASS in the background region) over $T_{RASS}$, in percent units, as a function of the $SOU/BKG$ ratio in the spectral region. Filled points show the individual measurements (error bars not shown for clarity), while the open squares represent the weighted mean ratio and its error in different intervals of $SOU/BKG$. Blue points refer to measurements in clusters with $R_{500}>9^\prime$, red are for the the remaining objects. The cyan triangle shows the median value for the regions with $SOU/BKG<0.2$ and clusters with $R_{500}>9^\prime$, significantly different than the mean with the same criteria. }
    \label{fig:TRASS}
\end{figure}
In Sec. \ref{sec:skybkg} we introduced our strategy to derive the parameters of the sky background components by performing a joint fit of the \xmm\ spectrum beyond $R_{200}$ with the RASS spectrum around the cluster and discussed the impact on the background parameters themselves. Here we want to assess the impact on temperature measurements, by comparing the best fit values  obtained with our pipeline ($T_{RASS}$) with those obtained turning off the joint fit with RASS data ($T_{NORASS}$), using \xmms\ data only to estimate the sky background parameters. We compare the temperature ratios as a function of $SOU/BKG$ in Fig. \ref{fig:TRASS}. We emphasize that we use the nomenclature $T_{RASS}$, but we never use RASS data for measuring the cluster temperature, only for determining the sky background model in a joint fit with \xmm. The differences in the temperature that we discuss here are the consequence of the different sky background parameters, presented in Sec. \ref{sec:skybkg} and Fig. \ref{fig:skybkg_compare}. \\
First, we focus on the six clusters of our sample with $R_{500}>9^\prime$, for which we do not have a clean region in the FOV and we used the ``closebkg'' annulus ($1.1R_{500}-14^\prime$), likely contaminated by cluster emission, as a background region (Sec. \ref{sec:spectral_extract}). In Sec. \ref{sec:skybkg} we noticed that the introduction of the joint fit with RASS reduces the CXB normalization by a factor of $17\%$ for these clusters. Concerning the ICM temperature, the effect of a reduced CXB normalization is negligible in regions with $SOU/BKG>1$: temperatures obtained with the two methods are always consistent within the error bars and the mean difference is less than $1\%$. In background-dominated regions ($SOU/BKG<1$), individual measurements may be significantly different, with variations up to $20-50\%$, larger than the statistical errors on the measurements. Nonetheless, there is not a clear trend (see blue points in Fig. \ref{fig:TRASS}): in most cases we find $T_{RASS} > T_{NORASS}$ but in others we find the opposite. We computed the weighted mean variations in different $SOU/BKG$ regimes and we find a significant difference only in the bin with $SOU/BKG<0.2$, where we find $(T_{NORASS}-T_{RASS})/T_{RASS}=(-22\pm4.5)\%$ (see Fig. \ref{fig:TRASS}). We note however that this mean value is driven by one single measurement with small uncertainty but with a variation of $\simeq50\%$ and may return an overestimate for this effect. The median value (also shown in Fig. \ref{fig:TRASS}) on the same sample is significantly lower ($-4.6\pm4.7\%$) and may better reflect the properties of the distribution. We note that, by construction of the CHEX-MATE sample, the clusters with $R_{500}>9^\prime$ are nearby low-mass objects, where the temperature in the external regions lies in the range $2-3$ keV. It is possible that the variations of the sky-background parameters, especially the CXB normalization, may have a larger impact in different temperature regimes. \\
We also checked that the joint fit with RASS did not introduce a bias in our temperature measurements, possibly due to the cross-calibration of the two instruments and to the different parameters of the foreground components, with the systematic underestimate of 1-4\% in the CXB normalization discussed in Sec. \ref{sec:skybkg_params}. To assess this, we compare $T_{RASS}$ with $T_{NORASS}$ also for the remaining 22 clusters with $R_{500}<9^\prime$ (red points in Fig. \ref{fig:TRASS}): in all cases temperatures are consistent within the two methods and start to differ only at $SOU/BKG<1$, but remain consistent within the error bars in most cases. We computed the mean weighted ratio to be    $(T_{NORASS}-T_{RASS})/T_{RASS}=(0.2\pm0.4)\%$ on the full sample. The difference increases in regions with low $SOU/BKG$ ($3.5\pm3.3\%$ in the range $0.2-0.5$ and $2.3\pm4.7\%$ when $SOU/BKG<0.2$), remaining nonetheless consistent with zero. We can conclude that if we introduce a systematic bias with our method, this is less than $5\%$. 

\subsubsection{Impact of residual focused component}
\label{sec:sys_SP}
\begin{figure}
    \centering \includegraphics[width=0.45\textwidth]{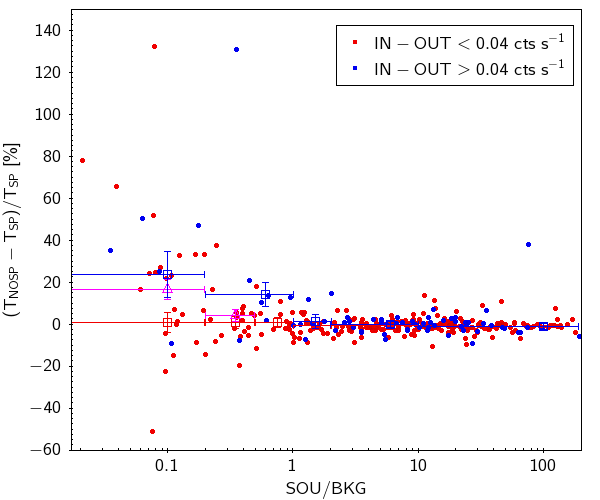}
    \caption{Ratio of the difference between the temperature values obtained neglecting the soft proton component ($T_{NOSP}$) and the one of our pipeline ($T_{SP}$, including a SP component) over $T_{SP}$ in percent units, as a function of the $SOU/BKG$ ratio in the spectral region. Filled points show the individual measurements (error bars not shown for clarity), while the open squares represent the weighted mean ratio and its error in different intervals of $SOU/BKG$. Blue points refer to measurements in observations with a significant residual contamination ($inFOV-outFOV>0.04$ cts/s), red are for observations with a low contamination. The magenta triangles shows the median value for clusters with low contamination, when the median is significantly different from the mean.}
    \label{fig:TSP}
\end{figure}
As discussed in Sec. \ref{sec:SPbkg}, we introduced a novel approach to model the residual focused component basing on the $IN-OUT$ indicator. This component, which we attribute to residual soft protons, is typically subdominant with respect to other background components in all DR1 observations  ($(IN-OUT)_{M2}<0.1$ cts s$^{-1}$), and becomes comparable to the intensity of the sky background and to the CRPB only in the cluster with the most contaminated observation (PSZ2 G$195.75-24.32$). Still, it can have an important impact in background dominated regions. To estimate this effect, we divide this sample in two subsamples: 24 objects with low residual contamination ($lowSP $ with $(inFOV-outFOV)_{M2}<0.04$ cts s$^{-1}$) and 6 with a significant residual contamination ($highSP$ with $0.04<(inFOV-outFOV)_{M2}<0.1$ cts s$^{-1}$) and we perform the fit turning off the SP correction. We then compare the temperatures obtained with this method with those obtained with our baseline pipeline and we show their ratio in Fig. \ref{fig:TSP} as a function of the $SOU/BKG$ in the fitting region. Individual measurements are typically consistent in source-dominated regions and start to show large variations (up to more than 100\%) when $SOU/BKG<1$. This effect is particularly large in the subsample with large residual contamination, where we measure a weighted mean $(T_{NOSP}-T_{SP})/T_{SP}=(14\pm6)\%$ in regions with $0.2<SOU/BKG<1$ and $(23\pm11)\%$ when $SOU/BKG<0.2$. Nonetheless, the effect is visible also on the individual measurements of clusters with a low contamination. While the weighted mean value for the $highSP$ sample does not seem to report significant differences (Fig. \ref{fig:TSP}), probably driven low by a few measurements with a small statistical error, the median difference (magenta square in Fig. \ref{fig:TSP}) in the $SOU/BKG<0.2$ range is $16\pm5\%$.

    The large impact of including or neglecting the SP component on the temperature measurements in external regions of galaxy clusters highlights the importance of modeling this component, even in ``clean'' observations. 
    We emphasize here however that with this comparison we are not measuring the real systematics associated to our SP modeling but rather measuring the potential effect if we neglect this component in our analysis, which we do not. 
    
\begin{figure}
    \centering \includegraphics[width=0.45\textwidth]{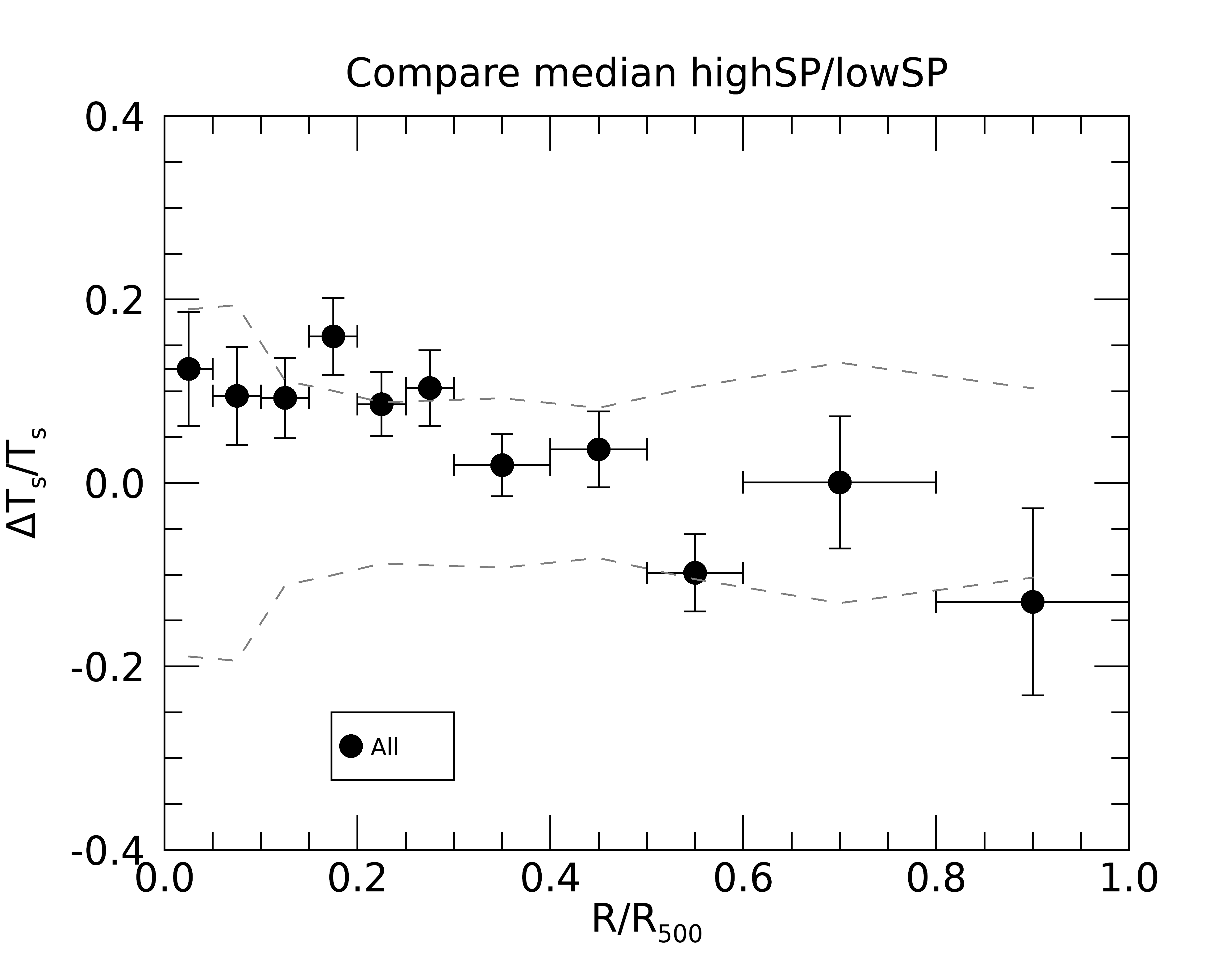}
    \caption{Ratio of the difference between the median scaled temperature profiles for the subsample with high contamination with respect to the low contamination sample. The points show the ratio between the median value in each bin, while the dashed lines mark the observed dispersion, which is typically larger than the measured difference, preventing its interpretation as a systematic bias.}
    \label{fig:TmedianSP}
\end{figure}

    An alternative method of measuring the systematic effect associated with the SP modeling is to compare the median scaled temperature profiles of the subsample with high contamination and those of the subsample with low contamination. In principle the profiles should not differ and any systematic difference could be interpreted as due to the SP modeling. We show their difference ($(T_{s,highSP}-T_{s,lowSP})/T_{s,lowSP}$, where the notation $T_s$ marks the scaled temperature $T/T_{[0.15-0.75]R500}$) as a function of the scaled radius $R/R_{500}$ in Fig. \ref{fig:TmedianSP}. In some radial bins, we measure a significant difference, which is, however, consistent with the observed dispersion of the profiles, preventing its interpretation as a systematic bias. We expect to derive better estimates when we analyze the full CHEX-MATE sample. For the moment, we can use the dispersion level ($10\%$ outside the core) as an upper limit to the systematics related to SP modeling.

\subsection{Error budget}

In the previous sections, we have quantified the impact of our choices for spectral analysis on the temperature estimate. In some cases, such as the choice of $N_H$ and the abundance table, we are able to measure a significant effect which allows us to associate a possible systematic error in our measurement. In other cases, the mean differences between the temperature under different assumptions are consistent with zero at one or two $\sigma$, in different $SOU/BKG$ or radial ranges. This means that the true systematic errors in our method are smaller than the statistical uncertainty for our sample of 30 clusters and we can only provide an upper limit for each of them. We estimated the upper limit at 95\% confidence level as twice the uncertainty on the weighted mean temperature difference and we report all of them in Table \ref{tab:syslimits}. As discussed before, in some cases, related to the background modeling, our estimate of systematics depends on the $SOU/BKG$ of the regions we are considering, with larger effects in more background dominated regions. Conversely, in other cases the effect is independent of the intensity of the source. In Table \ref{tab:syslimits}, we report the values in two ranges, for $SOU/BKG<0.2$ (where, as we will show in Sec. \ref{sec:meanprof}, temperature measurements are probably biased) and $0.2<SOU/BKG<0.5$, i.e. the most background dominated regions where we consider we have reliable temperatures. In both regimes, possible systematics related to the choices of our spectral analysis are of the order of a few per cent, with upper limit of the order of $10\%$ in the most background dominated regions. We note that this is the same order of magnitude of the cross-calibration between the three EPIC cameras, leading to significantly different temperatures when estimated with single detectors \citep[e.g.][]{schellenberger,nevalainen23}. We checked that for the DR1 sample, the mean difference between the value measured with a single detector and our baseline joint fit is $5\%$, consistent with literature results.   

\begin{table}[]
    \centering
    \caption{Potential systematic errors in temperature measurements in percent units. Upper limits are at 95\% c.l.}
    \begin{tabular}{l c c}
   \hline
   \hline
Effect & \multicolumn{2}{c}{Percent impact in $SOU/BKG$ range:}\\
       & $0.2-0.5$ & $<0.2$ \\
    \hline
    $N_H$  & 1.5\% & 1.5\% \\
    Ab. table & 3-6\% & 3-6\% \\
    ML-MCMC & $<5\%$  & $<7\%$ \\
    RASS    &  $<6.6\%$ & $<9.4\%$ \\
SP\tablefootmark{*} &  $<10\%$ &  $<10\%$ \\
\hline
    \end{tabular}
    \tablefoot{
    \tablefootmark{*}{Using the difference between the scaled profile in the high and low contamination subsamples}}
    \label{tab:syslimits}
\end{table}
\begin{figure}
    \centering
  \includegraphics[width=0.45\textwidth]{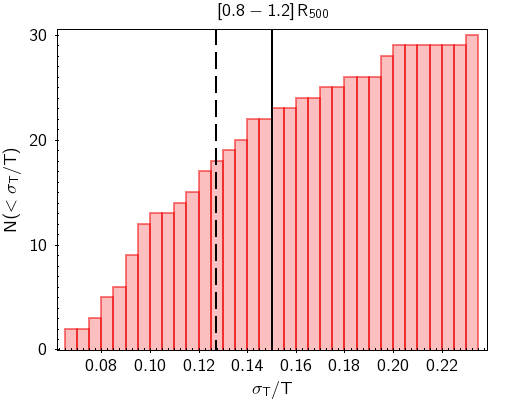}
  \caption{Cumulative distribution of clusters where the temperature measurement in the $[0.8-1.2]R_{500}$ bin reaches a nominal error as a function of the relative statistical error $\sigma_T/T$.  The black continuous line shows the goal in the CHEX-MATE feasibility (15\%), while the dashed line shows the median value on the sample ($12.7$\%) }
    \label{fig:error_budget}
\end{figure}
Besides the estimate of systematic errors induced by our method, it is also important to assess the statistical quality of our measurements. In Fig. \ref{fig:ML_vs_MC} we show the distribution of relative errors in the MCMC and ML fit: the bulk of the values is below 15\%, but in a few cases errors can reach values of 30-40\%. The size of the errors depends on several factors, such as the signal in the radial bin, the $SOU/BKG$ ratio and even the temperature value itself (with low temperature values featuring  smaller errors, even in relative terms, than higher temperatures). In the CHEX-MATE \xmm\ proposal preparation, we tuned the exposure time to reach an error of 15\% in the annulus $[0.8-1.2]R_{500}$, based on an empirical relation between the counts in the soft band and the error on temperature \citep[see][for more details]{paper1}. We can use our sample to verify this method {\it a posteriori}: we extracted spectra in this annulus for each DR1 clusters and fitted them with our pipeline. We used the ML fit, since we did not foresee using the MCMC approach at the time of the proposal. In Fig. \ref{fig:error_budget} we show the cumulative distribution of clusters with a symmetric relative temperature error smaller than a given $\sigma_T/T$. We reach the requirement of 15\% error for 22 clusters and the median error on the full sample is $12.7\%$. The maximum fractional error in this radial range is 23\%. We do not find any clear correlation between the size of the errors and the physical properties of the clusters. 

\section{Results}
\label{sec:results}
\subsection{The temperature profiles}
\label{sec:singleprofs}
\begin{figure*}[!t]
    \centering
    \includegraphics[width=\textwidth]{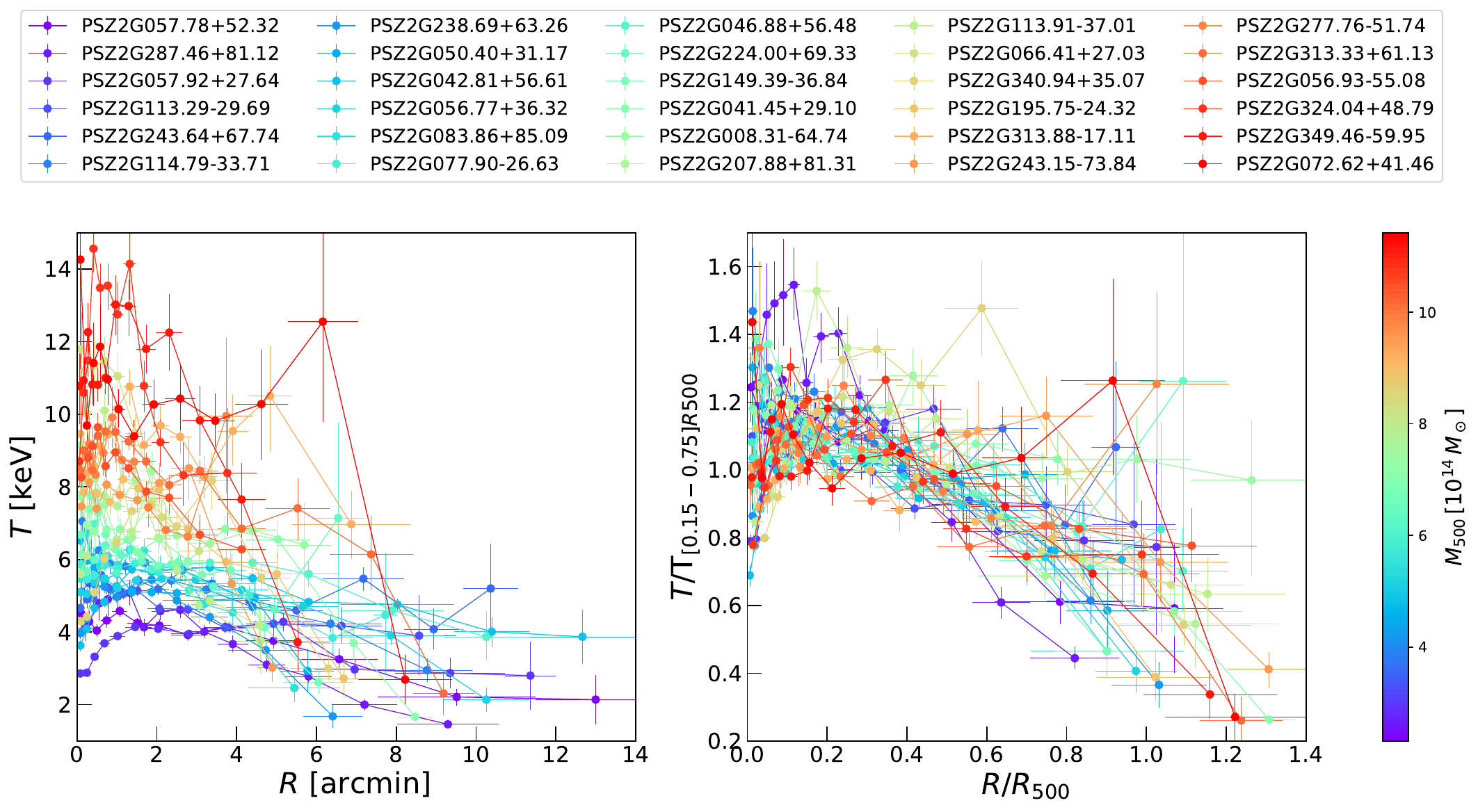}
    \caption{Individual temperature profiles of the 30 DR1 clusters, color-coded by the total mass. The left panel shows the temperature as a function of the projected position of the annuli in arcminutes, highlighting the large range of measured temperatures and extension of the DR1 sample. The right panel shows the rescaled profiles. }
     \label{fig:tprof_all}
\end{figure*}
We show the temperature profiles of the 30 DR1 clusters in Fig. \ref{fig:tprof_all}. The measured temperatures range from 2 to more than 14 keV, reflecting the large spread in mass of the CHEX-MATE and DR1 sample. The radial range of the profiles also reflects the presence of some nearby and very extended objects  and of high-redshift and compact clusters in the sample. The colour coding by total mass in the left panel of Fig. \ref{fig:tprof_all} highlights these dependencies. \\
We rescaled the profiles according to self-similar predictions 
 in the right panel of Fig. \ref{fig:tprof_all}. We divide the radial coordinate by $R_{500}$, estimated from the \planck{} SZ mass, reported in Table \ref{tab:allclusters}. We underline that these masses can be biased with respect to total masses and may scatter around the hydrostatic masses that would be derived from the X-ray data. As discussed in \citet{bartalucci23}, this may have an impact in the shape of the profiles and should be taken into account when comparing with simulations and with other samples. Concerning the y-axis, we rescaled the temperature by a mean value obtained by  fitting a spectrum extracted in an annulus in the range $[0.15-0.75]R_{500}$, i.e. excluding the core and the outer background- dominated regions. This scaling is not completely independent from the data in the profiles as it is extracted from the same data set. An alternative approach (which we will use for the scatter profile in Sec. \ref{sec:scatter} and for some literature comparison in Sec. \ref{sec:comparison}) is to rescale by an external value derived from the mass, through scaling relations.  
 The rescaling of the profiles reduces significantly the scatter in the profiles and the dependency on the cluster mass, as highlighted in the comparison of the two plots in Fig. \ref{fig:tprof_all}, both color-coded by cluster mass. Nonetheless, the scatter among the profiles is non-negligible, reflecting both the uncertainties on the measurements and the cluster to cluster variations (Sec. \ref{sec:scatter}). \\ 
 \begin{figure}
     \centering
     \includegraphics[width=0.49\textwidth]{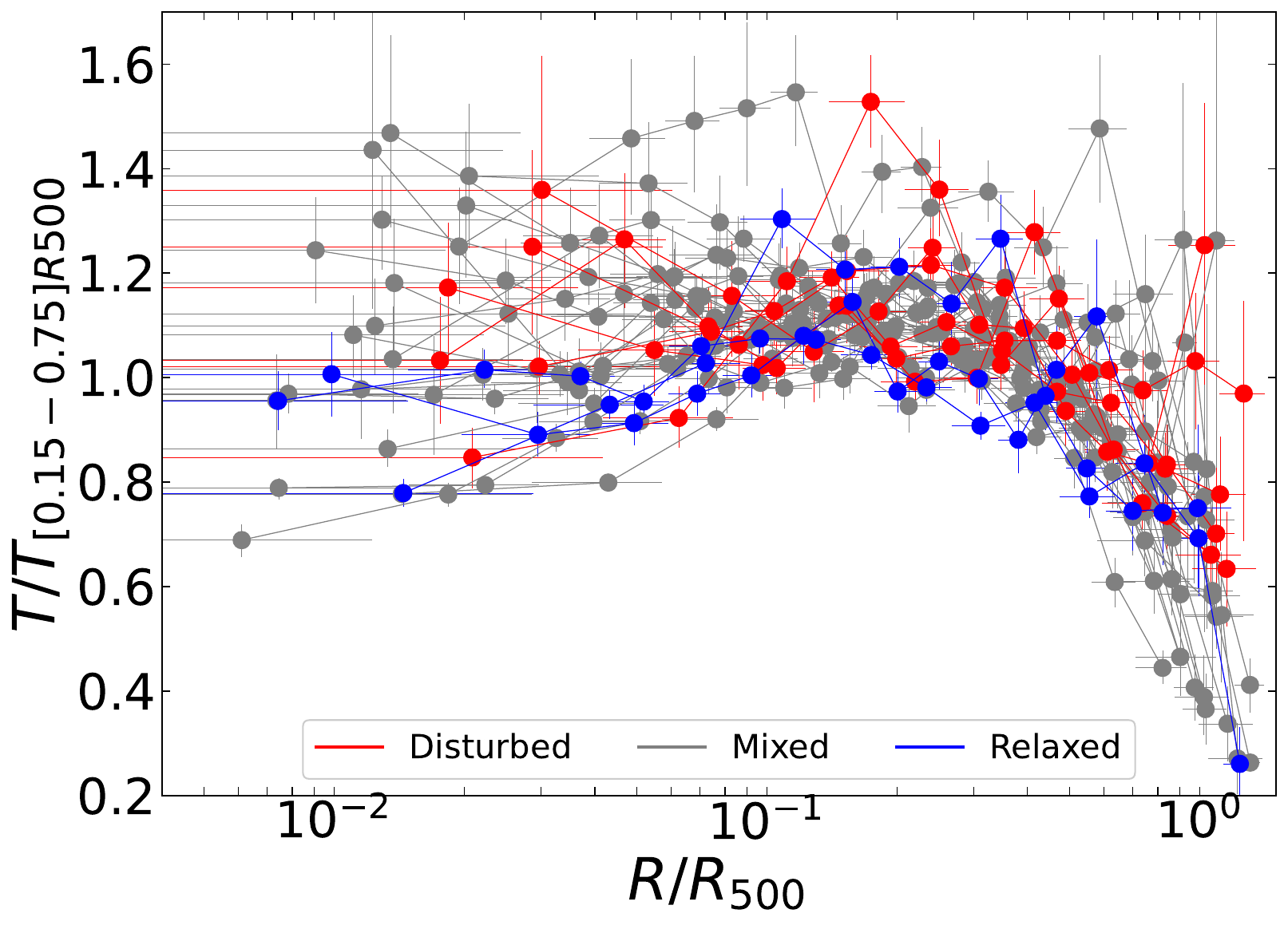}
     \caption{Rescaled temperature profiles of the DR1 sample: clusters belonging to the most relaxed class are shown in blue, to the most disturbed in red, and to the mixed class in gray, identified according to the $M$ parameter. We use the logarithmic scale on the x-axis to focus on the more central regions.}
     \label{fig:tprof-morpho}
 \end{figure}
 The temperature profiles in the central regions of galaxy clusters have been shown to correlate with the dynamical state of the cluster (e.g. \citealt{hudson10, leccardi10, pratt10}). In \citet{campitiello22}, we identified the most relaxed and the most disturbed clusters in the CHEX-MATE sample, according to the $M$ parameter, which is a combination of several dynamical indicators (surface brightness concentration, centroid shift, and power ratios). \citet{bartalucci23} showed that this classification segregates clusters with peaked emission measure profiles in the center from those showing a flatter slope. In Fig. \ref{fig:tprof-morpho}, we show the temperature profiles of the most relaxed, most disturbed and mixed clusters in the DR1 sample. We do not find a clear segregation of clusters, contrary to the emission measure profile, suggesting that our classification based on the $M$ parameter is not effective in separating clusters with different temperature profiles. 
 This may be due to the definition of the parameter itself, which combines indicators on different scales and is not tailored to measure the properties of the core. However, we should also note that we have only three clusters in DR1 belonging to the most relaxed class according to \citet{campitiello22} and none of them show very low values of the $M$ parameter. It is also, thus, possible that we are lacking the most extreme clusters which would drive the segregation, and this result should be verified in the full CHEX-MATE sample. Indeed, the KS test, used to check the consistency with the parent sample (Sec. \ref{sec:sample}), is not very sensitive to the tails of the distribution.\\

\subsection{The mean and median temperature profile}
\label{sec:meanprof}
\begin{figure*}
    \centering
    \subfloat{\includegraphics[width=0.49\textwidth]{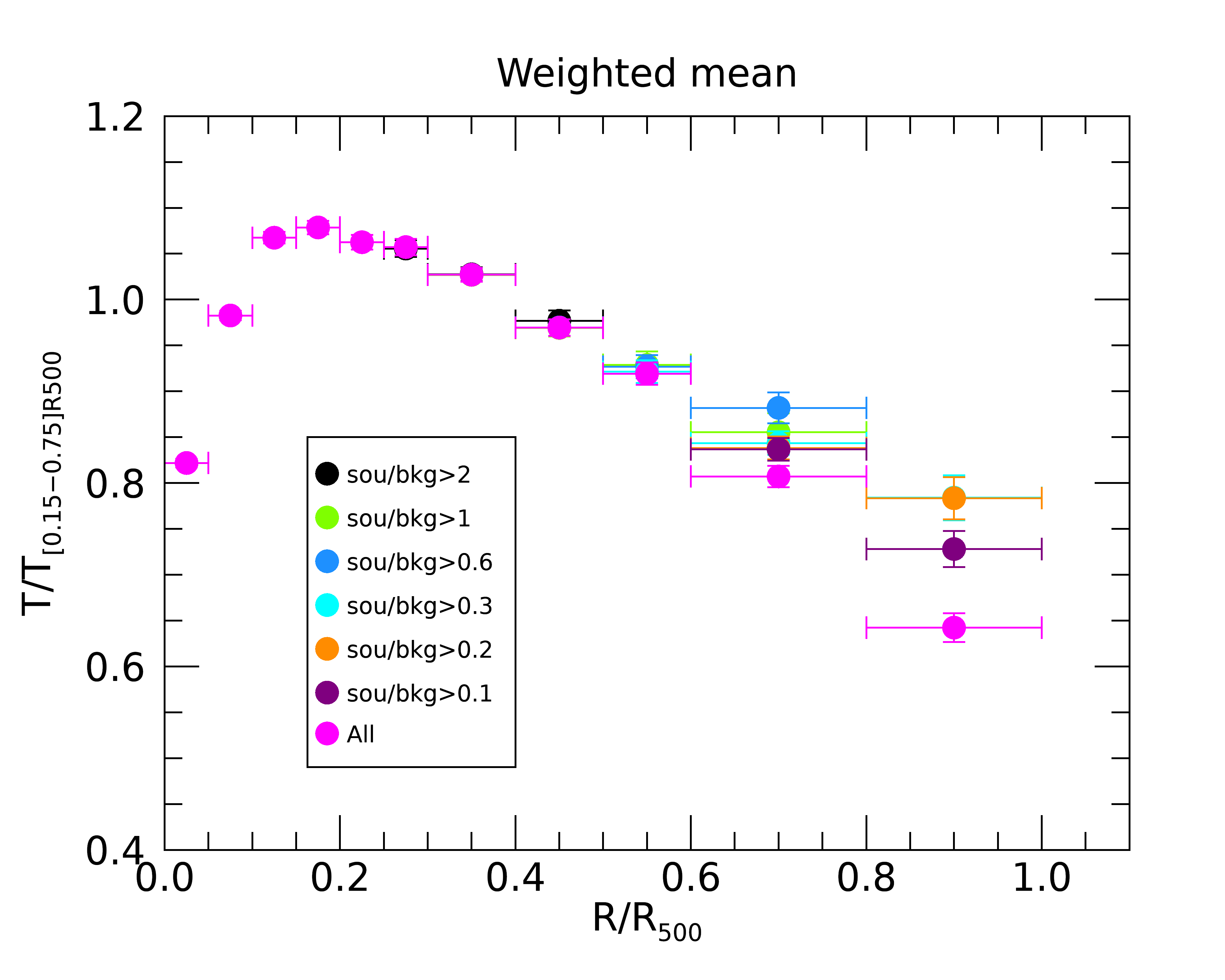}}
  \subfloat{\includegraphics[width=0.49\textwidth]{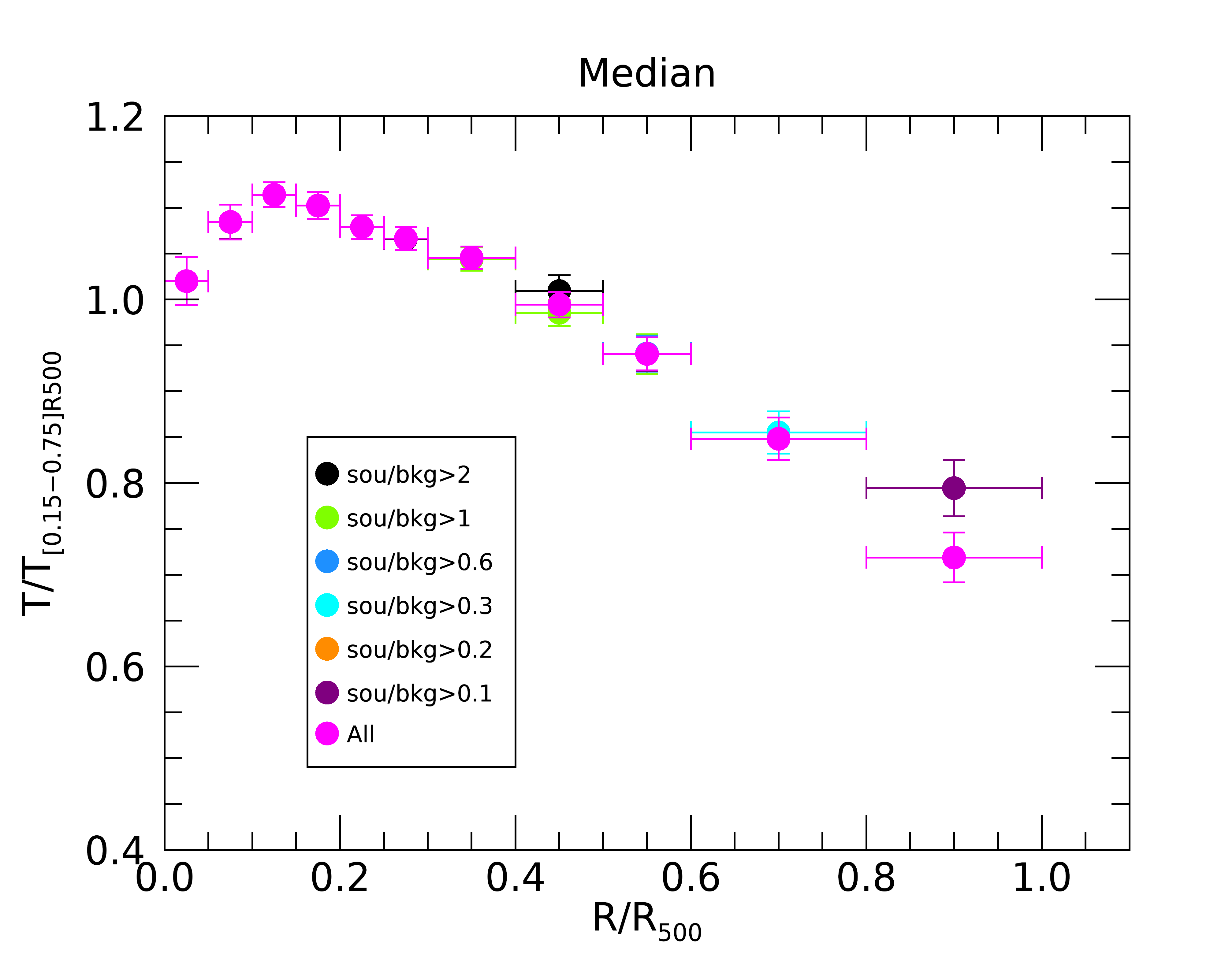}}
    \caption{Comparison of weighted mean ({\it left panel}) and median ({\it right panel})  temperature radial profiles applying different thresholds in the $SOU/BKG$ indicator. We plot only values with more than 15 (half sample) individual measurements.} 
     \label{fig:tprof_sbthresh}
\end{figure*}
Several methods can be applied to stack the rescaled temperature profiles of individual clusters (Sec. \ref{sec:singleprofs}) into a single ``average'' profile of the sample under study. For instance, \citet{leccardi08} report a weighted mean profile, obtained by weighting the $T_i$ individual measurements by their absolute errors  $dT_i$. They apply a further weight by considering the fraction of the original radial bin $r_{i}\pm dr_i$ which falls in the radial bin $R \pm dR$ used in the output ``average'' profile. This method has the advantage to account for the position and the extent of the original radial binning of the individual measurements with respect to the output mean profiles. Moreover, it considers the different errors of individual measurements, giving more weight to more precise estimates. However, it is important to note that the error bars on temperature measurements do not reflect only the statistical quality of the spectra from which they were derived, but also depend strongly on the temperature itself. Indeed, it is easier to measure low temperatures than high temperatures with \xmm, mainly because of the shape of the effective area and the higher intensity of the source with respect to the particle background at low energy. 
This is reflected into smaller uncertainties (also in relative terms) for low temperature values than for higher temperature measurements. It is easy to infer that the error weighting can bias low the average values in situations where we have a large spread in the temperature measurements, such as in the innermost bins, due to the presence of cool cores, and also in the external regions where measurements in background-dominated regions can lead to (biased) low-temperature values \citep{leccardi08}. An alternative approach used by other authors \citep[e.g.][]{arnaud10,lovisari19} is to compute the median of the individual measurements at a given radius rather than the weighted mean (see the comparison in \citealt{lovisari19} for the metal abundance profile). By construction, the median is less sensitive to the tails of the distribution and therefore should be less biased in the situations described above with low temperature values. Conversely, as it is not weighted, we do not account for the position of the initial radial binning with respect to the one of the final profiles. In our analysis, we estimate the error bars on the median profile by performing a randomization of the individual $T_i$ measurements in each bin within their error bars and recompute the median 1000 times. 

\begin{figure}
    \centering
    \includegraphics[width=0.5\textwidth]{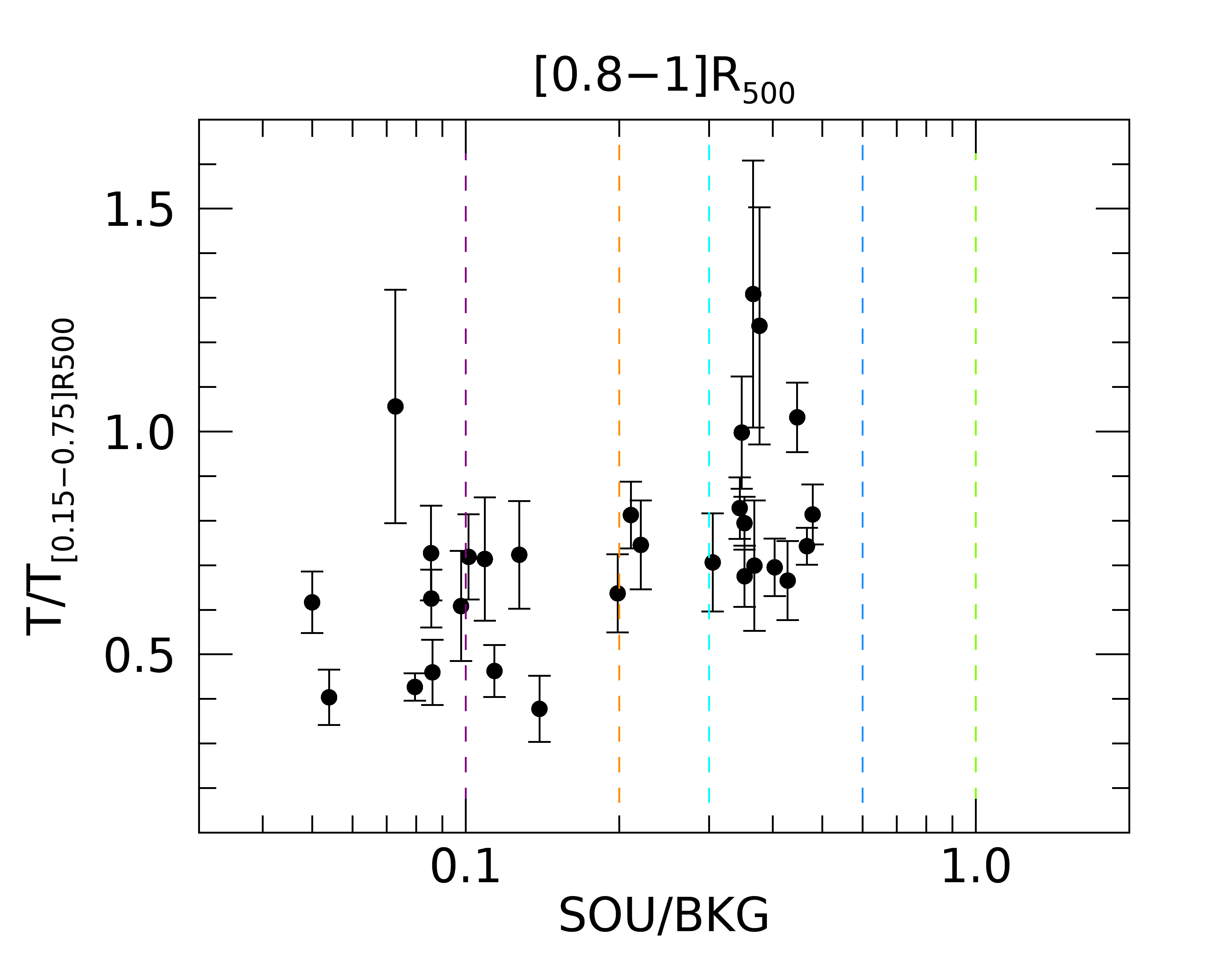}
    \caption{Distributions of the rescaled temperatures as a function of the $SOU/BKG$ in the extraction region for the measurements encompassing $0.9R_{500}$. The vertical dashed lines mark the thresholds used in Fig. \ref{fig:tprof_sbthresh}, with the same colour scheme.}
    \label{fig:distr_T_s2b}
\end{figure}

\begin{figure}
    \centering
    \includegraphics[width=0.5\textwidth]{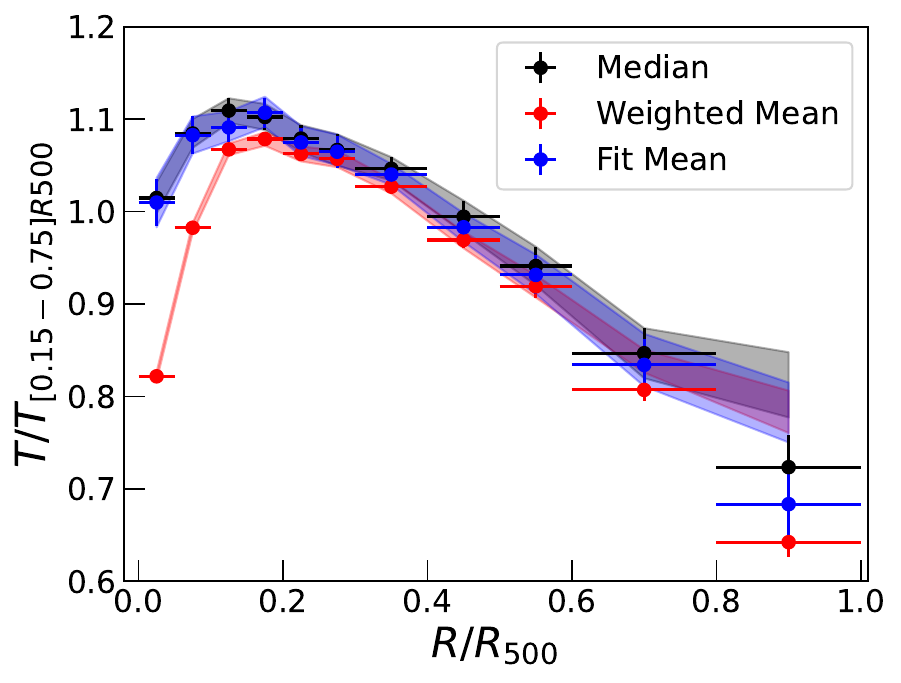}
    \caption{Comparison of the average profiles obtained with different methods: median (black points and grey shaded area), weighted mean (red) and fit mean (blue). Filled points refer to values derived by all measurements, while shaded regions show the values obtained when applying the $SOU/BKG>0.2$ selection.  Out of the core, the different methods return consistent results after excluding regions with low $SOU/BKG$, while if we use all measurements the weighted mean always returns lower values than the other methods.}
    \label{fig:compare_wmean_med_fit}
\end{figure}

In Fig. \ref{fig:tprof_sbthresh}, we show the average profile, both with the weighted mean (taking into account the radial bin fraction) and with the median. In both cases, following \citet{leccardi08}, we show the profiles obtained considering only temperature measurements in regions with the $SOU/BKG$ indicator above different thresholds (namely $2,1,0.5,0.3,0.2,0.1$ and no-threshold), plotting only the points where the average is obtained with more than 15 independent measurements (half sample). Focusing on the weighted mean profile (left panel), we note that profiles with different thresholds are almost identical up to $0.5R_{500}$ and start to show some differences only at larger radii. While up to $0.8R_{500}$ differences are not significant, 
in the radial bin between $0.8$ and 1 $R_{500}$, the mean values of the profile without applying any threshold and the one with $SOU/BKG>0.1$ (magenta and purple in the left panel of Fig. \ref{fig:tprof_sbthresh}) are significantly lower (at more than $3\sigma$) than the other values. This means that the mean profile is likely biased low, by a factor of about $15\%$. 
Conversely, the results obtained with higher thresholds do not deviate from one another, suggesting that a threshold as low as $SOU/BKG=0.2$ can be safely used to measure temperature profiles up to $R_{500}$, which is one of the goals of the CHEX-MATE project.\\ 
We underline that the value we consider here as a safe threshold for our measurements, $SOU/BKG>0.2$, is three times lower than the safe value suggested by \citet{leccardi08}, who introduced this diagnostic plot and also used the weighted mean for their profiles. This implies that the CHEX-MATE pipeline, allows to obtain reliable temperature profiles at much lower $SOU/BKG$ limit, where the cluster intensity is only 20\% of the background. 
We investigated if this improvement could be due to the MCMC fitting approach used here, but results obtained with the standard ML fitting do not differ significantly from what is shown in Fig. \ref{fig:tprof_sbthresh}. The improvement is most likely due to the more advanced physical background model introduced in the CHEX-MATE pipeline and to the high quality of our observations, tailored to reach a good accuracy in temperature measurements at $R_{500}$.\\ 
The median profile, shown in the right panel of Fig. \ref{fig:tprof_sbthresh}, does not show the steepening of the profiles obtained with low $SOU/BKG$ thresholds that was apparent in the weighted mean profile: except for the last bin, all measurements 
are very consistent within each other at a given radius. Even in the bin $[0.8-1]R_{500}$ the difference is less than $10\%$ and only appears if we include regions with $SOU/BKG<0.1$. This is qualitatively expected since the median is less sensitive to the low temperature values in the tails of the distribution, as it is apparent also in the inner two bins of the profile, where the median profile shows only a moderate decrease, while the weighted mean profile shows a much larger decrease, due to the presence of a few low temperature values, with small error bars, in cool-core clusters.  \\
To understand the effect of including low  $SOU/BKG$ measurements in our mean and median profiles we looked at the distribution of the individual measurements $T_{i}$ and $(SOU/BKG)_i$ values in different radial bins. Up to $0.6R_{500} $, the distribution of   $T_{i}$ as a function of $SOU/BKG$ is rather flat, therefore the exclusion of a few points at low $SOU/BKG$ does not alter significantly the distribution and its mean and median are not affected by the cut. However, at larger radii, we start to see a correlation between temperature and $SOU/BKG$, with lower temperatures measured in more background dominated regions. An example is shown in Fig. \ref{fig:distr_T_s2b} for the bin at $0.9R_{500}$: most $T_{i}/T_{[0.15-0.75]R500}$ values below $0.7$ are measured in regions with $SOU/BKG<0.2$. Therefore, if we exclude these regions the median changes from $0.72$ to $0.79$ ($\sim 9\%$) while the weighted mean moves from $0.63$ to $0.78$ ($\sim 18\%$). \\
In Fig. \ref{fig:compare_wmean_med_fit} we compare the weighted mean, the median and the mean fit profile. The latter is computed as the best-fit mean value in each radial bin, approximating the distribution with a Gaussian function, and is a byproduct of the scatter computation described in Sec. \ref{sec:scatter}, to which we refer for details. We show both the profiles obtained with all measurements and those excluding the regions with $SOU/BKG<0.2$ (shaded areas). In the central regions, the weighted mean is biased low because of the presence of cool cores,  while the mean and median return consistent results. If we consider the values obtained with all measurements (filled circles), the weighted mean tend to be always lower than the other methods and in the last bin at $0.9R_{500}$ the difference with the median is maximal, with the mean returning an intermediate value. Conversely, out of the core all profiles are consistent when we exclude the regions with $SOU/BKG<0.2$. We report the weighted mean and the median profiles under this condition in Table \ref{tab:mean_median}).
Since the median is the quantity less affected by the presence of low $SOU/BKG$, from now on we shall use it for the comparison with other samples in the literature (Sec. \ref{sec:comparison}), for which it is not possible to apply the  $SOU/BKG$ selection.

\begin{table}[]
    \centering
     \caption{Weighted mean and median temperature profiles (rescaled by $T_{[0.15-0.75]R500}$) and their ratio, after excluding regions with $SOU/BKG<0.2$.}
    \begin{tabular}{c c c c c}
    \hline
    \hline
    $R/R_{500}$ & $\Delta R/R_{500}$ & Weighted Mean & Median & Ratio \\
    \hline
0.025 &  0.025 &  $0.822 \pm 0.005$ &  $ 1.015 \pm 0.017 $ &  0.810 \\
0.075 & 0.025 & $0.983 \pm  0.005 $ &  $ 1.085 \pm 0.015 $ &   0.906 \\
0.125 & 0.025 & $1.067 \pm 0.006  $ &  $ 1.109  \pm 0.013  $ &  0.962  \\
0.175 & 0.025 & $1.079  \pm 0.007 $ &  $  1.102  \pm 0.014  $ &  0.978  \\
0.225 & 0.025 & $1.062  \pm 0.008 $ &  $  1.079  \pm 0.015  $ &  0.985  \\
0.275 & 0.025 & $1.057  \pm 0.009 $ &  $  1.067  \pm 0.017  $ &  0.991  \\
0.350 & 0.050  & $1.027  \pm 0.007 $ &  $ 1.046  \pm 0.013  $ &  0.982  \\
0.450  & 0.050 & $0.969  \pm 0.009 $ &  $ 0.994  \pm 0.017  $ &  0.975 \\
0.550  & 0.050 & $0.919  \pm 0.012  $ &  $0.941  \pm 0.021  $ &  0.977  \\
0.700  & 0.100 & $0.838  \pm 0.012  $ &  $ 0.847  \pm 0.026  $ &  0.990  \\
0.900  & 0.100 & $0.783  \pm 0.023  $ &  $0.813  \pm 0.036   $ & 0.964  \\
\hline
    \end{tabular}
    \label{tab:mean_median}
\end{table}

\subsection{The intrinsic scatter}
    \label{sec:scatter}
\begin{figure}
    \centering
    \includegraphics[width=0.5\textwidth]{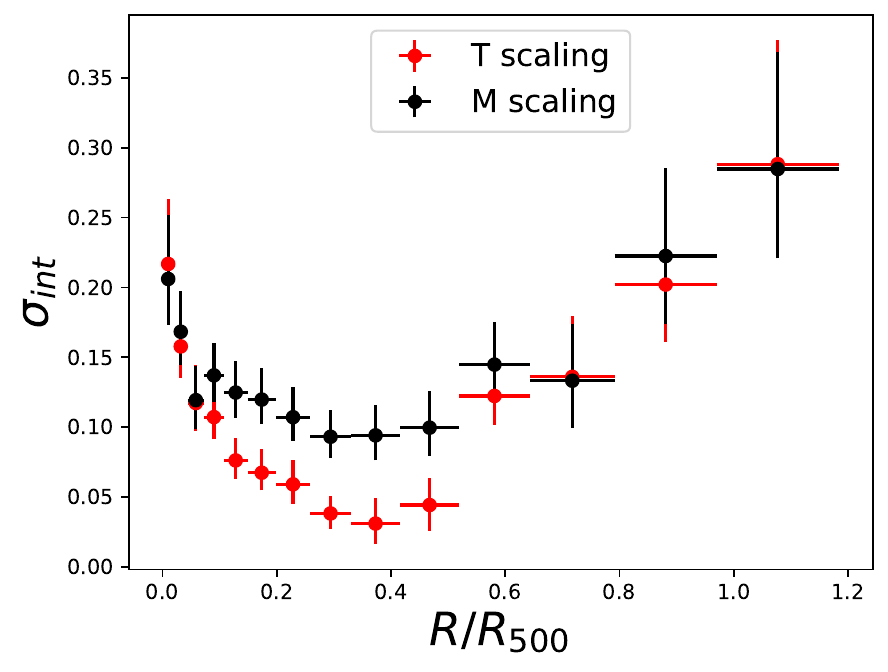}
    \caption{Profile of the intrinsic scatter in the scaled temperature measurements of DR1 clusters. The two profiles refer to the two different methods for scaling temperature: a global temperature measured in the $[0.15-0.75]R_{500}$ annulus (red points) and the temperature $T_{500}$ estimated from the mass (black points).}
    \label{fig:sigma_int}
\end{figure}
In Fig. \ref{fig:tprof_all}, we qualitatively show that the scatter in the profiles reduces significantly when we rescale temperatures by their mean value in the range $[0.15-0.75]R_{500}$ and radii by $R_{500}$. The residual scatter after scaling is  due both to the statistical uncertainties on the individual measurements and to the intrinsic scatter between the profiles, which reflects the variations from cluster to cluster and can shed light on the non self-similar processes affecting the ICM. 
We computed the radial profile of the intrinsic scatter by modeling the individual scaled $T_i^s=T_i/T_{[0.15-0.75]R500}$ measurements in a given radial bin with a Gaussian, whose width is the sum of the squares of the statistical and intrinsic scatter.
We then fit them in a Bayesian framework to derive the parameters of the Gaussian, i.e. the mean (used in Sec. \ref{sec:meanprof} and Fig. \ref{fig:compare_wmean_med_fit}) and the intrinsic scatter (see \citealt{ghirardini19} and \citealt{bartalucci23} for more details). The resulting intrinsic scatter profile for the DR1 temperatures is shown in Fig. \ref{fig:sigma_int}. We show two profiles, obtained with two different scalings of the temperature values. For the red one, we used the mean temperature estimated in the annulus $[0.15-0.75]R_{500}$, as shown in Sec. \ref{sec:singleprofs}, while for the black one we used an independent estimate of $T_{500}$, derived from the SZ $M_{500}$, using Eq. 10 in \citet{ghirardini19}.
We immediately notice that the scatter profile with the $T_{[0.15-0.75]R500}$ scaling is significantly lower than the one with the external mass scaling in the radial range $[0.1-0.6]R_{500}$. Indeed, the covariance between the measured points and the scaling quantities has been shown to induce suppression of the scatter \citep{pratt22}. Conversely, the scaling with the temperature derived from the mass relation is independent of the measured values, since we use the mass derived from the SZ signal, and the scatter profile is indeed larger (see Sec. \ref{sec:comparison}). For this reason, from now on, we shall use the mass scaling when measuring the scatter. \\
The scatter profile shows the typical behavior of thermodynamic quantities, highlighted also in \citet{ghirardini19}: it shows a high value in the center, reflecting the difference between cool-core and non cool-core clusters, then reaches a minimum between $0.2$ and $0.5R_{500}$ and starts increasing again at large radii.
We note that the scatter in the temperature profiles is low, around 10\% at its minimum value, smaller than the values observed in the Emission Measure profiles for the full CHEX-MATE sample \citep{bartalucci23}. Indeed, \citet{ghirardini19} showed that temperature is the thermodynamic quantity with the smallest intrinsic scatter.

\subsection{Fit of the temperature profile}
\label{sec:fitPL}
\begin{figure} 
    \centering
    \includegraphics[width=0.5\textwidth]{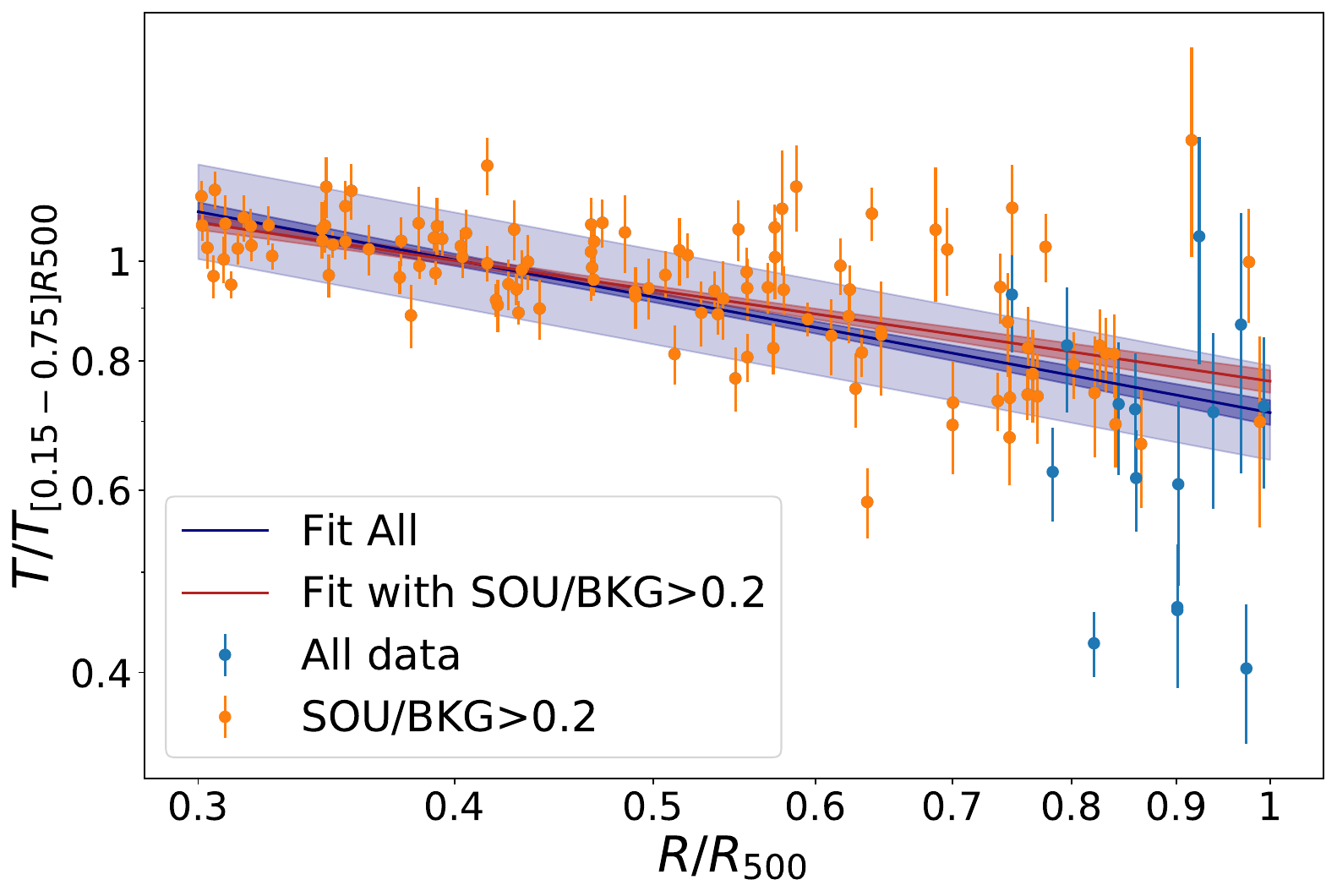}
    \caption{Fit of the temperature profile with a power-law for $R>0.3R_{500}$. Orange (blue) points represent regions where $SOU/BKG$ is greater (smaller) than $0.2$. The blue line and dark envelope represent the best fit function and its errors on the full dataset, while the light shaded area marks the intrinsic scatter. The red line and envelope show the best-fit for regions with $SOU/BKG>0.2$ (intrinsic scatter not shown for clarity)}.  
    \label{fig:fit_PL}
\end{figure}

The shape of the temperature profiles of galaxy clusters has been widely discussed in the literature, looking for a universal function  \citep[e.g.][]{baldi12,ghirardini19}. The most used function has been introduced by \citet{vikhlinin06}, originally applied to the three-dimensional deprojected profiles, but often used also to model the projected profiles with different choices of the parameters \citep[e.g.][]{baldi12,ghirardini19}. This model predicts a temperature gradient in the central regions, which however is typically present only in cool-core clusters, and it is thus customary to fit separately the profiles of the two classes of objects. As discussed in Sec. \ref{sec:singleprofs} we do not have a clear separation between the two classes in our sample and the number of objects is relatively small. We thus prefer to postpone this analysis to the full CHEX-MATE sample. Nonetheless, beyond the core regions, the temperature profiles of cool core and non-cool core clusters have been shown to be quite similar, thus allowing us to fit the entire sample with a single function. Moreover, in this radial regime the profile can be well approximated with a power-law with $T/T_{[0.15-0.75]R500}=N(R/R_{500})^{-\alpha}$. We decided to fit the profiles for $R>0.3R_{500}$ and derive the best fit parameters for the slope, the normalization and the intrinsic scatter, first using the full dataset and then considering only measurements with $SOU/BKG>0.2$. In both cases, we perform our fit in a Bayesian framework, as presented in Sec. \ref{sec:scatter}.
The points and best fit functions are shown in Fig. \ref{fig:fit_PL}. For the full dataset, we measure the slope $\alpha=0.37 \pm 0.04$ and the intrinsic scatter $\sigma_{int}=0.11\pm0.01$. If we exclude the regions with $SOU/BKG<0.2$, the slope becomes slightly flatter $\alpha=0.29\pm0.03$ and the scatter reduces to $\sigma_{int}=0.07\pm0.01$. We also split the radial range in two bins ($0.3<R/R_{500}<0.55$ and $0.55<R/R_{500}<1$) and fit each of them with a separate power-law \citep[see e.g.][]{ghirardini19}. In the $[0.3-0.55]R_{500}$ radial bin we do not have points with $SOU/BKG<0.2$, therefore we performed a single fit and measured $\alpha=0.25\pm0.05$. In the $[0.55-1]R_{500}$, the profile steepens and we find $\alpha=0.70\pm0.15$ for the full sample and $0.44\pm0.15$ for $SOU/BKG>0.2$. (see also Table \ref{tab:slopes})\\
The steepening of the profile while moving to external regions has been already observed in the literature \citep[e.g.][]{ghirardini19} and our results are in qualitative agreement (see also Sec. \ref{sec:comparison}). We note however that the steepening is less pronounced if we consider only measurements with $SOU/BKG>0.2$ and that even in the full radial range the best fit profile is flatter than when using all measurements (Fig. \ref{fig:fit_PL}). This is in agreement with the results already shown for the mean and median profile in Sec. \ref{sec:meanprof} and it is likely due to temperature measurements biased low in heavily background-dominated regions (Fig. \ref{fig:distr_T_s2b}). We checked the results presented in this section also using a $SOU/BKG$ estimator computed in the soft band and we find consistent results.  

\section{Discussion}
\label{sec:discussion}

\subsection{Comparison with previous results}
\label{sec:comparison}
\begin{figure*}
    \centering
    \subfloat{\includegraphics[width=0.5\textwidth]{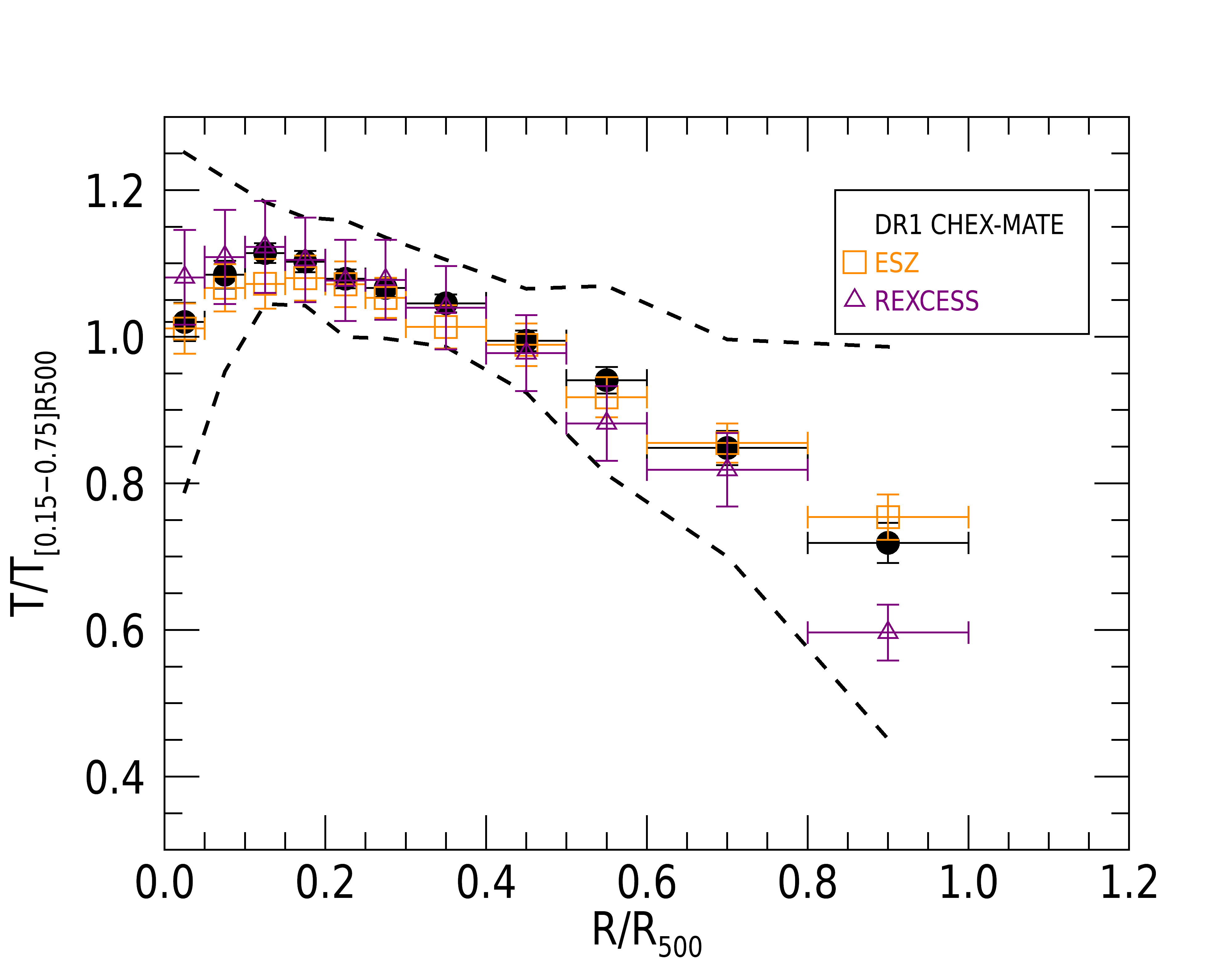}}
    \hspace{-0.5 cm}
    \subfloat{\includegraphics[width=0.5\textwidth]{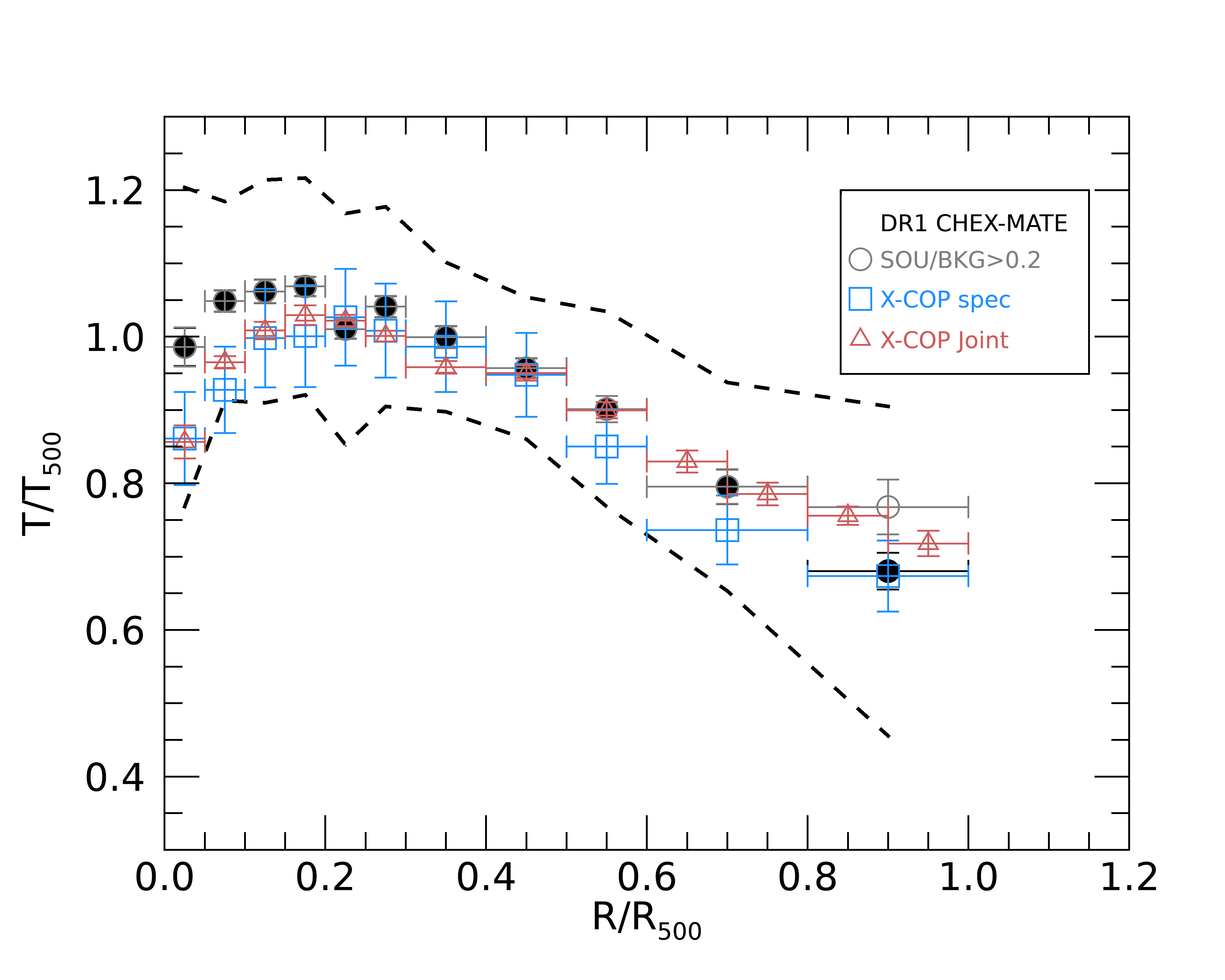}}
    \caption{Comparison of median temperature profiles of the CHEX-MATE DR1 sample (black circles) with other samples studied in the literature. {\it Left panel}: comparison with ESZ and REXCESS median profiles, where we have scaled temperature values by the best fit in a region $[0.15-0.75]R_{500}$. {\it Right panel}: comparison with X-COP, both the standard results from the X-ray analysis (blue squares) and the joint fit with SZ data (red triangles). The grey empty circles show the DR1 median profile after applying the $SOU/BKG>0.2$ selection. Here the temperature scaling $T_{500}$ is computed from the mass with the expression in \citet{ghirardini19}, both for X-COP and for CHEX-MATE data. Dashed lines show the total scatter in the DR1 profile.}
     \label{fig:median_literature}
\end{figure*}
In this section, we compare our average profiles with results from other samples with different selections. We tried to be consistent in our comparisons, computing the median from the individual profiles for the literature samples with the same code used for DR1 analysis and applying a similar scaling on temperature whenever possible. Since we cannot apply $SOU/BKG$ thresholds in the comparison samples, we do not apply it to the DR1 sample either. We focus only on cluster samples observed by \xmm\ and for which we had access to the individual profiles.\\
In the left panel of Fig. \ref{fig:median_literature}, we compare our median temperature profile with the one of REXCESS \citep{pratt10} and ESZ (a sample of 62 SZ-selected clusters described in \citealt{planck_pipV}). For these samples, we could rescale the temperature by the value measured in an annulus $[0.15-0.75]R_{500}$, as we do for our data, with the only difference that $R_{500}$ in the comparison samples is obtained from $M_{Y_X}$ while in DR1 from $M_{SZ}$. Moreover the binning scheme used to derive the temperature profiles in both samples is very similar to the one we used here.
REXCESS is a purely X-ray selected sample \citep{bohringer07}, covering a redshift range $0.05-0.18$ and a mass range $M_{500}$ in $[1-8]\, 10^{14} M_\odot$ \citep{pratt10}. Conversely, ESZ is the intersection of the 189 clusters detected by \planck{} in the Early SZ sample \citep{planck_esz} and the \xmm{} archive as of mid-2011 \citep{planck_esz_calib}. The DR1 median profile is consistent, at less than one sigma, with the other measurements in almost all radial bins. 
We note that the error bars
of the DR1 profiles are always smaller than in the other samples although with a similar sample size. This may be related to the better quality of the CHEX-MATE data, tailored to reach a 15\% error on the temperature measurements at $R_{500}$, with respect to the other samples. The only significant difference is with REXCESS in the $[0.8-1]R_{500}$ bin, where temperature measurements are more sensitive to the details of the background treatment. Moreover, the different selection as well as the lower median mass in the sample could also have a role in the difference. \\
In the right panel of Fig. \ref{fig:median_literature} we compare the DR1 median profile with that of the X-COP sample, both from the standard X-ray analysis (blue squares) and with the joint fit with SZ data (red triangles, see \citealt{ghirardini19} for more details). Here, the background treatment of the \xmm{} data is more similar to what we used, but the binning and observation strategy is different. Moreover, we could not scale the temperature by an external value in the same region, so we relied on rescaling by a predicted temperature from the mass, as introduced in Sec. \ref{sec:scatter}, using in both cases Eq. 10 in \citet{ghirardini19}. This could induce some difference, since for X-COP we used the hydrostatic masses rather than $M_{SZ}$. In the external regions, the agreement between the DR1 profile and the spectroscopic X-COP is very good. Only in the last radial bin, the joint X-SZ measurement is slightly larger than the DR1 profile, but it is interesting to note that it is consistent with our measurement once we exclude low $SOU/BKG$ regions (grey empty circles in Fig. \ref{fig:median_literature}). This is further indirect indication of a bias in measuring X-ray temperatures in strongly background dominated spectra.
There are some differences in the more internal regions which are at least partly due to the different fraction of cool cores in the two samples and to the different binning. \\
\begin{figure}
    \centering
    \includegraphics[width=0.5\textwidth]{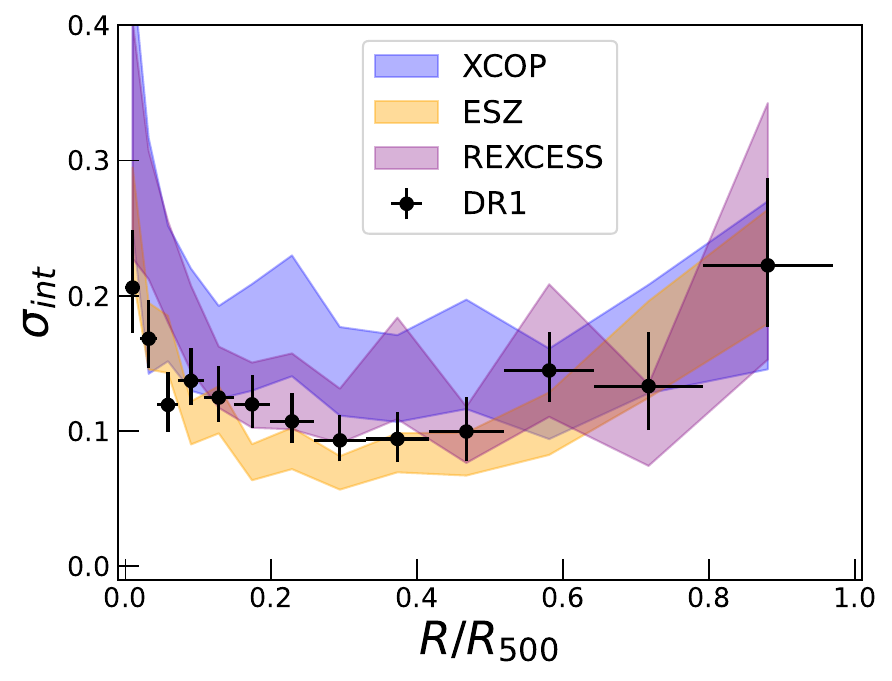}
    \caption{Comparison of the intrinsic scatter profiles in CHEX-MATE DR1 (black points) with other samples in the literature (colored shaded areas), showing consistent results at all radii.}
    \label{fig:compare_sigma}
\end{figure}
In Fig. \ref{fig:compare_sigma}, we compare the scatter profile of CHEX-MATE DR1 with the literature samples discussed above. The DR1 scatter profile is consistent with that of REXCESS and ESZ at all radii and only marginally smaller  than the one in X-COP between $[0.15-0.5]R_{500}$. We note that in all cases we used $T_{500}$ derived from the mass to rescale the temperature measurements, but only for the DR1 sample the scaling value is independent, as it is derived from the SZ masses. Indeed, for X-COP we used the hydrostatic masses, while for ESZ and REXCESS we used the mass derived from $Y_X$. In both cases they are not independent from the temperature measurements and this may in principle induce a covariance which suppresses the scatter.  \\
We fitted the profile with a power-law also for the comparison samples in the same radial ranges used for DR1 (Sec. \ref{sec:fitPL}) and report the best fit slopes in Table \ref{tab:slopes}. For the comparison samples, we always use all available measurements in our fits. 
The slope values are typically consistent within 2$\sigma$ with the values measured for our sample using all our measurements. It is interesting to note that, especially in the radial bin $[0.55-1]R_{500}$, all profiles show a steep slope and only our measurement with $SOU/BKG>0.2$ is significantly flatter, which is however consistent with the slope measured from the combination of X-ray and SZ data in X-COP by \citet{ghirardini19} in a similar radial range ($\alpha=-0.34\pm0.18$ for $0.56<R/R_{500}<0.86$). This suggests that once we exclude heavily background-contaminated regions, our analysis allows us to reproduce the shape of the temperature profiles as derived by other independent methods, in principle less affected by background systematic than X-ray measurements. 
During the fit procedure, we scale our profiles, as well as REXCESS and ESZ, with $T_{[0.15-0.75]R_{500}}$, while the X-COP profiles are rescaled based on the mass.

\begin{table}[]
    \centering
     \caption{Best fit slope in different radial range for our sample and comparison samples.}
    \begin{tabular}{c c c c}
    \hline
    \hline
      Sample   & $[0.3-1]$&  $[0.3-0.55]$& $[0.55-1]$\\
      \hline
         DR1 all & $0.37\pm0.04$ & $0.25\pm0.05$ & $0.70\pm0.15$ \\
         DR1 high $S/B$ & $0.29\pm0.03$ & $0.25\pm0.05$ & $0.44\pm0.15$ \\
         REXCESS & $0.50\pm0.05$ & $0.20\pm0.05$ & $0.99\pm0.19$ \\
         ESZ & $0.38\pm0.03$ & $0.16\pm0.04$ & $0.73\pm0.14$ \\
         X-COP spectro & $0.45\pm0.05$ & $0.16\pm0.14$ & $0.61\pm0.16$ \\
         simulations & $0.34\pm0.03$ & $0.24\pm0.07$ & $0.41\pm0.08$ \\
        \hline
    \end{tabular}
    \label{tab:slopes}
\end{table}

\subsection{Comparison with numerical simulations}
\begin{figure*}
    \centering
    \includegraphics[width=0.45\textwidth]{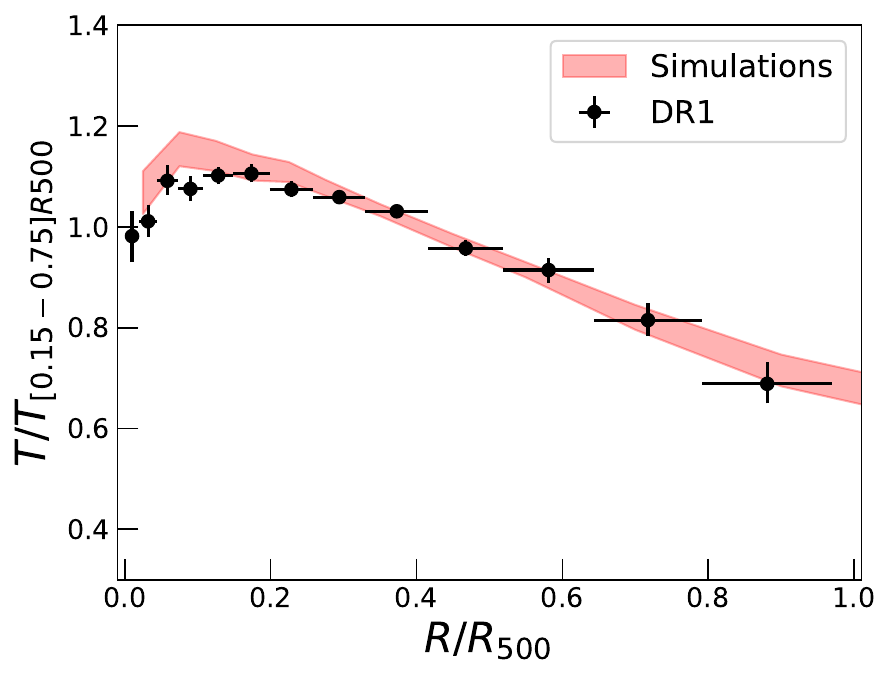}
    \includegraphics[width=0.45\textwidth]{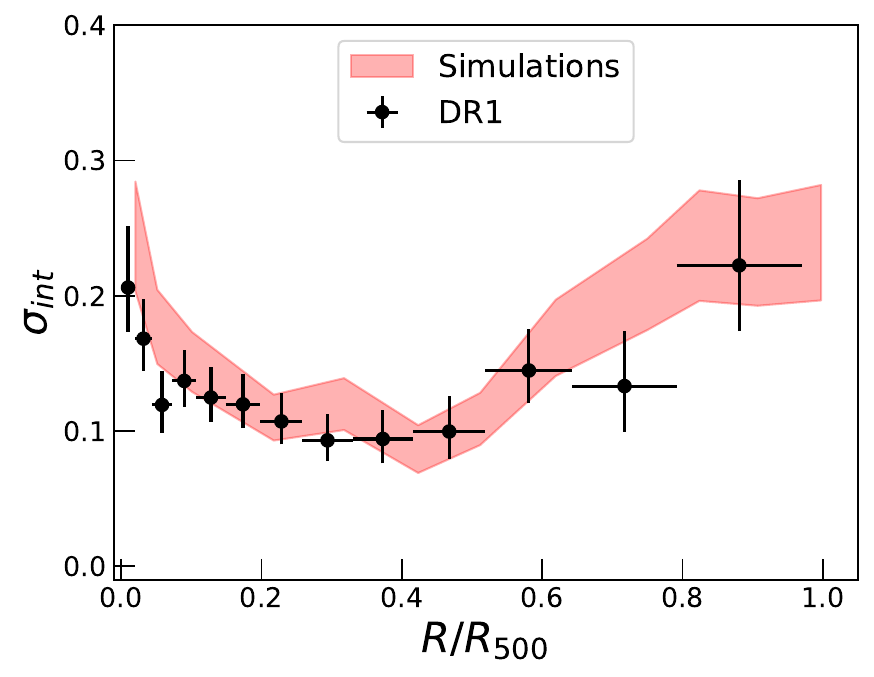}
    \caption{Comparison of the mean temperature ({\it left panel}) and intrinsic scatter ({\it right panel}) of the DR1 sample with those obtained from numerical simulations, using both spectroscopic-like  temperatures. } 
    \label{fig:compare_simul}
\end{figure*}
In this section, we compare our results with predictions from numerical cosmological simulations. More specifically, we used the GADGET-X version of the simulations from the \textsc{The Three Hundred} collaboration \citep{cui18}, which contains a relatively large set of massive clusters and is thus well-suited to compare with the CHEX-MATE sample. The baryon physics adopted in these simulations includes the star formation models and stellar and AGN feedback, described in \citet{rasia15}.
We built a simulated counterpart for the DR1 sample using a procedure similar to the one adopted in \citet{bartalucci23}, to which we refer for details. Briefly, we match each observed cluster with a simulated one, in the closest redshift snapshot and with similar mass, taking into account a $20\%$ possible underestimate of the mass $M_{SZ}$ of DR1 clusters due to hydrostatic bias. For each simulated cluster, we extract spectroscopic-like temperature \citep{mazzotta04} profiles in fixed bins of $R_{500}$, projected in three orthogonal directions.  We compute a temperature and an error in each radial bin for each object, using the mean and dispersion of the three projections.\\
We then analyze the profiles of simulated clusters as we do for observations (Sec. \ref{sec:scatter}), to obtain the mean profileshown in Fig. \ref{fig:compare_simul}. The mean scaled profile, obtained by rescaling temperatures by the value measured in the annulus $[0.15-0.75]R_{500}$, shows a nice agreement with the observed profile at $R>0.3R_{500}$. This is confirmed by the fit with a power-law of the simulated profiles, which returns a slope $0.34\pm0.03$ in the $[0.3-1]R_{500}$ radial range, consistent with the value for the DR1 sample (see Table \ref{tab:slopes}). It is interesting to note that for $R>0.55R_{500}$ we obtain a slope of $0.41 \pm 0.07$, which is similar to the value obtained using only the regions with $SOU/BKG>0.2$, thus supporting our choice of excluding the more background-dominated regions. We see some differences in the central regions, which may be due to the different morphological mix between the observed and simulated sample, but also to the difficulties of simulations to fully reproduce the complex physics of cool core clusters.
\\
To compute the profile of intrinsic scatter, also shown in Fig. \ref{fig:compare_simul}, both for observations and simulations we rescale the temperature by $T_{500}$ derived from the mass to minimize the covariance. For simulations, we reduce the true masses by a factor of 20\% to mimic "hydrostatic-like" masses.
The scatter in simulated clusters shows a very nice agreement with the observations, suggesting that our simulations and the DR1 sample are consistently sampling the variety of properties of the clusters population. Moreover, it suggests that these simulations capture the main physical processes (gravity-driven accretion and feedback processes) driving the shape of the profiles and their diversity.

\section{Summary and conclusions}
In this paper, we present the pipeline that will be used in the CHEX-MATE collaboration for spectral extraction in radial annuli, fitting and production of projected radial profiles and we apply it for the study of the temperature profile of a subsample of 30 objects, representative of CHEX-MATE. \\
The pipeline combines the best practices developed during previous projects, namely REXCESS \citep{pratt07,pratt10}, Planck \citep{planck_esz_calib,planck_pipV}, X-COP \citep{tchernin16,ghirardini19}, M2C \citep{bartalucci18,bartalucci19}, taking into account the peculiar characteristics of the CHEX-MATE sample, including both nearby extended objects and more distant compact clusters \citep{paper1}. The main novelties of our pipeline, presented in this paper for the first time, are the construction of a physical model for the particle background (both induced by high-energy Galactic cosmic rays and soft protons) for all EPIC detectors and the application of a Bayesian MCMC framework, allowing propagation of uncertainties on the background parameters up to the final results of the spectral analysis.  \\
We applied the pipeline to a subset of 30 CHEX-MATE clusters (DR1), which was built to cover the different observation properties of the full sample (background level, soft proton contamination) and, at the same time, to be representative of the physical properties of the parent sample, including the different angular sizes of the targets.  We measured temperatures ranging from $\sim 2$ to $\sim 14$ keV in different clusters and regions and we could map the profiles from the cores to the external regions around $R_{500}$ (Fig. \ref{fig:tprof_all}). We have shown that, with the CHEX-MATE pipeline and data quality, we are able to recover temperatures in regions where the source intensity is as low as 20\% of the background in the $0.7-10$ keV band, a significant improvement with respect to past analysis (\citealt{leccardi08} considered reliable only measurements in regions where $SOU/BKG>0.6$). We estimated that the choices that we inevitably had to take during the spectral fitting and background modeling impact the temperature measurements for less than 10\%, as long as we focus in regions with $SOU/BKG>0.2$. \\
We used our results to measure the weighted mean and median temperature profile and found that the latter is less affected by regions with low $SOU/BKG$. We thus use the median profile to compare our results with literature samples, where we cannot apply the $SOU/BKG$ cut, and found consistent results. We also measured the profile of the intrinsic scatter and showed that scaling by a mean temperature $T_{[0.15-0.75]R500}$ suppresses the scatter with respect to the scaling for an external quantity, such as the gravitational temperature derived from the mass(Fig. \ref{fig:sigma_int}). With the external scaling the agreement is very good, both with other samples (Fig. \ref{fig:compare_sigma}) and with simulations (Fig. \ref{fig:compare_simul}).
We modeled the temperature profile at $R>0.3R_{500}$ with a piece-wise power-law and noticed that the profile steepens going outwards, in agreement with other samples in the literature. It is interesting to note that in the radial range $[0.55-1]R_{500}$ the slope of the profile is steep and consistent with all other X-ray measurements in the literature. 
However, once we exclude heavily background-contaminated regions with $SOU/BKG<0.2$, our analysis returns a flatter profile, which reproduces the shape of the temperature profiles derived by the joint fit of SZ and X-ray measurements \citep{ghirardini19}, in principle less affected by background systematic than X-ray spectroscopic measurements, and matches the predictions of numerical simulations. \\
The temperature profiles presented in this paper will be used in combination with the surface brightness measurements shown in \citet{bartalucci23} to reconstruct the three dimensional radial profiles of most thermodynamic quantities of the ICM,  the total mass profile through hydrostatic equilibrium equation, and all the related integrated quantities, with the purpose of addressing the main scientific question of the CHEX-MATE projects \citep{paper1}.  We will apply the CHEX-MATE pipeline presented in this paper for spectral extraction and fitting to the full CHEX-MATE sample of 118 clusters to extract radial temperature profiles for all of them.
This will allow us to extend the analysis presented in this paper to the full sample and thus to perform the statistical comparison which was not possible with only 30 objects. The full sample will allow us to compare the temperature profiles in different mass and redshift ranges, as well as in different dynamical states. 

\begin{acknowledgements}
We thank the referee for her/his timely report and for providing constructive comments. 
Based on observations obtained with XMM-Newton, an ESA science mission with instruments and contributions directly funded by ESA Member States and NASA.
M.R., F.G., S.E., S.M., I.B., H.B., S.D.G.,  F.D.L., S.G., L.L., P.M. and G.R. acknowledge the financial contribution from the contracts Prin-MUR 2022, supported by Next Generation EU (n.20227RNLY3 {\it The concordance cosmological model: stress-tests with galaxy clusters}), ASI-INAF Athena 2019-27-HH.0, ``Attivit\`a di Studio per la comunit\`a scientifica di Astrofisica delle Alte Energie e Fisica Astroparticellare'' (Accordo Attuativo ASI-INAF n. 2017-14-H.0), and from the European Union’s Horizon 2020 Programme under the AHEAD2020 project (grant agreement n. 871158).
H.B., P.M., and F.D.L. acknowledge also the financial contribution from INFN through the InDark initiative, from Tor Vergata Grant ``SUPERMASSIVE-Progetti Ricerca Scientifica di Ateneo 2021'', and from Fondazione ICSC, Spoke 3 Astrophysics and Cosmos Observations, National Recovery and Resilience Plan (Piano Nazionale di Ripresa e Resilienza, PNRR) Project ID CN\_00000013 ``Italian Research Center on High-Performance Computing, Big Data and Quantum Computing'' funded by MUR Missione 4 Componente 2 Investimento 1.4: Potenziamento strutture di ricerca e creazione di ``campioni nazionali di R\&S (M4C2-19 )'' - Next Generation EU (NGEU).
L.L. also acknowledges support from INAF mini grant 1.05.12.04.01.
M.S. acknowledges financial contributions from contract ASI-INAF n.2017-14-H.0 and INAF Theory Grant 2023 ``Gravitational lensing detection of matter distribution at galaxy cluster boundaries and beyond'' (1.05.23.06.17).
R.M.B. acknowledges support by Centre National d'\'Etudes Spatiales (CNES) under the EPICSSC project for XMM-Newton.
G.W.P. acknowledges support from CNES, the French space agency.
J.S. was supported by NASA Astrophysics Data Analysis Program (ADAP) Grant 80NSSC21K1571
This research was supported by the International Space Science Institute (ISSI) in Bern, through ISSI International Team project \#565 ``{\it Multi-Wavelength Studies of the Culmination of Structure Formation in the Universe}''.
We also made use of the following packages: FTOOLS (\citealt{1995ASPC...77..367B}), DS9 (\citealt{2003ASPC..295..489J}), and several  python libraries, such as numpy \citep{numpy}, matplotlib \citep{matplotlib}, Astropy (\citealt{2013A&A...558A..33A,2018AJ....156..123A}), PyMC \citep{pymc}.
\end{acknowledgements}

\bibliographystyle{aa} 
\bibliography{Chexmate_DR1_Tprof} 

\appendix
\section{Calibration of the background model}
\label{sec:app_bkg}
This section presents the calibration of the physically motivated background model introduced in Sect. \ref{sec:physmodel} on a large set of more than 500 individual blank-sky pointings. The selected blank fields were mainly collected in the framework of the XMM-XXL survey \citep{pierre16} and following surveys. The typical duration of the observations is in the range 10-40 ks, which matches well the typical duration of CHEX-MATE observations. The blank-sky pointings were reduced using the exact same procedure as devised in Sect. \ref{sec:data_reduction}, including the extraction of CRPB spectra (Sect. \ref{sec:particlebkg}) and of the SP contamination indicator $inFOV-outFOV$.

\subsection{Stacked residual spectra}
\label{sec:app_bkg1}
We extracted spectra from four annuli centered on the telescope's aim point covering the radial ranges $[0^\prime-5^\prime]$, $[5^\prime-9^\prime]$, $[9^\prime-12^\prime]$, and $[12^\prime-15^\prime]$. For each spectrum, we modeled the CRPB contribution using the method devised in Sect. \ref{sec:particlebkg} and used the ROSAT all-sky survey spectra to model the sky background (see \ref{sec:skybkg}). These two components are well characterized and their origin is well understood, such that the difference between the measured spectra and the predicted model gives us a handle on the residual focused component, at least partly due to SP. We then sorted the observations in term of their residual SP contamination as traced by the $inFOV-outFOV$ indicator and stacked the residual spectra in bins of $inFOV-outFOV$. 

\begin{figure*}
    \centering
    \resizebox{\hsize}{!}{\includegraphics{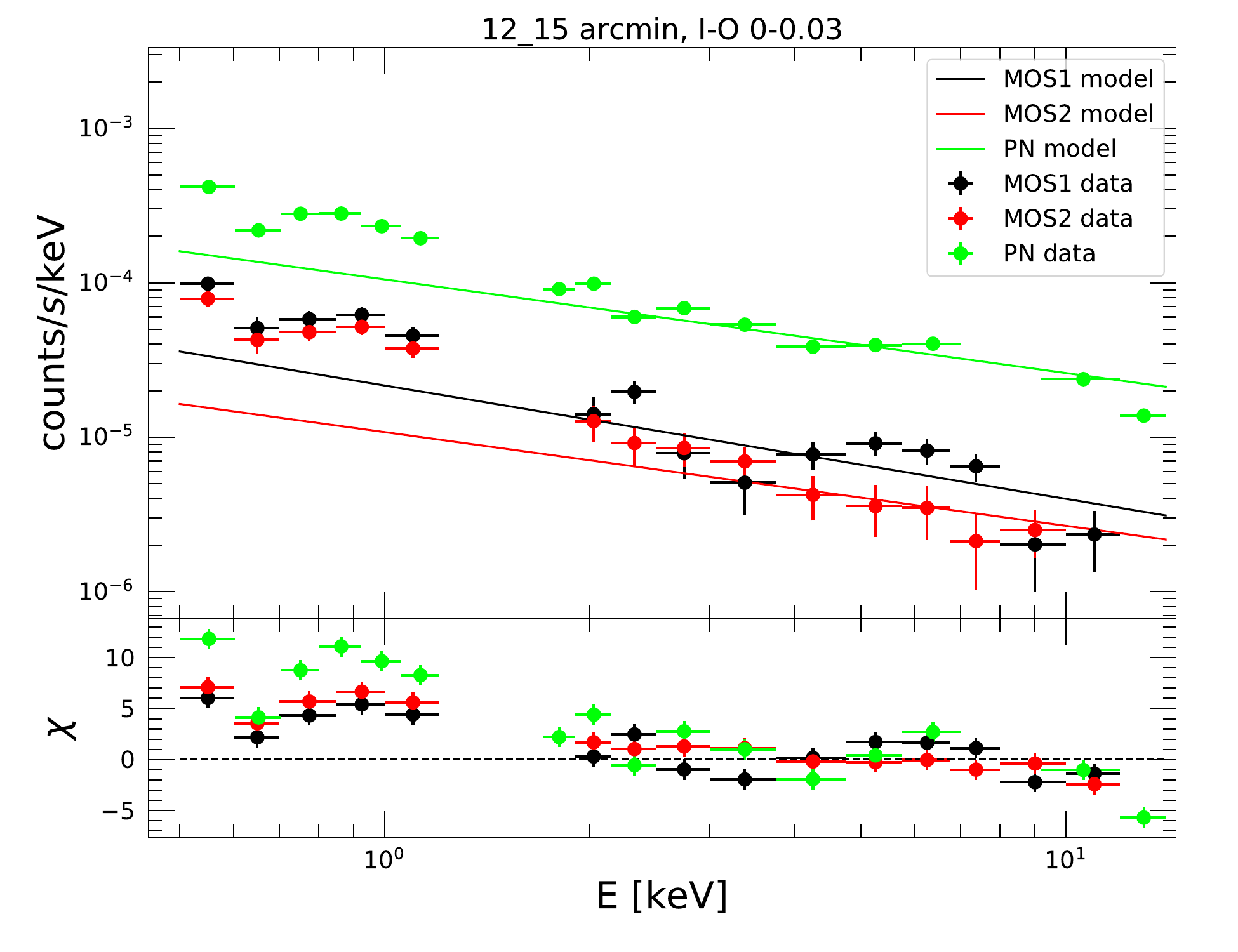}\includegraphics{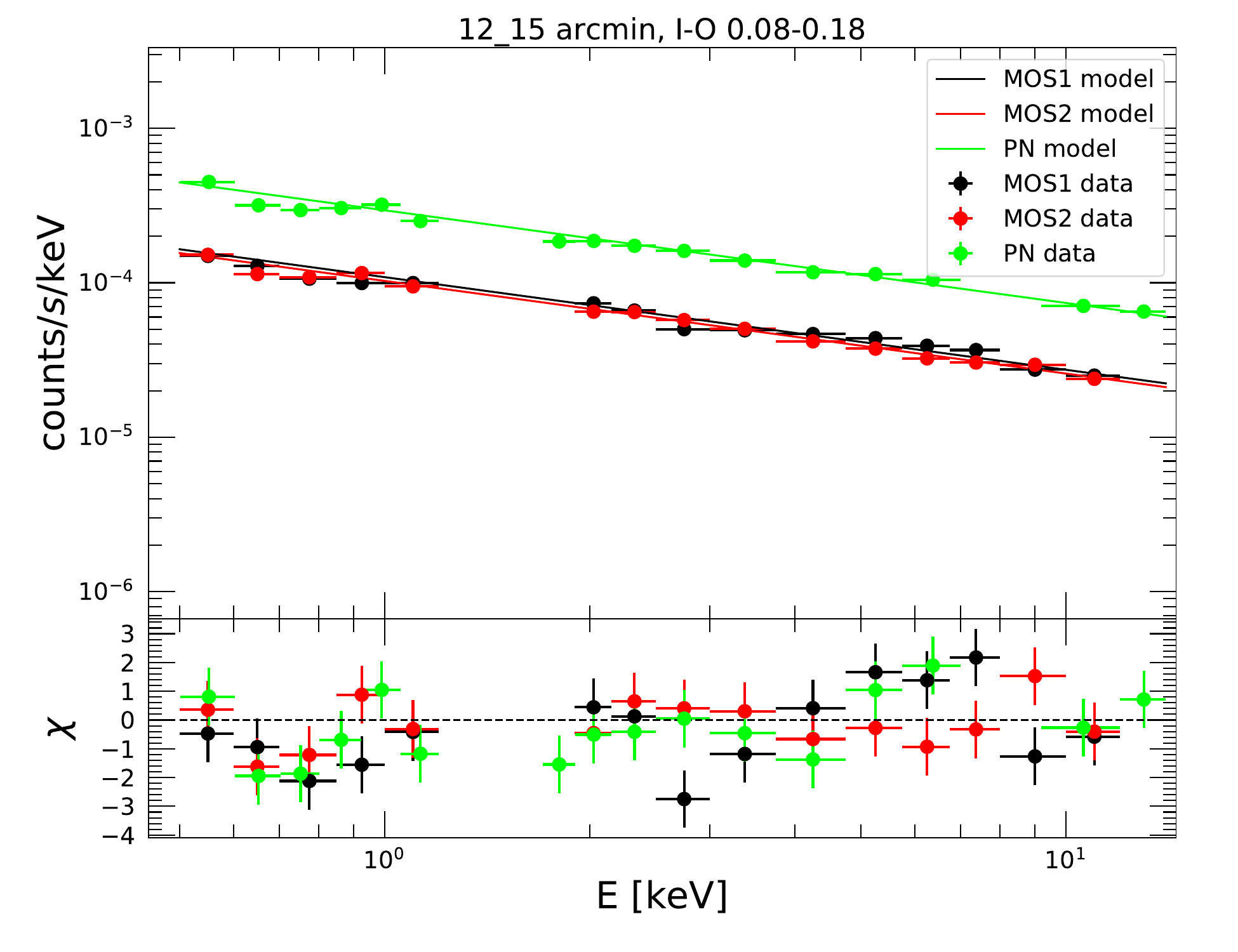}}
    \caption{Stacked residual spectra in the $[12^\prime-15^\prime]$ annulus for two sets of blank-sky observations sorted by their values of $inFOV-outFOV$. The left-hand panel shows the low-contamination dataset ($inFOV-outFOV<0.03$ cts/s) whereas in the right-hand panel we are selecting only highly contaminated observations ($0.08<inFOV-outFOV<0.18$ cts/s). The residuals were fit beyond 2 keV with a power law indicated by the solid lines. The bottom panels show the residuals compared to the model.}
    \label{fig:stack_residuals}
\end{figure*}

In Fig. \ref{fig:stack_residuals} we show the stacked residuals in the $[12^\prime-15^\prime]$ annulus for low ($inFOV-outFOV<0.03$ cts/s) and high SP contamination ($0.08<inFOV-outFOV<0.18$ cts/s). We see a significant excess over the model in all cases, although the amplitude of the excess strongly depends on $inFOV-outFOV$. At low energy ($E<2$ keV), where the dominant background component is the sky emission, the residuals are essentially independent of $inFOV-outFOV$. This points toward a systematic calibration offset between \xmm{} and ROSAT, as already shown by \citet{eckert11}. The ROSAT spectral normalization appears on average 14\% lower than the \xmm{} one, which we correct for the remainder of the analysis by applying a renormalization factor of 1/0.88 to the ROSAT spectra (see Sect. \ref{sec:skybkg}). Conversely, beyond 2 keV the residual spectra strongly depend on $inFOV-outFOV$, indicating that the model residuals scale well with the SP contamination. Therefore, we fitted the residual spectra beyond 2 keV with a power law $I(E)\propto E^{-\Gamma}$, which is indicated with the solid lines in Fig. \ref{fig:stack_residuals}. The spectral shape of the residual SP component appears universal, with a slope of $\Gamma=0.6$ that is nearly independent of $inFOV-outFOV$. 

\subsection{Relation between SP normalization and $inFOV-outFOV$ for MOS}
\label{sec:app_bkg2}

Given the finding that the excess component seems to correlate with SP contamination and that its spectral shape is nearly universal, we fit the residual spectra of each observation with a power law $I(E)=N_{QSP}E^{-\Gamma}$ with a fixed slope $\Gamma=0.6$. We then study how the normalization of the SP component, $N_{QSP}$, can be predicted on the basis of $inFOV-outFOV$. Assuming that the additional component considered here can be entirely attributed to a residual SP contamination, we expect this component to vary across the detector with a vignetting curve that is consistent with the known SP vignetting, which is flatter than that of photons. To take the SP vignetting into account, for each spectrum we used the SAS task \texttt{protonscale} to determine an effective region area (hereafter PROTSCAL) and normalized the fitted SP normalization by the corresponding PROTSCAL value, such that in case the excess component is induced by a residual SP component its pattern across the detector is correctly take into account.

\begin{figure*}
\centering
\resizebox{\hsize}{!}{\includegraphics{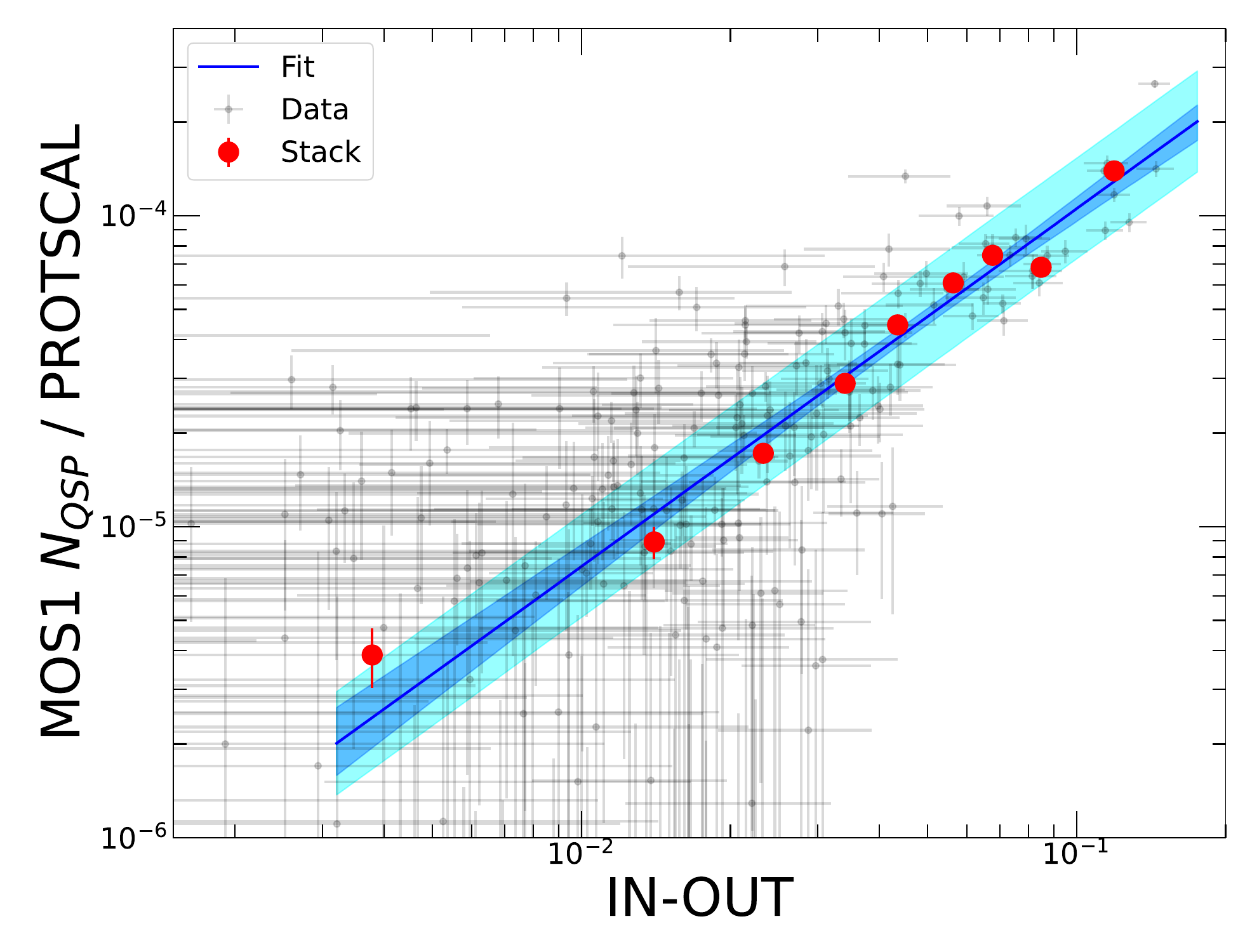}\includegraphics{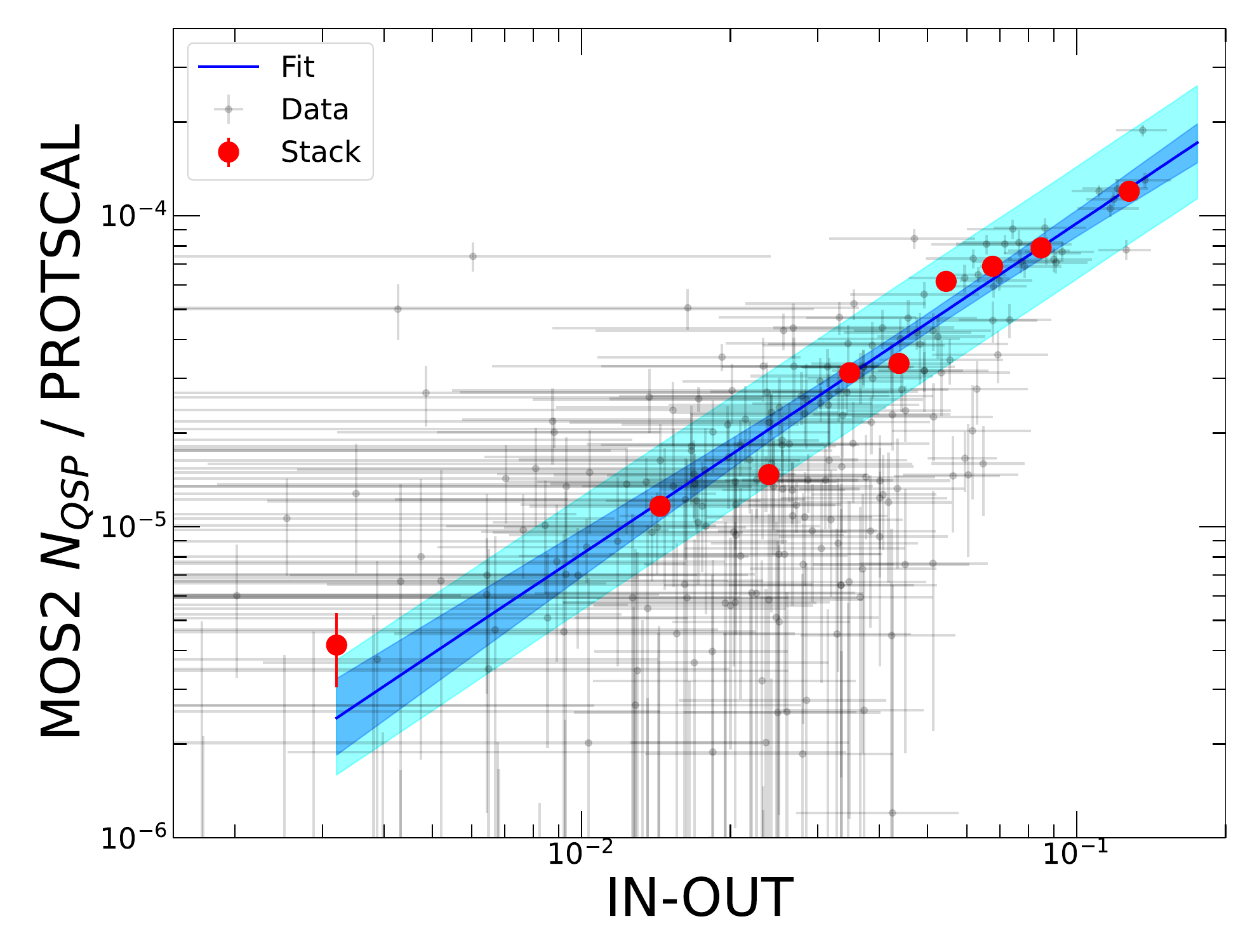}}
\caption{Normalization $N_{QSP}$ of the excess component as a function of $inFOV-outFOV$ for MOS1 (left) and MOS2 (right). The grey points show individual blank-sky observations whereas the red points show the stacked results in bins of $inFOV-outFOV$. The blue solid line and shaded area show the fit to the data and the $1\sigma$ error envelope around the model, whereas the cyan shaded area shows the intrinsic scatter of the relation. }
\label{fig:nqsp_io_mos}
\end{figure*}

In Fig. \ref{fig:nqsp_io_mos} we show the fitted SP normalization scaled by the SP vignetting factor PROTSCAL as a function of $inFOV-outFOV$ for the two MOS detectors. We observe a clear correlation between the two quantities, with a Pearson correlation coefficient of 0.82 (MOS1) and 0.88 (MOS2). For comparison, we also show the normalization of the excess component obtained when fitting stacked spectra in bins of $inFOV-outFOV$, which allows us to better highlight the relation. We then fit the relation with a power law,

\begin{equation}\frac{N_{QSP}}{PROTSCAL} = {\rm A_{SP}} \left(\frac{inFOV-outFOV}{0.05\mbox{ cts/s}}\right)^{{\rm B_{SP}}} 
\label{eq:nqsp_vs_io}\end{equation}

and a free log-normal intrinsic scatter. For MOS1 we measure ${\rm A_{SP,MOS1}}=(4.73 \pm 0.29)\times 10^{-5}$ and ${\rm B_{SP,MOS1}}=1.15\pm0.10$ with an intrinsic scatter $\sigma_{\log N}  =0.16_{-0.05}^{+0.07}$ dex. We obtain very similar values for MOS2, with ${\rm A_{SP,MOS2}}=(4.53 \pm 0.31)\times 10^{-5}$, ${\rm B_{SP,MOS2}}=1.06\pm0.09$ and an intrinsic scatter $\sigma_{\log N}  =0.18_{-0.05}^{+0.06}$ dex. The consistent results obtained with the two MOS detectors are probably unsurprising, given that the detector and telescope properties are very similar. 

\subsection{Relation between SP normalization and $inFOV-outFOV$ for pn}

\begin{figure}
    \centering
    \resizebox{\hsize}{!}{\includegraphics{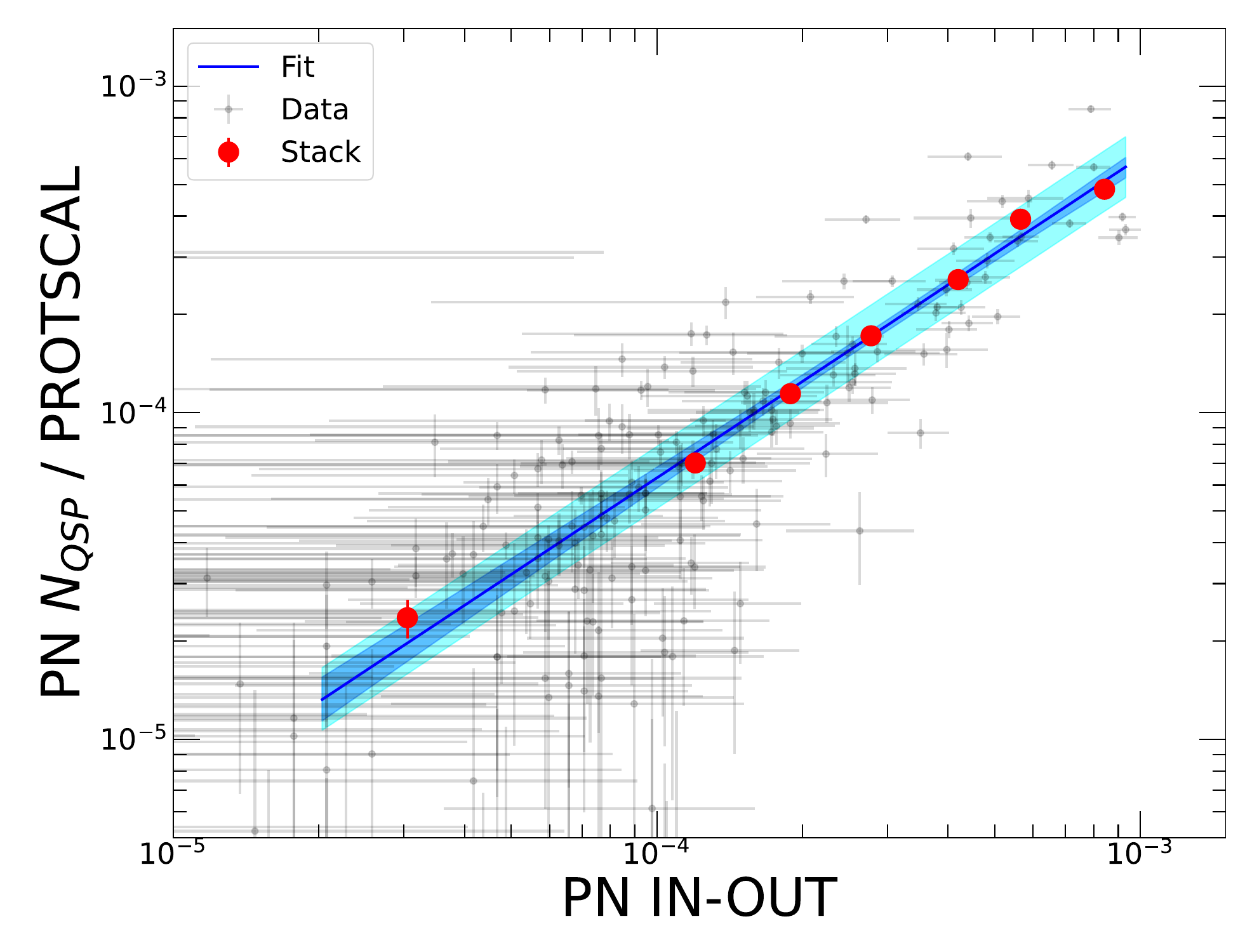}}
    \caption{Normalization $N_{QSP}$ of the excess component as a function of the pn $inFOV-outFOV$ indicator, computed as in Eq. \ref{eq:in_out_pn}. As in the previous plot, grey points represent individual measurements and red points are stacked results, the blue solid line and area show the best fit relation and its 1$\sigma$ envelope, while the light blue shows the intrinsic scatter.}
    \label{fig:nqsp_vs_pn_io}
\end{figure}

As in the previous section, we attempted to correlate the normalization of the excess component with the $inFOV-outFOV$ indicator for the pn detector as well. While we also found a relation between the pn excess component and $inFOV-outFOV$ (computed for the MOS2 detector as in \citealt{marelli17}), the relation appears to have a much larger intrinsic scatter of $\sim0.4$ dex, such that it is much more difficult to predict the value of $N_{QSP}$ for the pn data on the basis of $inFOV-outFOV$. The main reason for the large intrinsic scatter is the difference in the selection of good time intervals between the MOS and pn detectors. Indeed, given the higher sensitivity of the pn to SP, the light curve filtering procedure (see Sect. \ref{sec:data_reduction}) typically excludes a higher fraction of the total observation time for pn than for MOS, such that the selection of good time intervals need not coincide. Given the highly variable nature of the SP component, the relation between the excess component and the $inFOV-outFOV$ indicator based on MOS2 data exhibits a substantial level of scatter. To alleviate this issue, we considered the possibility of defining its own $inFOV-outFOV$ indicator for pn as well. This is rendered difficult by the lack of shielded area in the pn. Following the procedure devised in Sect. \ref{sec:particlebkg}, we predict the expected intensity of the pn CRPB on the basis of the MOS2 $outFOV$ value, which can be done in spite of the different selection of good time intervals as the CRPB is known to vary over timescales that are typically much longer than our observation. We can then determine the high-energy count rate inside the pn FoV, scale this value per unit area, and compute an equivalent of $inFOV-outFOV$ in the following way:
\begin{equation}
    (inFOV-outFOV)_{PN} = CR_{ann} - {\rm A_{CRPB}} ~ outFOV_{MOS2} 
\label{eq:in_out_pn}
\end{equation}
with ${\rm A_{CRPB}}$ the CRPB proportionality constant defined in Eq. \ref{eq:pn_qpb} and $CR_{ann}$ the [10-14] keV count rate in the $[12-15]$ arcmin annulus centered on the aim point. The use of the external annulus allows us to exclude any potential photon contribution in case a bright, hot cluster is observed on axis. We then attempted to correlate the new indicator defined in Eq. \ref{eq:in_out_pn} with the normalization of the pn excess component. The resulting relation is shown in Fig. \ref{fig:nqsp_vs_pn_io}.  We found a much tighter correlation than with the MOS-based quantity, with a correlation coefficient of 0.85. Fitting the data with the same model as in Eq. \ref{eq:nqsp_vs_io} albeit with $(inFOV-outFOV)_{PN}$ instead of $inFOV-outFOV$, we found ${\rm A_{SP,PN}}=(1.85 \pm 0.08)\times 10^{-4}$, ${\rm B_{SP, PN}}=0.98\pm0.05$, and $\sigma_{\log N}=0.10_{-0.04}^{+0.05}$ dex.  We can see that the normalization of the excess component for pn  is about 4 times higher than that of the MOS detectors. In all cases, the slope of the relation is close to one.

\section{Effects of the different abundance tables on the best-fit spectral measurements of the parameter of the {\tt phabs(apec)}}
\label{app:ab_phabs}
We have estimated the effects of changing the adopted abundance table in XSPEC on the best-fit spectral parameters of a thermal model, either absorbed or not. We refer to the spectral model  {\tt phabs(apec)} set as a reference and in which both components depend on the assumed table for the metal abundance\footnote{about the dependence of the {\tt phabs} model see https://heasarc.gsfc.nasa.gov/xanadu/xspec/manual/node249.html}.

We simulate spectra in XSPEC convolved with the {\it XMM} pn response using the {\tt fake} command with no background, and defining an exposure time ($t_{\rm exp}=10^6$ sec) and a normalization of the thermal model ($K_{apec}=1$) high enough that the count statistic is not a limitation.
We evaluate the impact of the assumed table for the metal abundance in two cases, 
one with $N_H = 0.05 \times 10^{22} {\rm cm}^{-2}$ and fixing $N_H=0$.
The latter condition allows us to quantify the impact on the thermal model only.
We simulate the spectrum with a metallicity of $Z=0.3$ times the solar values in {\it aspl} \citep{asplund09}, 
and investigate how the best-fit parameters change by changing the abundance table with the  XSPEC command {\tt abund}.
We consider all the abundance tables available in XSPEC\footnote{see https://heasarc.nasa.gov/xanadu/xspec/manual/XSabund.html also for references}, 
plus \citealt{lodders09} (hereafter {\it lodd09}) that is not available in XSPEC and is loaded with the command {\tt abund file lodders09.tab}, where "lodder09.tab" is a one-column file with the number density relative to hydrogen of the elements with atomic number  $Z_{\rm elem}$ from 1 to 30.
We run the procedure using a range of temperatures (2, 5, 8 keV) and redshifts (0.05, 0.2, 0.6) defined to enclose the properties of the systems that will be studied in the {\it CHEX-MATE} analysis.
Each spectrum is rebinned to have at least 1 count in each bin and fitted with Cash statistics over the range 0.5--8 keV.

\begin{figure}[hbt]
\begin{center}
\includegraphics[page=2,trim=0 0 0 250,clip,width=0.5\textwidth, keepaspectratio]{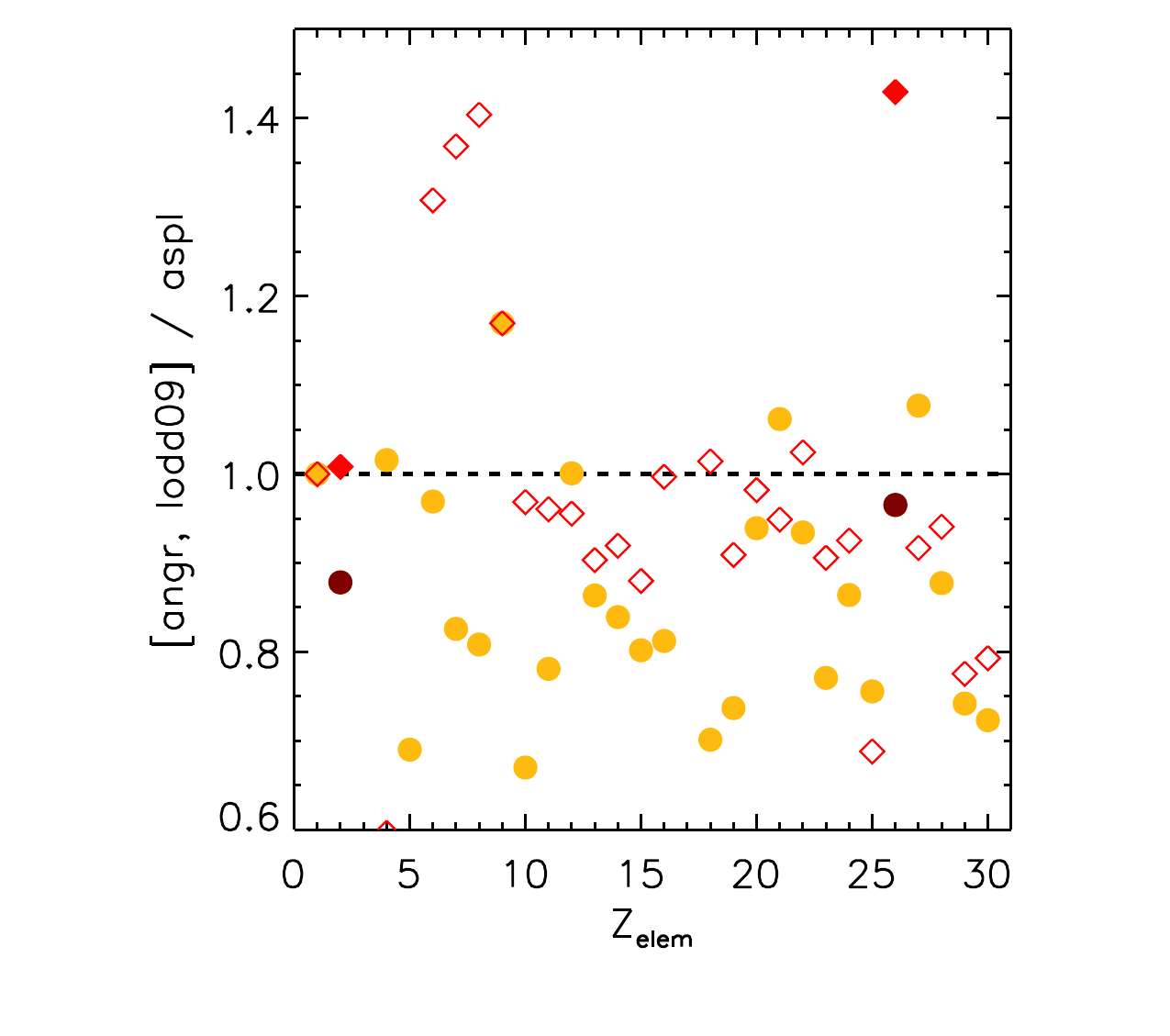}
\end{center} \vspace*{-1.3cm}
\caption{\small{
The ratio of the abundances of the different elements (indicated with their atomic number $Z_{\rm elem}$) between the values quoted in {\it angr} and {\it lodd09} and the values of reference in {\it aspl}. 
In brown ({\it lodd09}) and filled diamonds ({\it angr}), 
we show the relative abundance of helium ($Z_{\rm elem} = 2$) and iron ($Z_{\rm elem}=26$).
} } \label{fig:elem}
\end{figure}

We focus our analysis on the changes due to the use of the abundance tables {\it angr} and {\it lodd09} 
in fitting simulated spectra produced with the abundance table {\it aspl}. 
The values in the tabulated metal abundances are plotted in Fig.~\ref{fig:elem} as a function of the atomic number and normalized to the ones in {\it aspl}.
The main differences with respect to {\it aspl} are the higher (by $\sim$40\%) Fe abundance in {\it angr} 
and the lower (by 12\%) He abundance in  {\it lodd09} \citep[see also][]{ettori20}.

\begin{figure*}[htb]
\begin{center}
\includegraphics[page=1,trim=0 10 0 200,clip,width=0.48\textwidth, keepaspectratio]{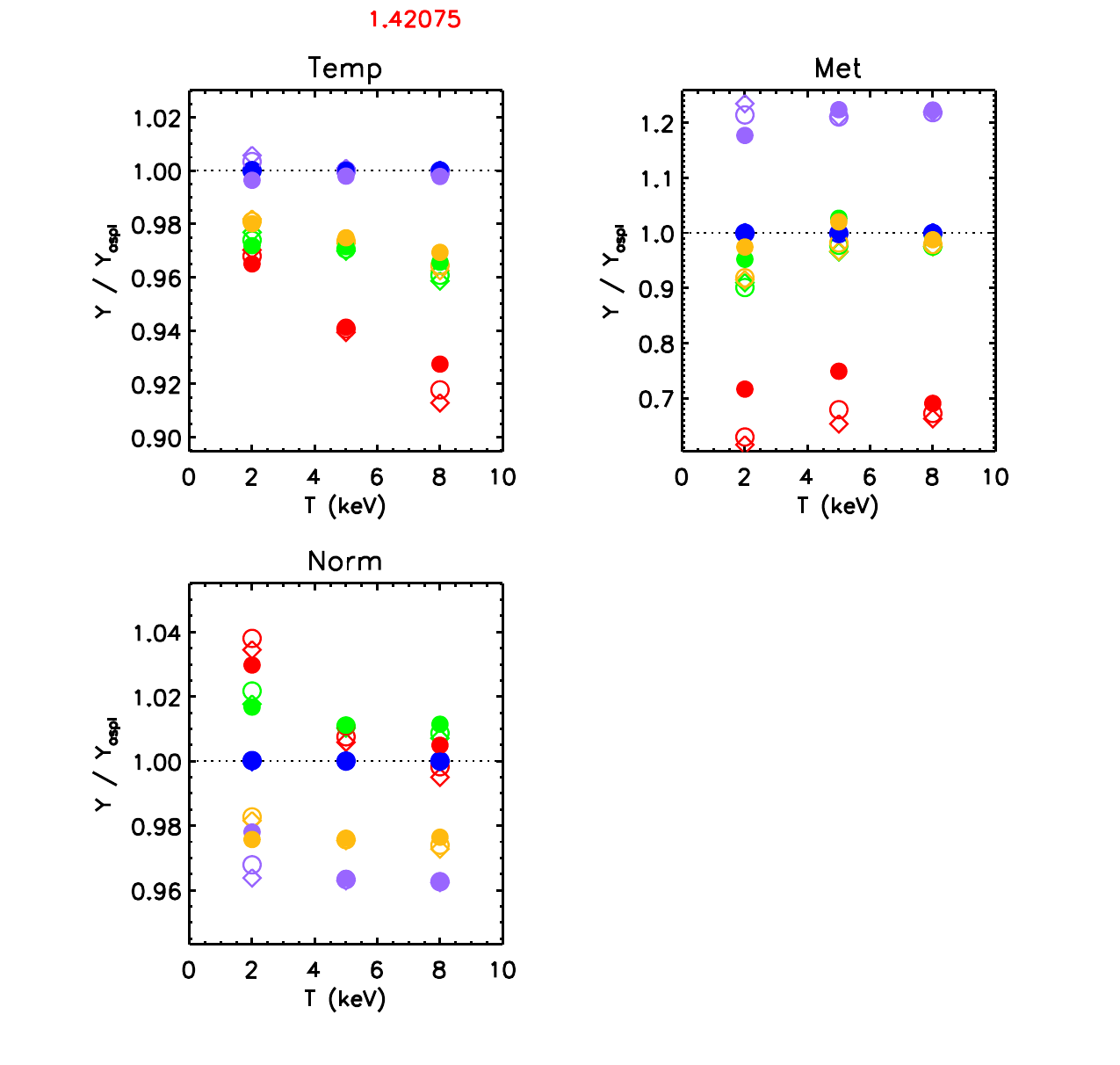}
\includegraphics[page=1,trim=0 10 0 200,clip,width=0.48\textwidth, keepaspectratio]{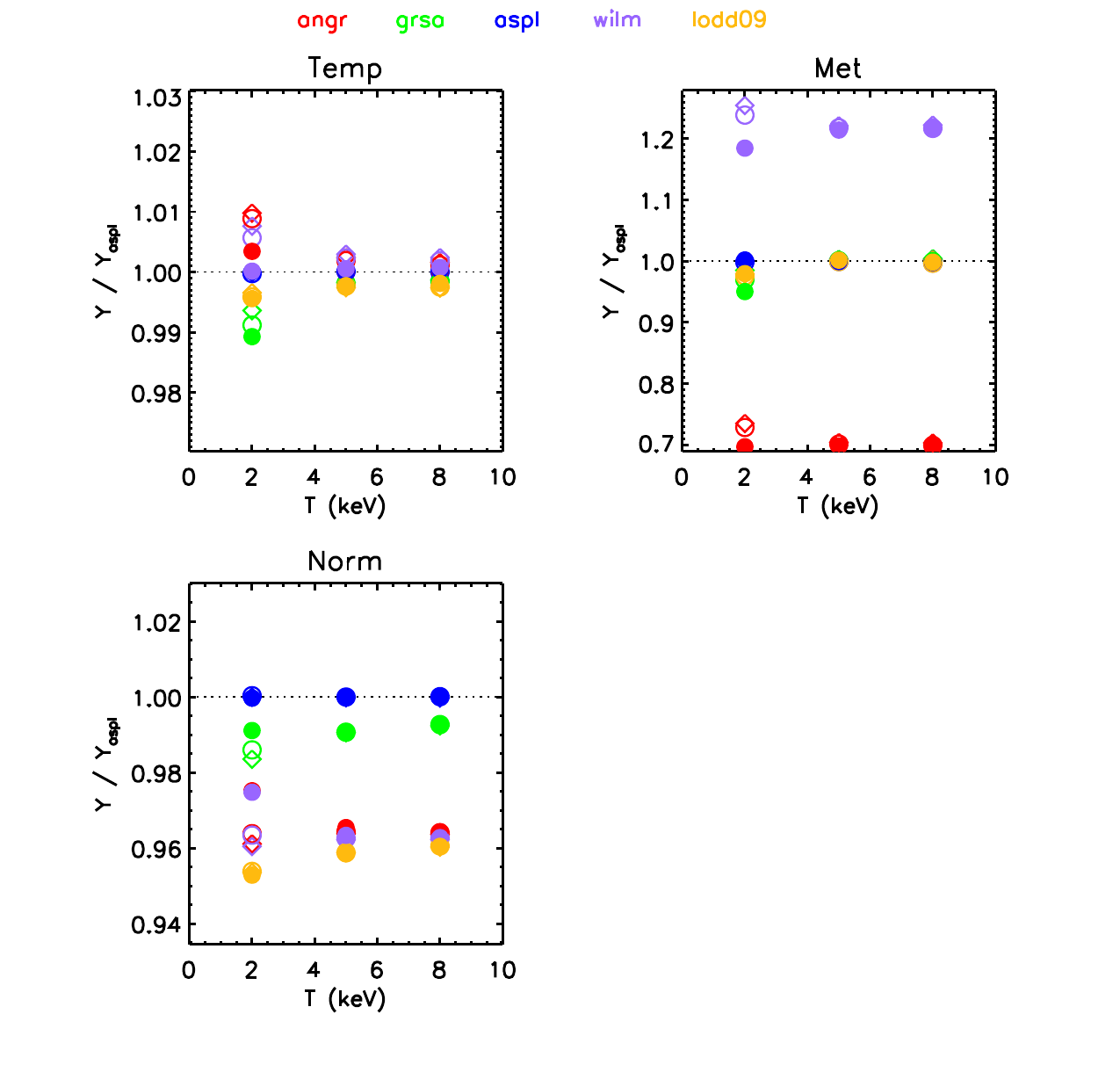}
\end{center} \vspace*{-1cm}
\caption{Ratios between the best-fit parameters and the input values. Points are colour-coded according to the adopted abundance table ({\it aspl}: red, {\it lodd09}: yellow, {\it aspl}: blue, {\it grsa}: green, {\it wilm}: purple).
Different symbols refer to different redshifts ($z=0.05$: open diamonds; $z=0.2$: open circles; $z=0.6$: filled circles).
(Left) Assuming $N_H=0.05 \times 10^{22} {\rm cm}^{-2}$; (right) with $N_H=0$.
} \label{fig:var}
\end{figure*}

\begin{table}[ht]
\caption{Variation in percentage with respect to {\it aspl}. 
The quoted values indicate the minimum, mean and maximum over the range investigated in temperature and redshift.
The best-fit parameters are the {\tt apec} normalization $K$, the temperature $T$ and the metal abundance $Z$,
and are estimated with different $n_H$.
} \label{tab:variation}
\vspace*{0.3cm}
\centering
\begin{tabular}{c c c} 
\hline \hline \multicolumn{3}{c}{$N_H = 0.05 \times 10^{22} {\rm cm}^{-2}$} \\
\hline
& {\it angr} & {\it lodd09} \\
K & $-0.5, 1.4, 3.8$ & $-2.7, -2.3, -1.7$ \\
T & $-8.7, -5.7, -3.0$ & $-3.8, -2.7, -1.8$ \\
Z & $-38.5, -32.6, -25.1$ & $-8.5, -3.1, 2.0$ \\
\\
\hline
\multicolumn{3}{c}{$N_H = 0$} \\
\hline
& {\it angr} & {\it lodd09} \\
K & $-3.9, -3.5, -2.5$ & $-4.7, -4.2, -3.9$ \\
T & $0.0, 0.3, 1.0$ & $-0.4, -0.3, -0.2$ \\
Z & $-30.3, -29.3, -26.5$ & $-2.4, -0.7, 0.2$ \
\\
\hline 
\end{tabular}
\end{table}

The outputs of the spectral analysis are saved and the results are plotted in Fig.~\ref{fig:var} and summarized in Table~\ref{tab:variation}.
We conclude that the differences are more severe (with an average bias of $-30$\%) on the estimates of the metal abundance in {\it angr} because of its higher reference abundance of Fe, 
whereas they are in the order of few per cent on the measurements of $K$ and $T$.
When an $N_H = 0.05 \times 10^{22} {\rm cm}^{-2}$ is considered, the mean bias in normalization $K$ is $-2 (+2)$\% for {\it lodd09} ({\it angr}).
On the temperature $T$, it is $-3$\% for {\it lodd09} and $-6$\% for {\it angr}. The metallicity $Z$ is biased by $-3$\% ($-30$\%) in {\it lodd09} ({\it angr}).
When no Galactic absorption is considered and just the effect on the thermal model is quantified, 
the biases using {\it lodd09} with respect to the adopted {\it aspl} are in the order of $<1$\% on $T$, $-1$\% on $Z$, 
and about $-4$\% on $K$; using {\it angr}, they are similar on $T$ and $K$, and are still $-30$\% on $Z$.\\
We emphasize that the estimates performed here for $N_H\neq0$ are valid only for the {\tt phabs} model, which uses the abundance table set for the XSPEC fit. Other models, such as {\tt wabs} and {\tt tbabs} use their own hard-coded abundance table. 

\section{Data and figures of individual clusters}
\label{app:gallery}
In this Appendix, we report the data, the plots of the temperature profiles, and the images with masked point sources and annuli used for spectral extraction for all the clusters in our sample. The full table with the temperature profiles is available in electronic format at CDS. Here we show an excerpt in Table \ref{table:excerpt}. The first column shows the name of the cluster, while the second and third ones show the sky extension $R_{500}$ in arcminutes and the mean temperature $T_{[0.15-0.75]R500}$, estimated in the annulus between $0.15$ and $0.75R_{500}$. Columns 4 and 5 report the coordinates of the center used for spectral extraction (the peak in \citealt{bartalucci23}), while columns 6 and 7 show the center of the radial bin $R$ and the width $dR$ (annulus included between $R-dR$ and $R+dR$). Columns 8, 9, and 10 report the best fit temperature and its errors (downwards and upwards) obtained with our baseline MCMC procedure, while column 11 shows the $SOU/BKG$ ratio in the radial bin.

\begin{table*}[!h]
\begin{tabular}{c c c c c c c c c c c}
  \hline
  \hline
   Name  & $R_{500}$ & $T_{mean}$ & R.A.  & Dec. & radius & width & $T$ & $dT_{down}$ & $dT_{up}$ & $SOU/BKG$ \\
    & arcmin & keV & deg & deg & arcmin & arcmin & keV & keV & keV & \\
  \hline
  PSZ2 G008.31-64.74 &   4.51 &   6.83  &  344.7010 &  -34.8005 & 0.0825  &  0.0825 &  7.74 & 0.91 &  0.82   &  29.69  \\ 
  PSZ2 G008.31-64.74 &   4.51 &   6.83  &  344.7010 &  -34.8005 &   0.2475 &   0.0825 & 6.95 & 0.58 & 0.57   &  29.72  \\ 
  PSZ2 G008.31-64.74 &   4.51  &  6.83  &  344.7010 &  -34.8005 &  0.4400  &  0.1100   &6.76 & 0.45 &  0.47  &  18.40   \\ 
  PSZ2 G008.31-64.74 &   4.51  &  6.83  &  344.7010 &  -34.8005 &  0.6875  &  0.1375 & 7.51  & 0.47 &   0.46 &  11.29  \\  
  PSZ2 G008.31-64.74 &   4.51  &  6.83  &  344.7010 &  -34.8005 &  0.9900    &  0.1650  & 6.55 & 0.38 & 0.36  & 7.28   \\  
  PSZ2 G008.31-64.74 &   4.51  &  6.83  &  344.7010 &  -34.8005 &  1.3750   &  0.2200   & 6.60  & 0.33 &  0.30  & 5.01   \\  
  PSZ2 G008.31-64.74 &   4.51  &  6.83   & 344.7010 &  -34.8005 &  1.8700    &  0.2750  & 8.44 & 0.54 &  0.53   & 3.38  \\   
  PSZ2 G008.31-64.74 &   4.51  &  6.83  &  344.7010 &  -34.8005 &   2.5025  &     0.3575 & 6.66 &  0.36 & 0.38  & 1.98  \\  
  PSZ2 G008.31-64.74 &   4.51  &  6.83  &  344.7010 &  -34.8005 &  3.3275   & 0.4675 & 6.44  & 0.50 &  0.51   & 1.01   \\  
  PSZ2 G008.31-64.74 &   4.51  &  6.83  &  344.7010 &  -34.8005 &  4.4000   &  0.6050  & 6.81 & 0.86 & 0.86   &  0.35   \\  
  PSZ2 G008.31-64.74 &  4.51  &  6.83  &   344.7010 &   -34.8005 &   5.6925 &    0.6875 & 6.40  & 1.86 & 1.17  &  0.12    \\  
  PSZ2 G041.45+29.10 &   6.48  &  6.35  &  259.4366 &  19.6766   & 0.1100    &  0.1100   & 6.14 & 0.73 & 0.71  &  12.85  \\  
  PSZ2 G041.45+29.10 &   6.48   & 6.35  &  259.4366 &  19.6766   & 0.3025  &  0.08250 & 7.37  & 0.66  & 0.67   &  13.4    \\ 
  \hline
\end{tabular}
\caption{Temperature profiles for the clusters of our sample. Full table is available at CDS.}\label{table:excerpt}
\end{table*}

\clearpage



\section*{Supplementary material}

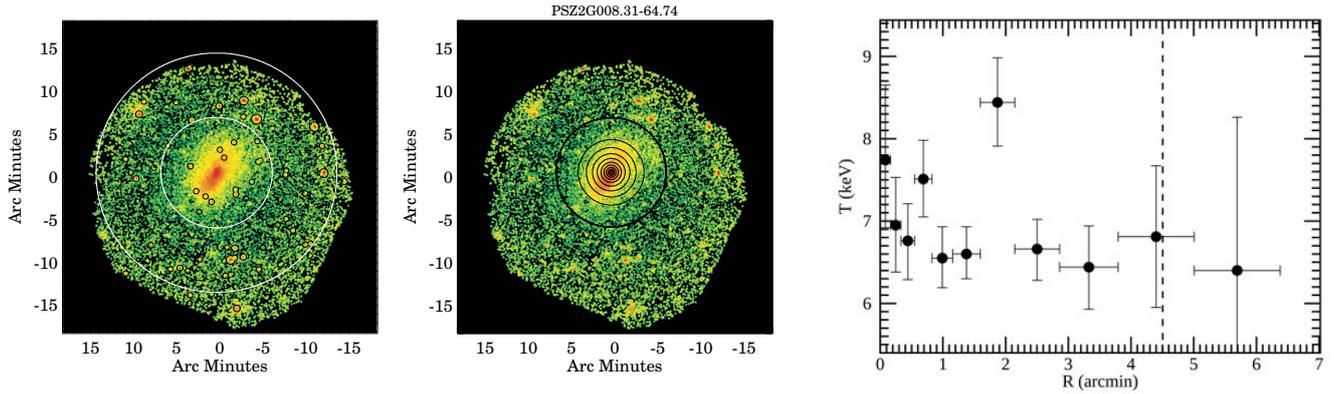

Fig. C.1: Image and temperature profile of PSZ2G008.31 − 64.74. The left panel shows the cluster image in the [0.7 − 1.2] keV band, with overlaid masked point sources and the annulus used for estimating the sky background. In the central panel we show the annular regions used for the extraction of the temperature profile, which is represented in the right panel, where the dashed line marks the position of $R_{500}$.

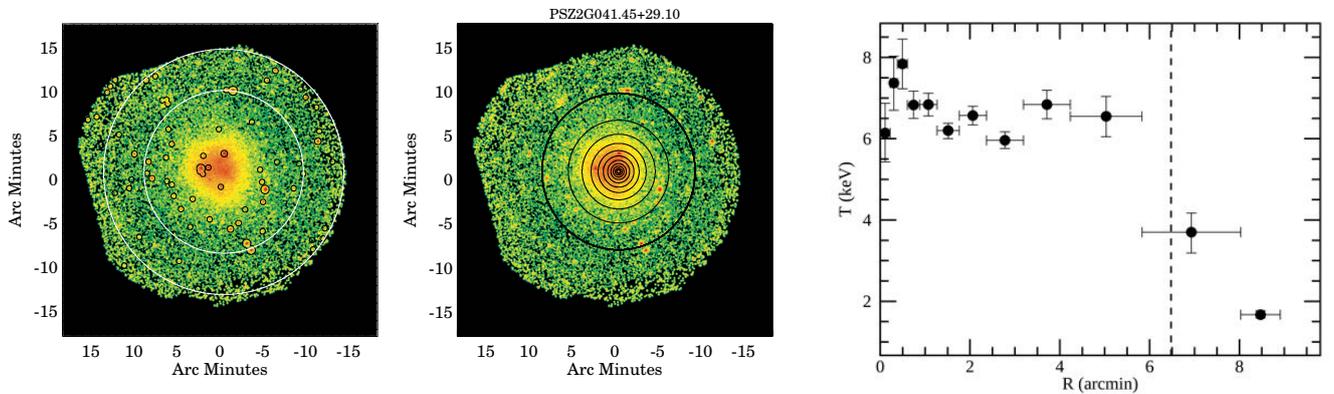

Fig. C.2: Same as Fig. C.1 for cluster PSZ2G041.45+29.10.

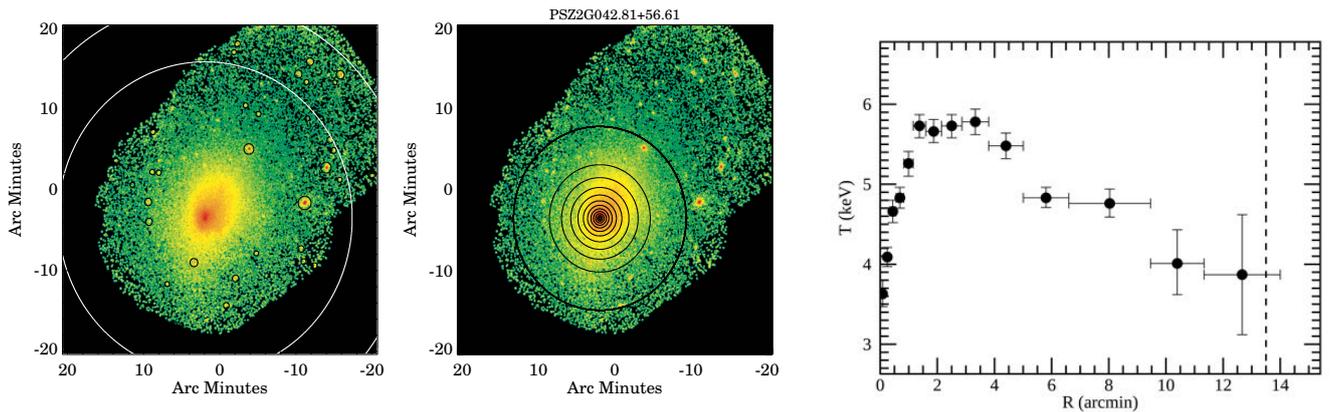

Fig. C.3: Same as Fig. C.1 for cluster PSZ2G042.81+56.61.

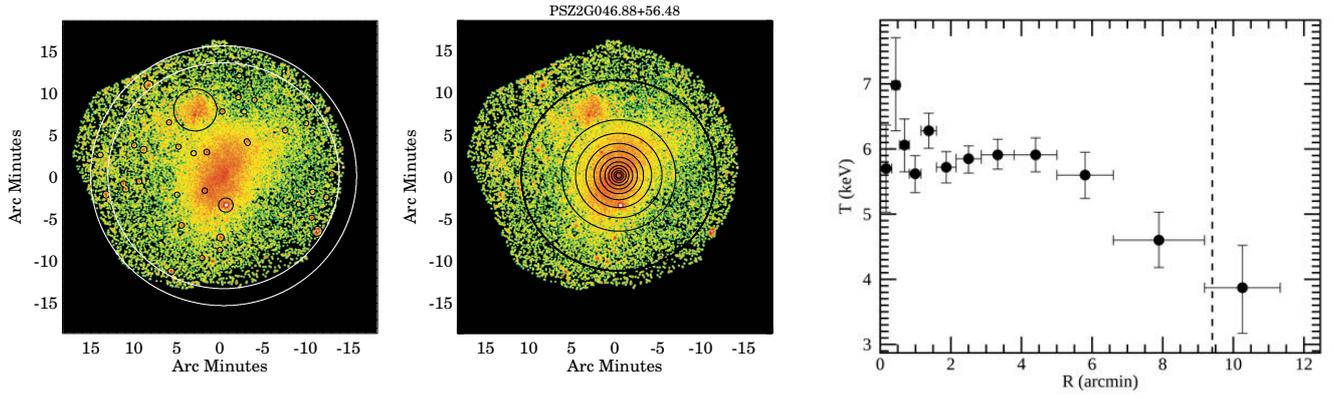

Fig. C.4: Same as Fig. C.1 for cluster PSZ2G046.88+56.48.

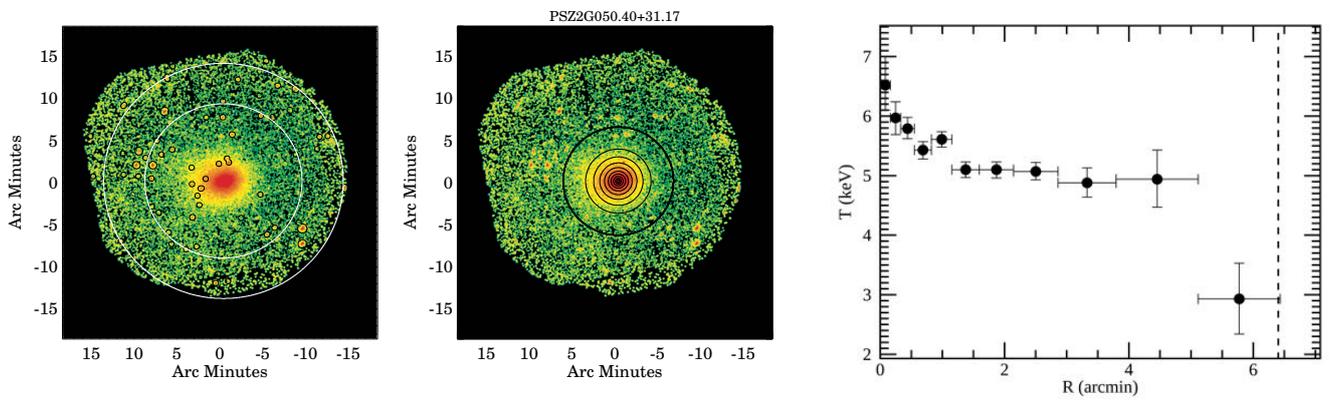

Fig. C.5: Same as Fig. C.1 for cluster PSZ2G050.40+31.17.

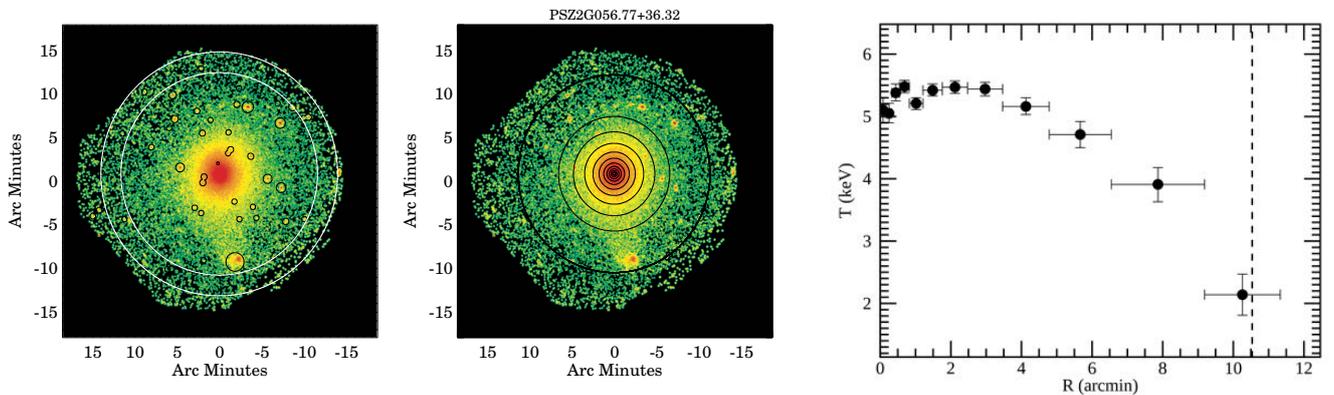

Fig. C.6: Same as Fig. C.1 for cluster PSZ2G056.77+36.32.

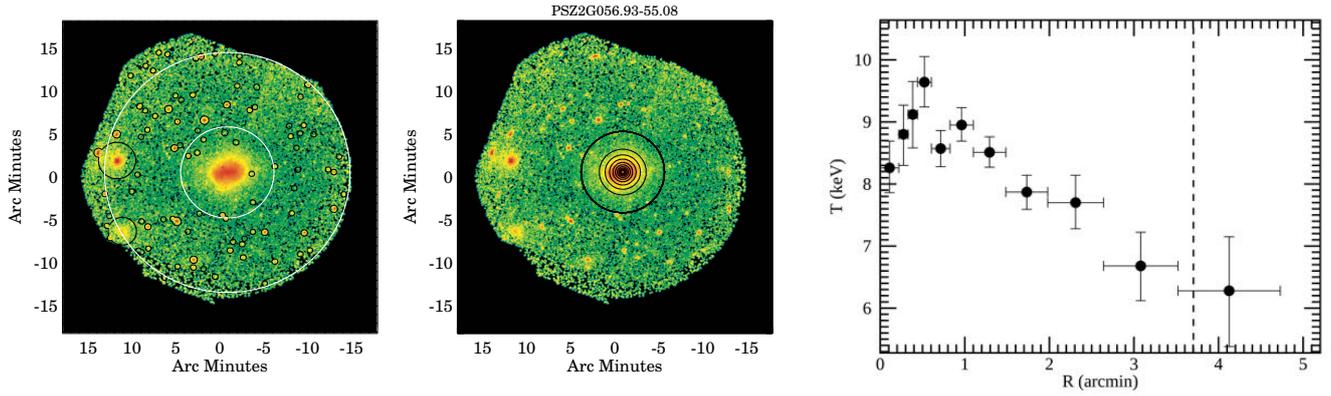

Fig. C.7: Same as Fig. C.1 for cluster PSZ2G056.93-55.08.

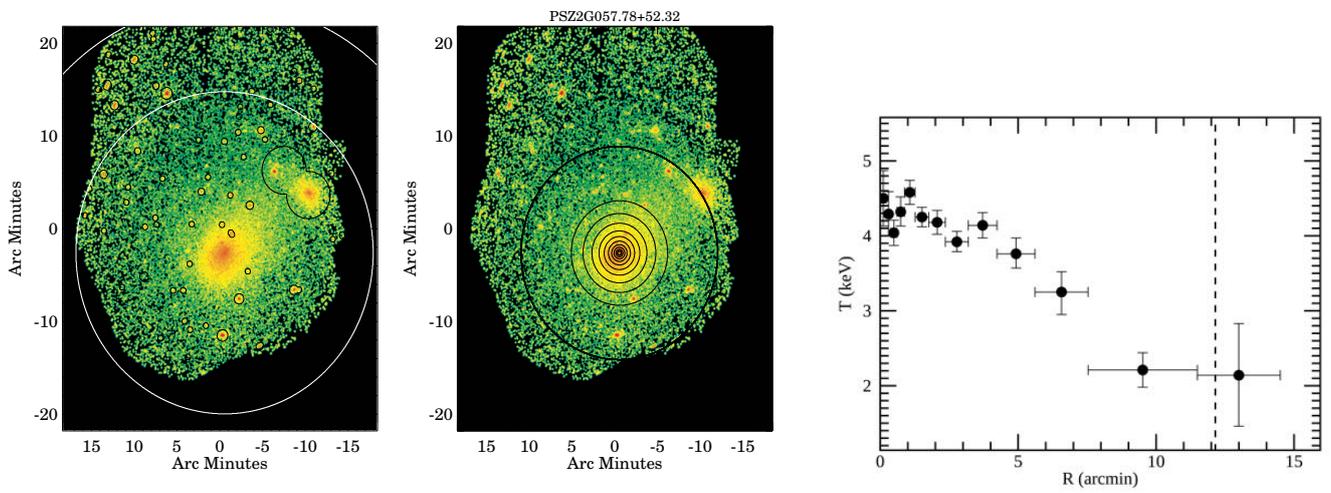

Fig. C.8: Same as Fig. C.1 for cluster PSZ2G057.78+52.32.

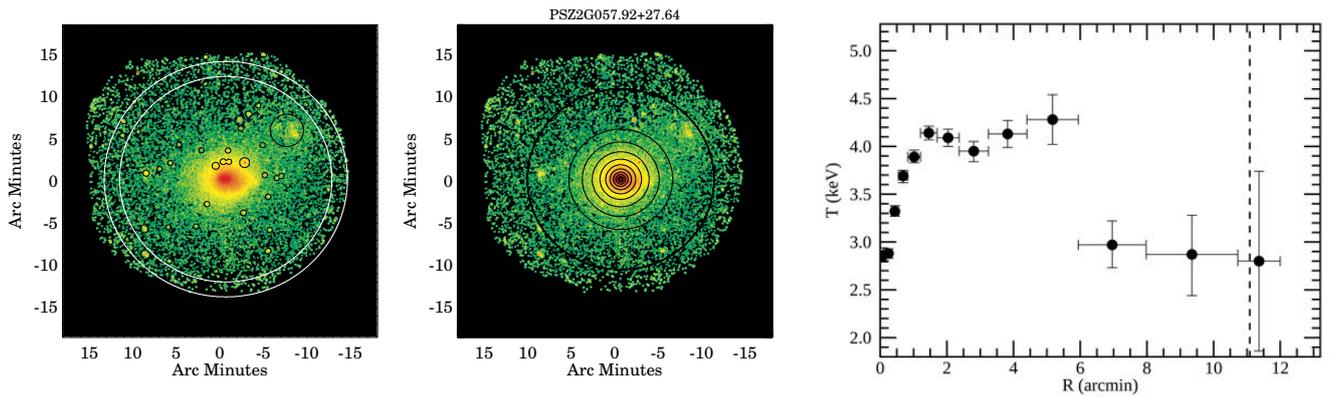

Fig. C.9: Same as Fig. C.1 for cluster PSZ2G057.92+27.64.

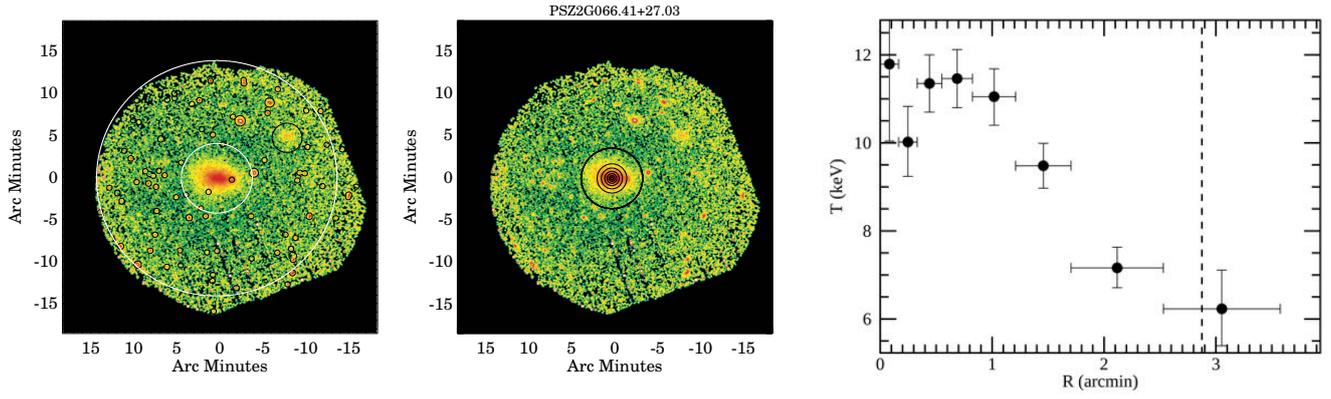

Fig. C.10: Same as Fig. C.1 for cluster PSZ2G066.41+27.03.

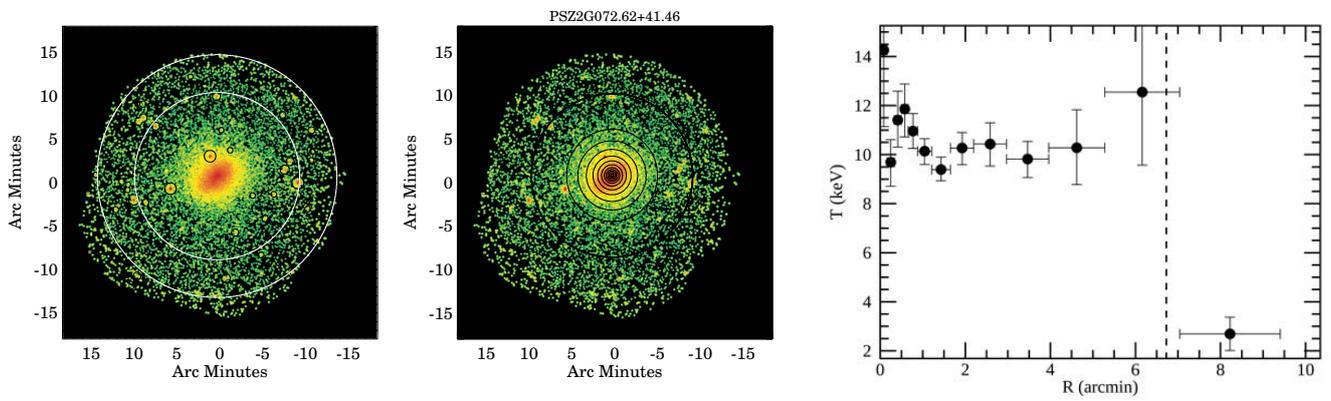

Fig. C.11: Same as Fig. C.1 for cluster PSZ2G072.62+41.46.

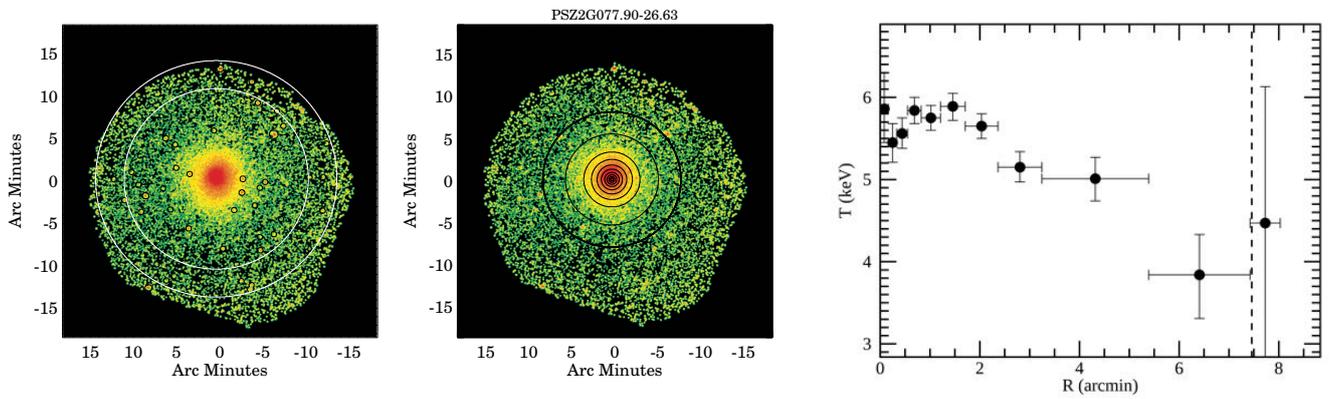

Fig. C.12: Same as Fig. C.1 for cluster PSZ2G077.90-26.63.

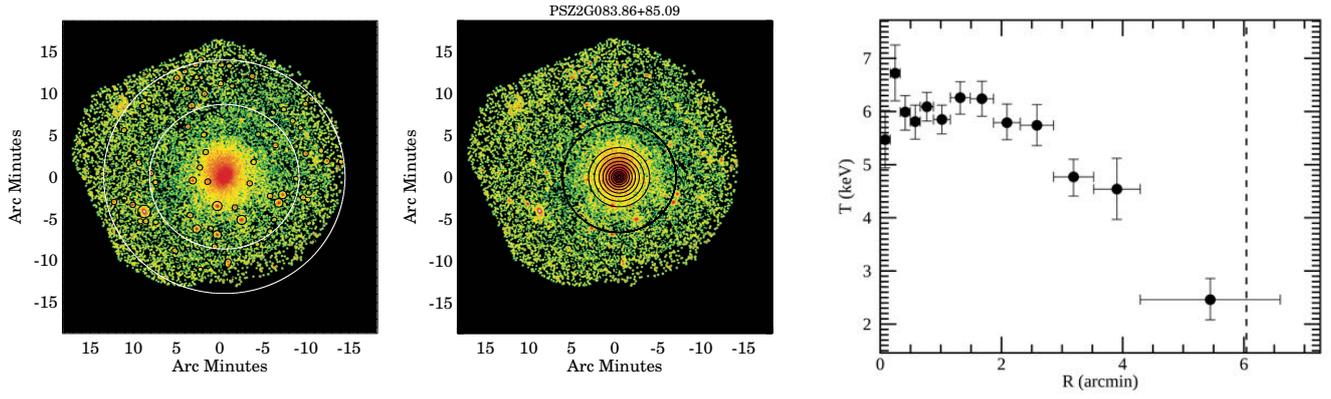

Fig. C.13: Same as Fig. C.1 for cluster PSZ2G083.86+85.09.

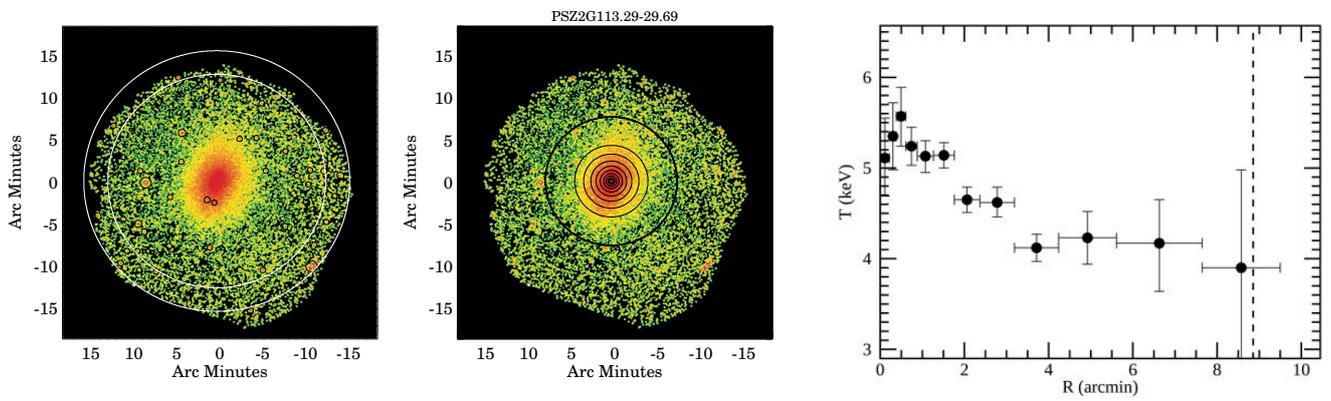

Fig. C.14: Same as Fig. C.1 for cluster PSZ2G113.29-29.69.

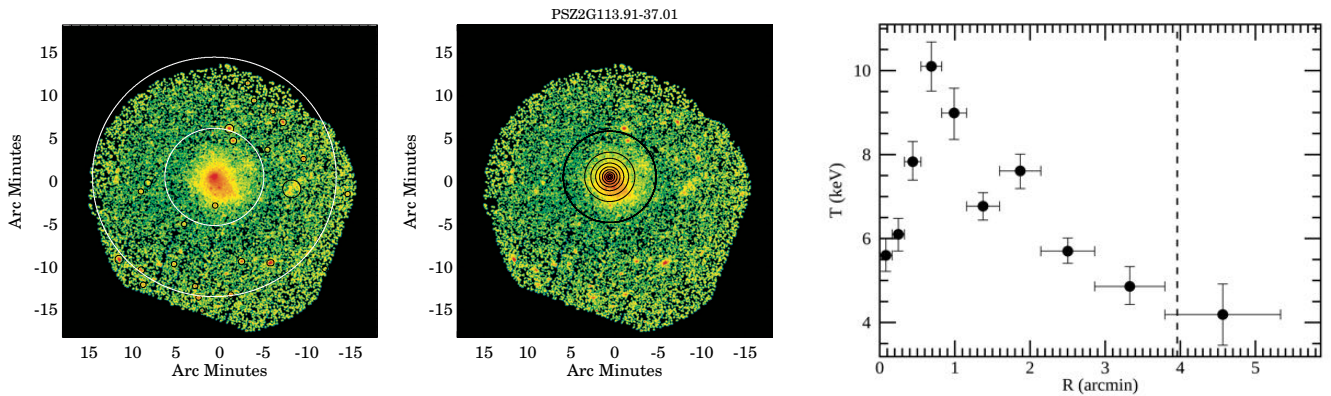

Fig. C.15: Same as Fig. C.1 for cluster PSZ2G113.91-37.01.

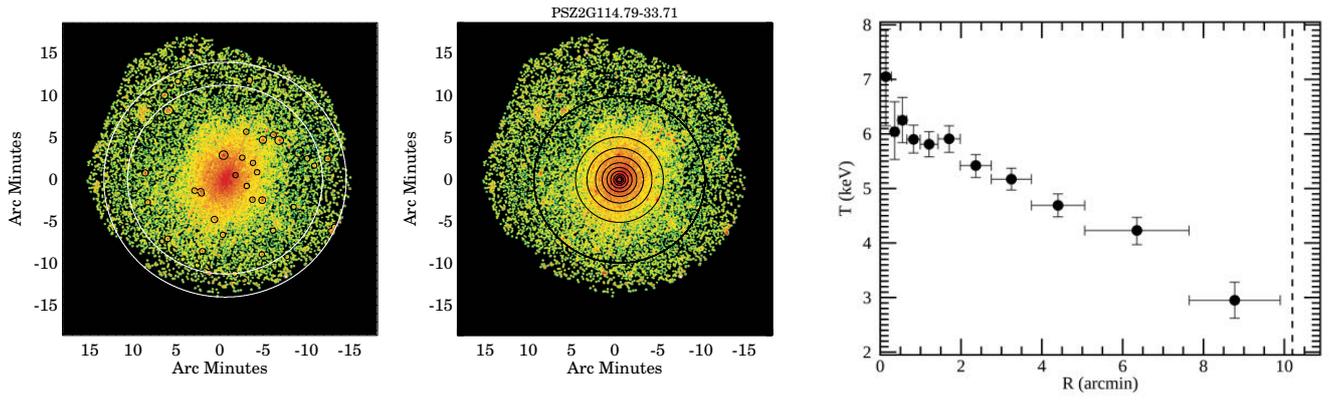

Fig. C.16: Same as Fig. C.1 for cluster PSZ2G114.79-33.71.

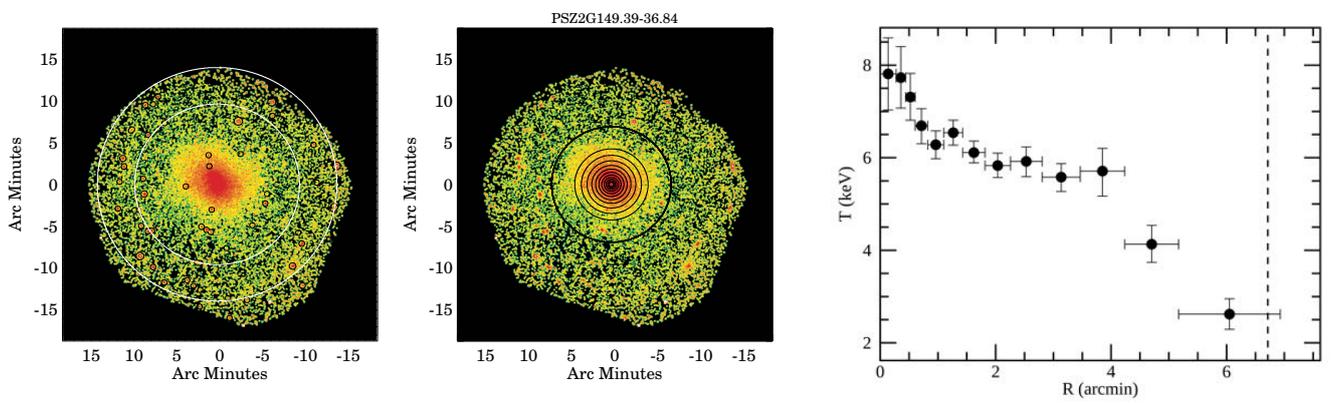

Fig. C.17: Same as Fig. C.1 for cluster PSZ2G149.39-36.84.

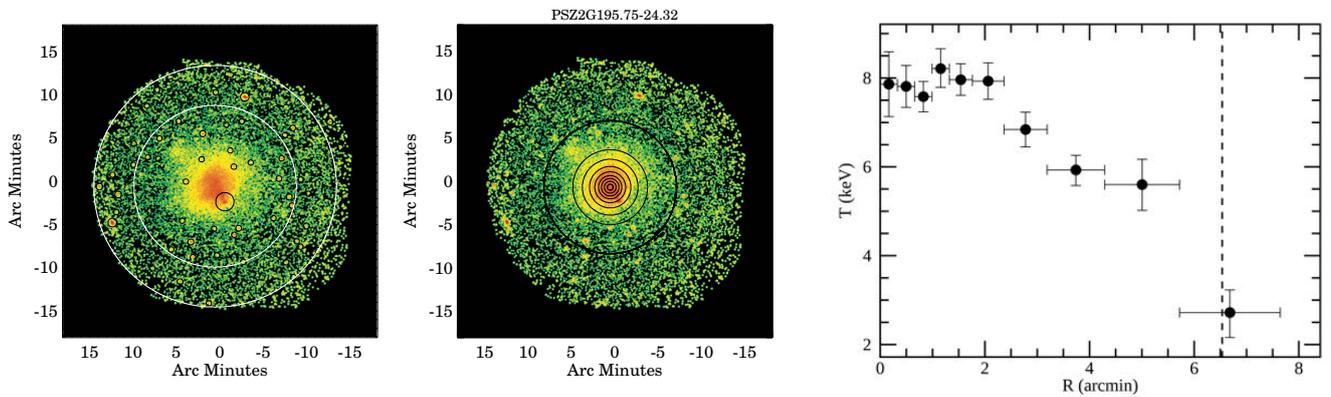

Fig. C.18: Same as Fig. C.1 for cluster PSZ2G195.75-24.32.

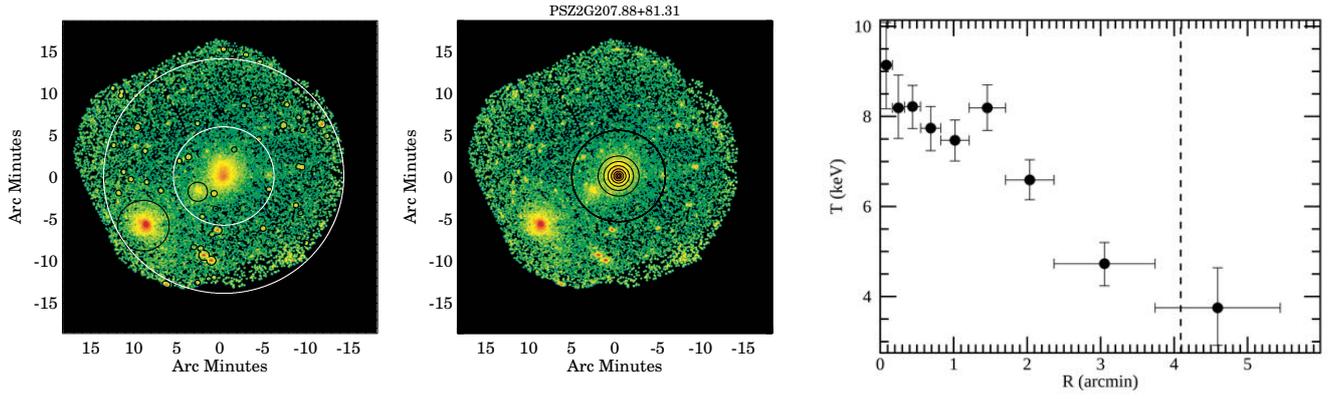

Fig. C.19: Same as Fig. C.1 for cluster PSZ2G207.88+81.31.

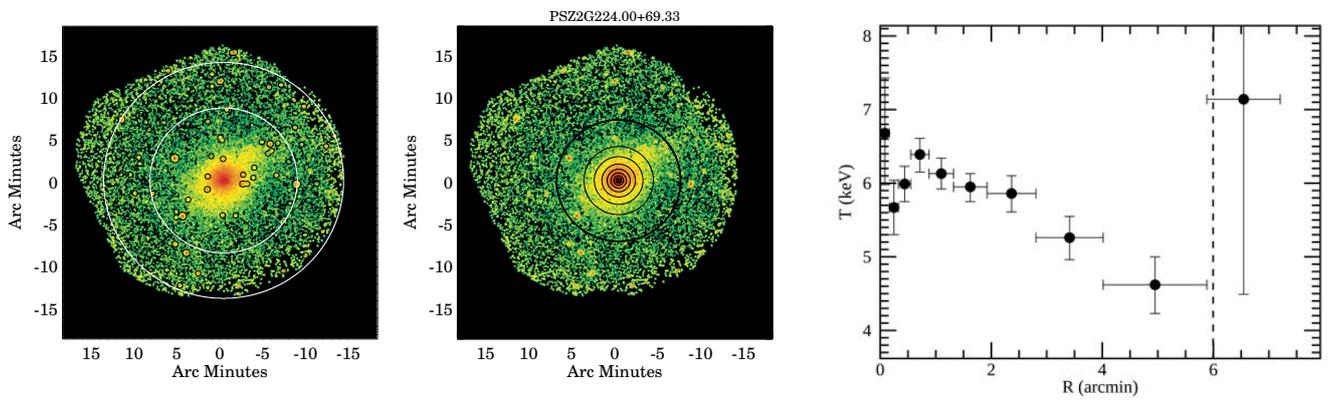

Fig. C.20: Same as Fig. C.1 for cluster PSZ2G224.00+69.33.

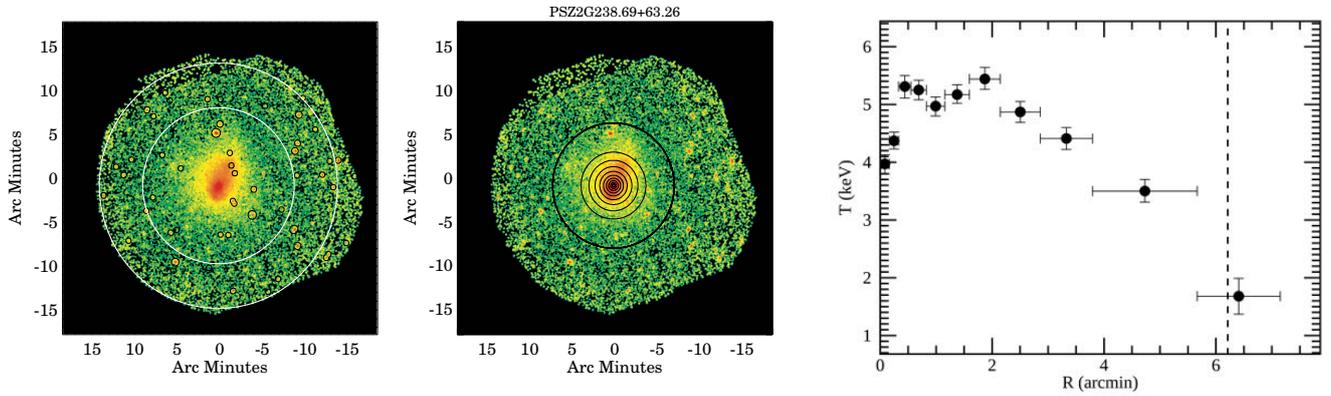

Fig. C.21: Same as Fig. C.1 for cluster PSZ2G238.69+63.26.

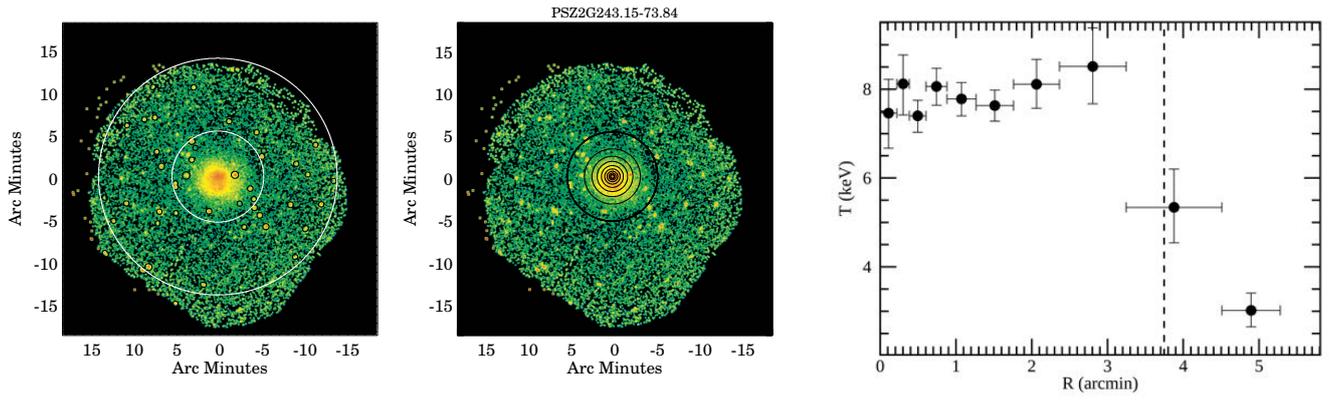

Fig. C.22: Same as Fig. C.1 for cluster PSZ2G243.15-73.84.

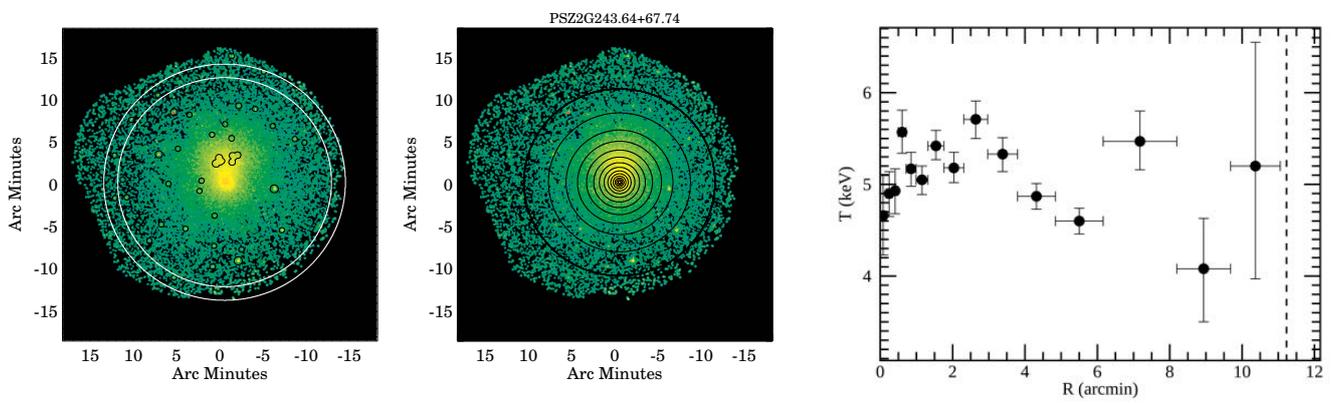

Fig. C.23: Same as Fig. C.1 for cluster PSZ2G243.64+67.74.

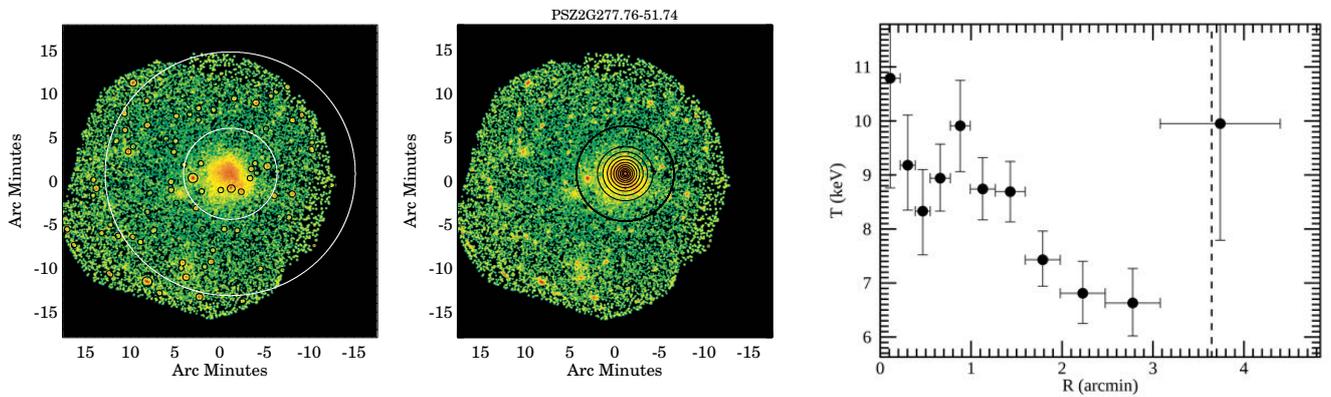

Fig. C.24: Same as Fig. C.1 for cluster PSZ2G277.76-51.74.

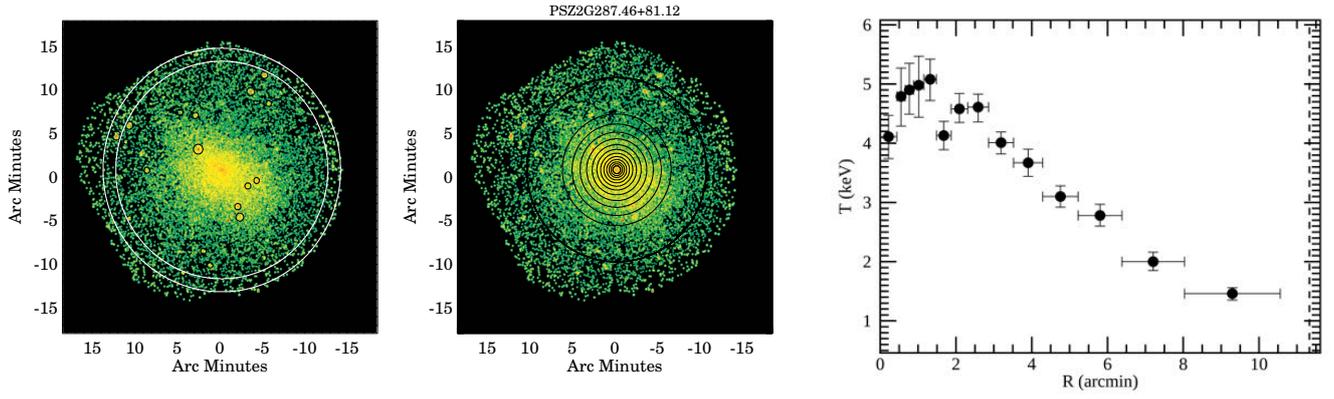

Fig. C.25: Same as Fig. C.1 for cluster PSZ2G287.46+81.12.

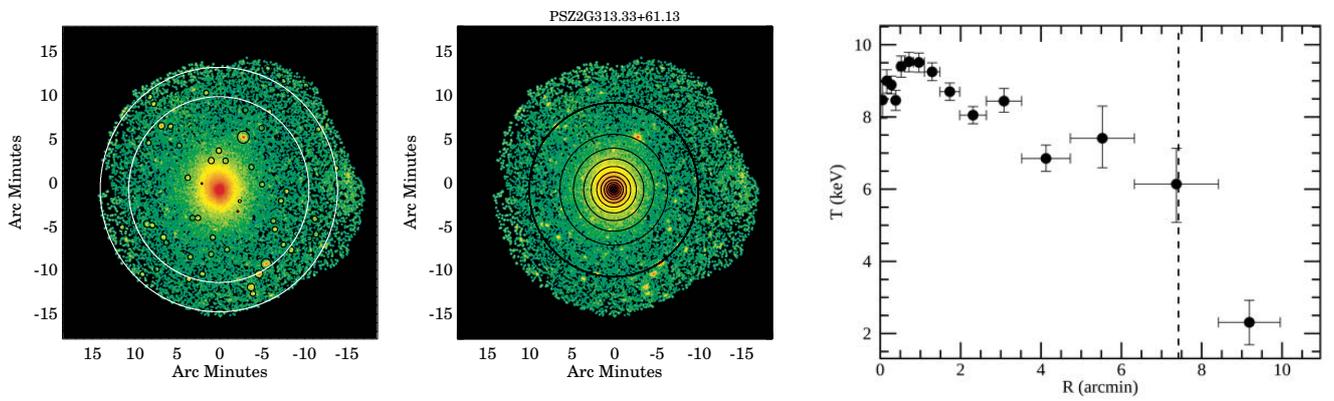

Fig. C.26: Same as Fig. C.1 for cluster PSZ2G313.33+61.13.

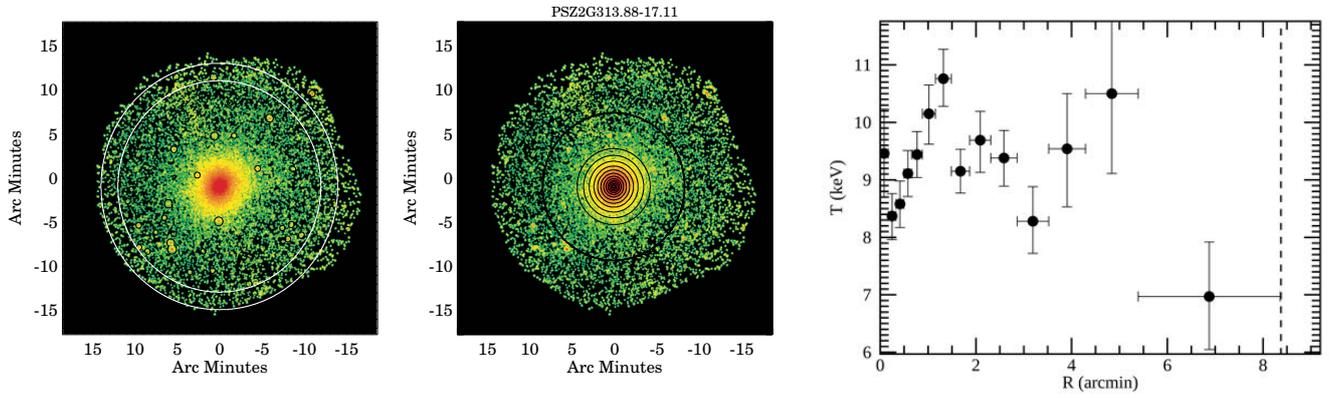

Fig. C.27: Same as Fig. C.1 for cluster PSZ2G313.88-17.11.

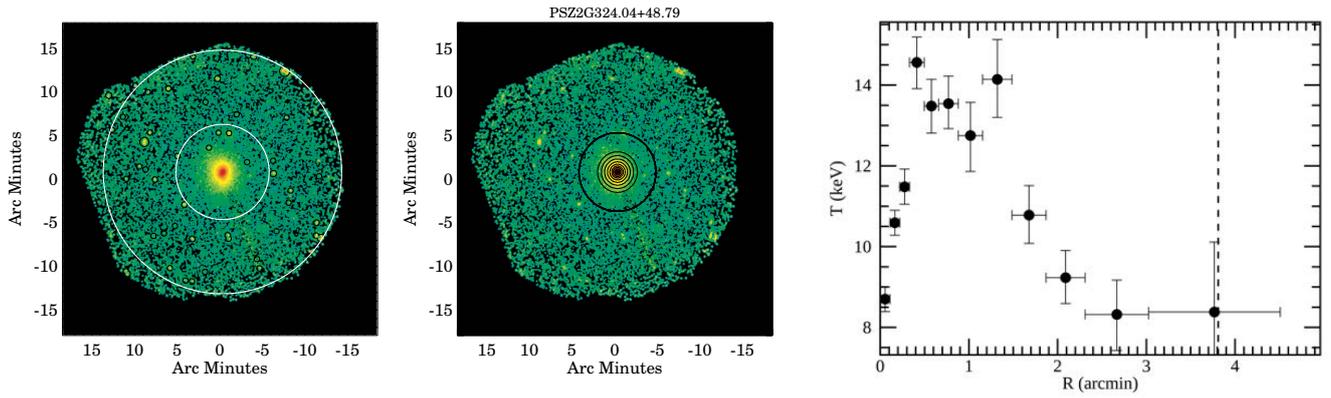

Fig. C.28: Same as Fig. C.1 for cluster PSZ2G324.04+48.79.

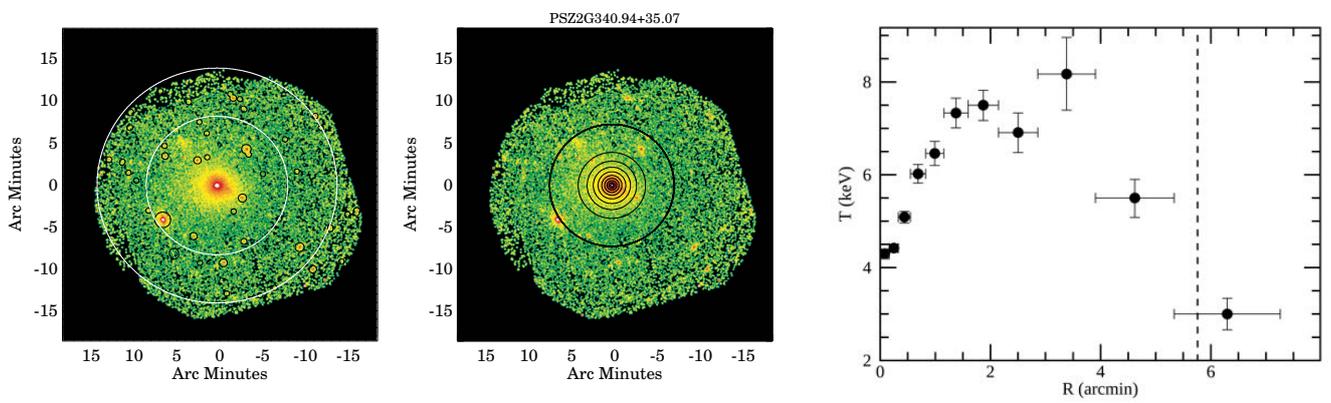

Fig. C.29: Same as Fig. C.1 for cluster PSZ2G340.94+35.07.

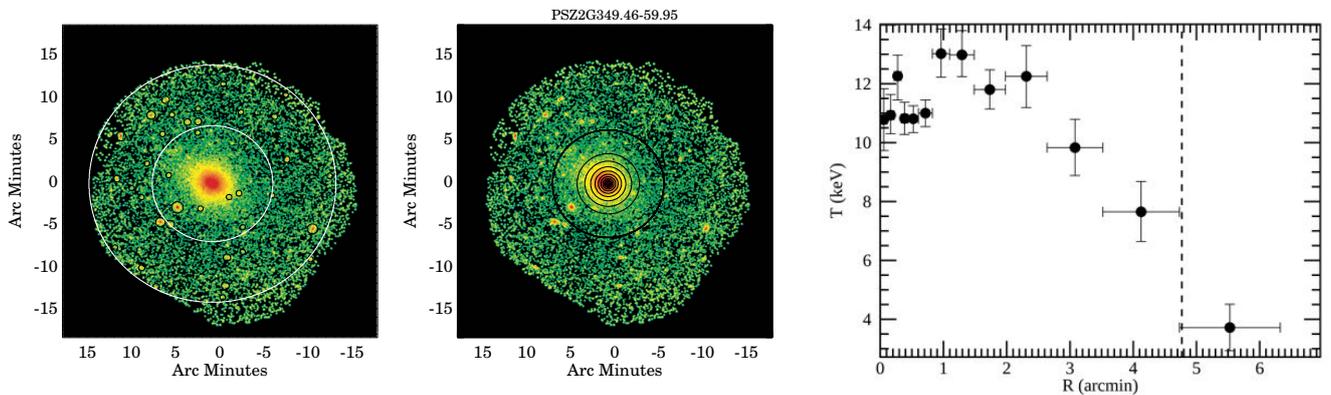

Fig. C.30: Same as Fig. C.1 for cluster PSZ2G349.46-59.95.




\end{document}